\documentclass[aps,prd,twocolumn,nofootinbib,eqsecnum,superscriptaddress,longbibliography]{revtex4-2}

\usepackage[centertags]{amsmath}
\usepackage{amssymb}
\usepackage{latexsym}
\usepackage{enumerate}
\usepackage{graphicx}
\usepackage{mathrsfs}
\usepackage[hidelinks]{hyperref}
\usepackage{stmaryrd}
\usepackage{import}
\usepackage{tensor}
\usepackage{xcolor}
\usepackage{bm}
\usepackage{multirow}
\usepackage{breakurl}
\usepackage{float}
\usepackage{lipsum}
\usepackage{siunitx}
\usepackage{comment}
\usepackage{mathtools}
\usepackage{autobreak}
\usepackage{subfig}
\usepackage{orcidlink}
\mathtoolsset{centercolon}
\usepackage{comment}
\DeclareMathOperator*{\FP}{FP}
\DeclareMathOperator*{\Pf}{Pf}

\newcommand{\vinf}{v_\infty}
\newcommand{\gammaE}{\gamma_{\rm E}}

\newcommand{\e}{\varepsilon}

\newcommand{\tb}{{\bar t}}
\newcommand{\rb}{{\bar r}}

\include{preamble}

\allowdisplaybreaks

\begin{document}

\title{Post-Minkowskian self-force in the low-velocity limit: scalar field scattering}

\begin{abstract}
     In this paper we present an approach to compute analytical post-Minkowskian corrections to unbound two-body scattering in the self-force formalism. Our method relies on a further low-velocity (post-Newtonian) expansion of the motion. We present a general strategy valid for gravitational and non-gravitational self-force, and we explicitly demonstrate our approach for a scalar charge scattering off a Schwarzschild black hole. We compare our results with recent calculations in [Barack et al., PRD 108, 024025 (2023)], showing complete agreement where appropriate and fixing undetermined scale factors in their calculation. Our results also extend their results by including in our dissipative sector the contributions from the flux into the black hole horizon.
\end{abstract}

\author{Donato Bini\,\orcidlink{0000-0002-5237-769X}}\affiliation{Istituto per le Applicazioni del Calcolo ``M. Picone,'' CNR, I-00185 Rome, Italy}\affiliation{INFN, Sezione di Roma III,  Via della Vasca Navale 84, 00146 Rome, Italy}

\author{Andrea Geralico\,\orcidlink{0000-0001-9608-3784}}\affiliation{Istituto per le Applicazioni del Calcolo ``M. Picone,'' CNR, I-00185 Rome, Italy}

\author{Chris Kavanagh\,\orcidlink{0000-0002-2874-9780}}
\affiliation{School of Mathematics \& Statistics, University College Dublin, Belfield, Dublin 4, Ireland, D04 V1W8}

\author{Adam Pound\,\orcidlink{0000-0001-9446-0638}}
\affiliation{School of Mathematical Sciences and STAG Research Centre, University of Southampton, Southampton, United Kingdom, SO17 1BJ}
\author{Davide Usseglio\orcidlink{0000-0003-2427-9547}}
\affiliation{Scuola Superiore Meridionale, Largo San Marcellino 10, 80138, Naples, Italy}
\affiliation{INFN, Sezione di Napoli, Complesso Universitario di Monte S. Angelo, Via Cintia Edificio 6, 80126, Naples, Italy}
\maketitle

\section{Introduction}

The detection of gravitational waves by the LIGO and Virgo \cite{LIGOScientific:2016aoc, LIGOScientific:2017vwq} collaborations has given to the scientific community a new way to investigate physics in the most extreme regimes, for instance around black holes. In the coming years, upgrades to ground-based gravitational wave detectors as well as the construction of new instruments, such as LISA  and the Einstein Telescope \cite{LISA:2017pwj, Punturo:2010zz}, will require more refined theoretical models of gravitational waveforms from compact binary coalescences. In order to achieve the required accuracy, various complementary techniques have been developed, including numerical relativity \cite{Pretorius:2005gq, Damour:2014afa,Hopper:2022rwo}, gravitational self-force theory~\cite{Mino:1996nk,Poisson:2011nh,Pound:2021qin}, 
and weak-field expansions such as post-Newtonian (PN) or post-Minkowskian (PM) theory~\cite{Blanchet:2013haa, Damour:2017zjx, Damour:2016gwp}. These direct approaches to producing gravitational waveforms by solving the Einstein field equations have also been combined using interpolating models such as the Effective-One-Body (EOB) approach~\cite{Buonanno:1998gg,Buonanno:2000ef} and the Phenom family of models~\cite{Ajith:2007kx, Santamaria:2010yb}. 

Historically,  waveform modelling has been devoted to the case of binaries on bound inspiralling trajectories resulting in merger events. But in the last few years, increasing attention has been paid to  scattering events. In this case the two objects are initially at a very large separation in a spacetime that is asymptotically Minkowski, then upon approaching each other on hyperbolic orbits, they interact gravitationally, emitting a burst of gravitational radiation before separating again. While such burst signals are potentially of astrophysical interest, the feasibility of detecting them is uncertain~\cite{Kocsis:2006hq,Codazzo:2022aqj,Mukherjee:2020hnm}.  
However, scattering events are of wide theoretical interest. In particular, weak-field, PM calculations of two-body scattering have proven to be a powerful tool to inform models of bound orbits, using a variety of new methods via EOB Hamiltonians~\cite{Bini:2019nra,Bini:2020wpo,Bini:2020nsb,Antonelli:2020aeb,Antonelli:2020ybz,Siemonsen:2019dsu}, mappings between bound and unbound observables~\cite{Kalin:2019rwq,Kalin:2019inp,Cho:2021arx}, or direct calculation of bound waveforms from scattering amplitudes~\cite{Adamo:2024oxy}.

The frontier of research describing weak-field scattering has been largely dominated by methods developed for high-energy particle physics, using  
effective field theory \cite{Kalin:2020mvi, Kalin:2020fhe, Kalin:2022hph, Dlapa:2021npj, Dlapa:2021vgp} and modern scattering amplitudes methods \cite{Kosower:2018adc,Bern:2019nnu, Bern:2019crd,Mogull:2020sak}.  
These methods 
have enabled calculations at orders that were unreachable with standard techniques in classical PM theory~\cite{Bern:2021yeh,Driesse:2024xad} and have been in agreement with results coming from PN theory~\cite{Bini:2021gat}. With these methods it is also possible to compute waveforms and compare them against standard PN results~\cite{Bini:2024rsy,Bautista:2021inx}.

Meanwhile, gravitational self-force theory, while traditionally developed to model extreme-mass-ratio events for the LISA mission \cite{LISAConsortiumWaveformWorkingGroup:2023arg}, has been showing strong potential for modelling binaries with mass ratios close to unity either directly \cite{Wardell:2021fyy} or by informing and calibrating EOB models \cite{Albertini:2022rfe,vandeMeent:2023ols}. Self-force theory, in the context of the two-body problem, is an expansion with respect to the mass ratio of the two black holes. This allows a formulation of the problem within the well-established framework of black hole perturbation theory~\cite{Pound:2021qin}, without resorting to approximations for low velocities, such as PN, or small gravitational potentials, such as PM. Put another way, at each order in the mass ratio, self-force theory is fully relativistic, containing all orders of the weak-field expansions (as well as nonperturbative effects that are never encountered at any order in a weak-field expansion~\cite{Long:2021ufh}). At first order in self-force theory, this has enabled a rich synergy with PN theory, as first-order self-force calculations can be written in a semi-closed form in terms of the so-called Mano-Suzuki-Takasugi (MST) solutions \cite{Mano:1996vt}, which have a natural  expansion in the PN limit. In the context of bound orbits, this MST method has been leveraged extensively to produce PN expansions to extremely high PN order at linear order in the mass ratio; see, e.g., \cite{Fujita:2012cm,Bini:2013rfa,Shah:2013uya,Kavanagh:2015lva,Forseth:2015oua,Isoyama:2021jjd}. PN-expanded self-force results also played an important role in completing the 4PN conservative binary dynamics~\cite{Damour:2014jta} and resolving discrepancies in 4PN results~\cite{Damour:2016abl}.

Synergies between self-force theory and PM theory are potentially even stronger but are relatively unexplored compared to PN-SF interactions. The potential for self-force theory to inform PM theory was highlighted in \cite{Damour:2019lcq}, where it was shown that linear-order self-force information can fully determine the PM Hamiltonian at 4PM precision, and second-order self-force information can fully determine the PM Hamiltonian through to 6PM. Even more recently, it was shown how reformulating scattering amplitudes approaches in a way analogous to the self-force picture could be a powerful new approach \cite{Cheung:2023lnj,Cheung:2024jpo,Kosmopoulos:2023bwc}. Motivated by this, and in the wake of progress in PM theory, the self-force community has also begun to tackle the scattering problem; see \cite{Hopper:2017qus,Hopper:2017iyq,Gralla:2021qaf} and \cite{Long:2021ufh,Barack:2022pde,Whittall:2023xjp,Long:2024ltn}. Notably, the recent work \cite{Barack:2023oqp} showed fruitful direct comparison between a numerical self-force computation and an analytical PM Amplitudes derivation of the scattering angle in the context of a scalar toy model. Even more recently, Ref.~\cite{Long:2024ltn} showed how analytical PM expressions can be accurately resummed in the strong field using only a single data point from numerical self-force calculations.

A key problem in dealing with scattering orbits is that self-force calculations often rely on frequency-domain approaches. These are useful because bound geodesic orbits in Kerr are at most tri-periodic, leading to fields with discrete Fourier spectra. 
However, for unbound orbits, the Fourier spectrum is continuous and integrals (inverse Fourier transforms) must be performed in order to return to the time domain.
 
In this paper, we give the first analytical self-force calculations for scattering orbits, providing a systematic approach that could be applied, in principle, at any PM order. Our method relies on a low-velocity PN expansion to exploit MST methods and to simplify the evaluation of integrals; this PN expansion can readily be taken to high order. 
To demonstrate our method we work to 5PM order and relative 4.5PN accuracy in the scalar model of Ref.~\cite{Barack:2023oqp}. Comparing our results against the ones obtained in Refs.~\cite{Barack:2023oqp} and~\cite{Jones:2023ugm}, we find, where applicable, complete agreement up to the PN order we compute. We also provide new information at 5PM order, specifically determining unknown coefficients in the tidal action of Ref.~\cite{Barack:2023oqp}. We validate our methodology with bounded orbits and we find perfect agreement with the literature both for the local and non-local sector.

Our discussion is organized as follows. 
In Sec. \ref{sec3} we describe geodesic scattering, giving explicit PN and PN-PM expansions for relevant geodesic quantities, and we then recall the Barack-Long formulation of self-force--corrected scattering. In Sec. \ref{sec4}
 we present our computational methods for generic spin $s$, highlighting how the solutions for the Teukolsky equation split into two contributions: local and non-local. The local terms, in PN theory, are associated with instantaneous effects in the gravitational potential, while the non-local terms are associated with hereditary effects such as tails and tails-of-tails. Notably, all our integrals are finite, but intermediate steps in the calculation of hereditary effects require a careful distributional treatment, which we compare to a Hadamard ``finite part'' regularization procedure, as described in Appendices~\ref{appendix_B} and~\ref{appendix_A}. \,
In Secs.~\ref{sec5} and~\ref{sec6}, after a brief review of the scalar problem, we  apply our methods to a scalar charge in a hyperbolic orbit around a Schwarzschild black hole, obtaining (i) the Detweiler-Whiting regular field~\cite{Detweiler:2002mi} from the solution of the Klein-Gordon equation, and (ii) the components of the self-force computed from that solution. Finally in Sec.~\ref{sec7} we present and discuss our results, comparing them with the known PM expressions obtained up to 4PM.

\textit{Conventions.} We use $G=c=1$, but we also use a placeholder $\eta=1/c$ to clarify for the reader the nature of some PN series. The signature for the Schwarzschild metric is $(-,+,+,+)$.
We will refer throughout our paper to a PN expansion when it is valid at any PM order, while a PN-PM expansion will be valid only up to a certain PM order.

\section{Equations of motion}\label{sec3}
\subsection{Geodesic motion} 

Consider the Schwarzschild spacetime with line element $ds^2= g_{\alpha\beta}dx^\alpha dx^\beta$ written in standard Schwarzschild coordinates $(t,r,\theta,\phi)$ as
\begin{equation}
    ds^2 =-f(r)dt^2 + \frac{1}{f(r)}dr^2 + r^2(d\theta^2+\sin^2\theta d\phi^2)\,,
\end{equation}
where $f=1-2M/r$ .
A generic timelike geodesic orbit on the equatorial plane ($\theta=\pi/2$) is written in parametric form as $x^\mu =x_p^{\mu}(\tau)$, with  proper time $\tau$ and 4-velocity $\hat u=\hat u^\alpha\partial_\alpha=\dot x^\alpha_p\partial_\alpha$, such that
\begin{equation} \label{eq:geodeq}
    \dot{t}_p = \frac{E}{f(r_p)}, \hspace{0.5cm} \dot{\phi}_p = \frac{L}{r^2_p}, \hspace{0.5cm} \dot{r}^2_p=E^2 - V(r_p;L),
\end{equation}
where the overdot indicates $d/d\tau$. $E$ and $L$ are respectively the orbit's (specific) energy and angular momentum, and the radial potential is
\begin{equation}
V(r;L)=f(r)\left(1+\frac{L^2}{r^2}\right).
\end{equation}
The equation for $\dot{r}_p$ is quadratic, and we can write
\begin{equation}\label{rdot}
    \dot{r}_p = \epsilon_r \sqrt{E^2-V(r_p;L)},
\end{equation}
where $\epsilon_r=\pm1$ is a sign indicator keeping track of increasing/decreasing radial coordinate. We note that Eq.~\eqref{eq:geodeq} follows solely from the normalization $\hat u_\alpha\hat u^\alpha=-1$ and the definitions $E=-\hat u_t$ and $L=\hat u_\phi$; this implies that the same equations will hold for self-forced orbits if we adopt analogous definitions of $E$ and $L$ in terms of the perturbed 4-velocity.

Hyperbolic orbits are characterized by $E>1$ and $L>L_{{\rm{crit}}}(E)$, with
\begin{equation}
    L_{\rm{crit}}(E) = \dfrac{M}{vE}\sqrt{\dfrac{27E^4 + 9\beta E^3 - 36E^2 - 8\beta E + 8}{2}},
\end{equation}
where $\beta:= \sqrt{9E^2-8}$; see \cite{Barack:2022pde,Whittall:2023xjp}. Alternatively, rather than using $E$ and $L$, we can characterize hyperbolic orbits using the impact parameter $b$ and initial velocity at infinity $v$. These are related to $E$ and $L$ by 
\begin{equation} \label{relation_EL_bv}
    E = \dfrac{1}{\sqrt{1-v^2}}, \hspace{0.5cm} L = bvE,
\end{equation}
and they range over $0<v<1$ and $b>b_{\rm crit}$, where $b_{\rm crit}=L_{\rm crit}/(v E)$. \footnote{Other authors instead denote the velocity $\vinf$, since it is the relative velocity of the small body as $r_p\rightarrow\infty$. We will drop the subscript here for simplicity. }

The solution to the radial equation is also often given in terms of the relativistic anomaly $\xi$:
\begin{equation} \label{radial_motion_wrt_xi}
    r_p(\xi) = \frac{M p}{1+e\cos\xi},
\end{equation}
where $p$ is the semi-latus rectum and $e$ the eccentricity. The sign difference between the ingoing and outgoing directions  is then encoded in $\xi$, which takes values in the interval $(-\xi_{\infty},\xi_\infty)$, where $\xi_\infty = \arccos(-1/e)$. The parameters $p$ and $e$ are related to $E$ and $L$ by 
\begin{align}\label{EL vs pe}
E^2 = \frac{(p-2)^2 -4e^2}{p(p-3-e^2)},\hspace{0.5cm}  L^2 = \frac{p^2 M^2}{p-3-e^2}\,,
\end{align}
and for hyperbolic orbits they range over $e>1$ and $p>6+2e$. The complete treatment of hyperbolic orbits in terms of all these different orbital parameters is discussed in Ref.~\cite{Barack:2022pde}.

There are several ways to solve the geodesic equation exactly. For instance by using the inverse radial distance $u=M/r$ and combining the equation for $\phi_p$ and $r_p$, one finds
\begin{equation}
\label{eqphidiugeo} 
\left(\frac{du}{d\phi}\right)^2=2(u-u_1)(u-u_2)(u-u_3)\,,
\end{equation}
where $u_1$, $u_2$ and $u_3$ are the ordered roots of the equation 
\begin{equation}
2u^3-u^2+\frac{2M^2u}{L^2}+\frac{M^2(E^2-1)}{L^2}=0\,,
\end{equation}
and we have omitted subscript $p$'s to avoid confusion with the subscripts 1,2,3. For scattering orbits one has $u_1<0<u\leq u_2<u_3$, with $u_2$ corresponding to the closest-approach distance $ r_{\rm min}=Mu_2$. The condition $u_2=u_3$ gives the critical value of the angular momentum corresponding to capture by the black hole for fixed energy value.

By asking that
\begin{equation}
    \phi(u_2)=0
\end{equation}
(i.e., vanishing $\phi$ at periastron), one can express the solution to Eq.~\eqref{eqphidiugeo} in terms of elliptic integrals as follows:
\begin{equation}
\phi(u)=\frac{\sqrt{2}}{\sqrt{u_3-u_1}}[K(n)-F(\psi,n)]\,,
\end{equation}
with 
\begin{equation}
n=\sqrt{\frac{u_2-u_1}{u_3-u_1}}\,,\qquad 
\psi=\sqrt{\frac{u-u_1}{u_2-u_1}}\,,
\end{equation}
where $F(\varphi,k)$ and $K(k)$ are the incomplete and complete elliptic integrals of the first kind, respectively, defined by
\begin{equation}
\label{ellipintdef}
F(\varphi,k)=\int_0^{\varphi}\frac{dz}{\sqrt{1-k^2\sin^2z}}\,,\quad
K(k)=F(\pi/2,k)\,.
\end{equation}

We are then able to formally define the so-called scattering angle $\chi$ as the total change of $\phi$ during the scattering process:
\begin{equation}\label{chigeo}
    \chi = 2\int_{0}^{u_2}du\dfrac{d\phi_p}{du}-\pi\,,
\end{equation}
where we have restored the subscript $p$ for clarity.

Despite the simplicity of the analytical solutions for the geodesic motion, we are actually interested in solving it perturbatively. In particular, we will propose two equivalent methods to get approximate solutions: in the first, one chooses $u$ as the independent variable and a rescaled version of $(E,L)$ as orbital parameters; in the second, one uses a rescaled time as the independent variable and uses $(b,v)$ as orbital parameters. 

More importantly, the key difference between these two methods is that they rely on different expansions: the first method follows a pure PN approximation for the orbit, while the second uses a large-$b$, PM expansion. We have checked that, if one applies to one of the two solutions the expansion done for the other, the two solutions are equivalent.

\subsubsection{PN expansion of the geodesic orbit}

Using $u$ as the parameter along the geodesic, we write Eq.~\eqref{eq:geodeq} as
\begin{align}
\frac{dt_p}{du}&=-\epsilon_r\frac{M\sqrt{1+2\bar E}}{ju^2(1-2u)}\left(2u^3-u^2+\frac{2u}{j^2}+\frac{2\bar E}{j^2}\right)^{-1/2}
\,,\nonumber\\
\frac{d\phi_p}{du}&=-\epsilon_r\left(2u^3-u^2+\frac{2u}{j^2}+\frac{2\bar E}{j^2}\right)^{-1/2}\,,\label{eq:dtdu,dphidu}
\end{align}
where we have introduced the dimensionless energy and angular momentum parameters $\bar E$ and $j$ defined by 
\begin{equation}
E=\sqrt{1+2\bar E},\qquad L=Mj,
\end{equation}
respectively.

We now wish to solve the geodesic equation in a low-velocity limit i.e. with a PN-like expansion. To do this we require scaling estimates for $\bar{E}$, $j$ and $u$. From their definitions and from Eq.~\eqref{relation_EL_bv}, we clearly have $\bar{E}\sim v^2$ and $j\sim \frac1v$. On the other hand, $u$ is constrained as $u<1/b_{\rm crit}\sim v^2$. To track these scalings we will introduce the standard PN small parameter $\eta=1/c$ (essentially restoring powers of $c$ and using it to track small velocities with respect to $c$) and make the replacements 
\begin{equation}\label{PN_rescalings}
    u\to u\eta^2, \hspace{0.4cm} j\to j/\eta, \hspace{0.4cm} \bar E\to \bar E\eta^2\,.
\end{equation}
Interestingly, these are the same scalings one uses in the PN description of bound orbits.

Solving Eq.~\eqref{eq:dtdu,dphidu} for small $\eta$, we obtain 
\begin{align}
\label{solgeoschw}
t_p(u)&=\epsilon_r M\biggl\{\frac{\alpha j}{u}{\mathcal S}(\alpha,j;u)
-\alpha^3j^3{\mathcal L}(\alpha,j;u)\nonumber\\*
&\quad\left.
+\eta^2\left[
\frac{1+\alpha^2+2\alpha^2j^2(\alpha^2+2)u+j^4\alpha^2(\alpha^2-1)u^2}{2u\alpha j(\alpha^2+1){\mathcal S}(\alpha,j;u)}
\right. \right.\nonumber\\*
&  \qquad\quad\left.+\frac32\alpha j{\mathcal L}(\alpha,j;u)\right]
+O(\eta^4)
\biggr\}
\,,\\
\phi_p(u)&=\epsilon_r\biggl\{B(\alpha,j;u)
+\frac{\eta^2}{j^2}\biggl[3B(\alpha,j;u)\nonumber \\*
&\quad+\frac{2+3\alpha^2+u\alpha^2(6\alpha^2+5)j^2-u^2\alpha^2(\alpha^2+1)j^4}{\alpha(\alpha^2+1){\mathcal S}(\alpha,j;u)}
\biggr] \nonumber \\*
&\quad +O(\eta^4)\label{phip(u) PN}
\biggr\}\,,
\end{align}
where initial conditions have been chosen as $t_p(u_2)=0=\phi_p(u_2)$ at the closest approach, with 
\begin{equation}
u_2(\bar E,j)=\left(\!1\!+\!\sqrt{1+2\bar Ej^2}\right)/j^2\equiv\left(1\!+\!\sqrt{1+1/\alpha^2}\right)/j^2
\end{equation}
at the lowest order, and
\begin{align}
{\mathcal S}(\alpha,j;u)&=\sqrt{1+2 u j^2\alpha^2-u^2 j^4\alpha^2}\,,\nonumber\\
{\mathcal L}(\alpha,j;u)&=\ln\left(\frac{1+u j^2\alpha^2+{\mathcal S}(\alpha,j;u)}{j^2u\alpha\sqrt{\alpha^2+1}}\right)
\,,\nonumber\\
B(\alpha,j;u)&=\frac{\pi}{2}+\arctan\left(\frac{(1-u j^2)\alpha}{{\mathcal S}(\alpha,j;u)}\right)\,.
\end{align}
Note that in the above expressions $\alpha j$ and $uj^2$ are Newtonian quantities, with 
\begin{equation}
\alpha\equiv1/\sqrt{2\bar Ej^2}= 1/(p_\infty j),
\end{equation}
where we defined $p_\infty=\sqrt{2\bar E}$. 

The first few terms of the PN expansion of the geodesic scattering angle \eqref{chigeo} are given by 
\begin{align}
\frac{\chi}{2}&=B(\alpha)-\frac{\pi}{2}
+\frac{\eta^2}{j^2}\left[3B(\alpha)+\frac{2+3\alpha^2}{\alpha(\alpha^2+1)}
\right]
+O(\eta^4)
\,,
\end{align}
where $B(\alpha)=B(\alpha,j;0)=\frac{\pi}{2}+\arctan(\alpha)$.
Taking the series expansion for large values of the angular momentum parameter $j$ gives the PM expansion at each PN order,
\begin{multline}\label{chigeoPN}
\chi=\frac{2}{j}\left(\frac{1}{p_\infty}
+2p_\infty\eta^2\right)\\
+\frac{3\pi\eta^2}{j^2}\left(1+\frac{5}{4}p_\infty^2\eta^2\right)
+O\!\left(\frac{1}{j^3}\right).
\end{multline}

\subsubsection{PM expansion of the geodesic orbit}

To obtain the PM expansion, without a PN expansion, we restart our discussion by recalling Eq.~\eqref{radial_motion_wrt_xi} and rewriting Eq.~\eqref{eq:geodeq} in terms of the relativistic anomaly, keeping $(p,e)$ as parameters that specify the orbit:
\begin{align}
    \label{eq:dtdchi}\frac{dt_p}{d\xi} &= \frac{ M p^2}{(p-2-2e\cos\xi)(1+e\cos\xi)^2} \sqrt{\frac{(p-2)^2-4e^2}{p-6-2e\cos\xi}},
    \\ \label{eq:dphidchi}\frac{d\phi_p}{d\xi} &= \sqrt{\frac{p}{p-6-2e\cos\xi}}.
\end{align}
We change the parametrization from $(p,e)$ to $(b,v)$ by recalling the relations~\eqref{EL vs pe}, which can be inverted and then combined with Eq. \eqref{relation_EL_bv}.

Expanding Eq.~(\ref{eq:dtdchi}) for large $b$, we get
\begin{align} \label{sol:chifuncofT}
    \xi(\bar{t})& = \arctan \bar{t} +  \dfrac{M}{bv^2} \left[\frac{\bar{t}}{\sqrt{1+\bar{t}^2}}+\frac{\left(1-3 v^2\right) {\rm arcsinh}\, \bar{t}}{1+\bar{t}^2}\right]\nonumber \\
    &\quad + \frac{M^2}{(1+\bar{t}^2)b^2v^4}\Biggl[2 \bar{t} v^2+\frac{\left(2-9 v^2+9 v^4\right) {\rm arcsinh}\,\bar{t}}{\sqrt{1+\bar{t}^2}}\nonumber\\ 
    &\qquad\qquad -\frac{\bar{t} \left(1-3
   v^2\right)^2 {\rm arcsinh}^2\,\bar{t}}{1+\bar{t}^2}-\frac{15}{2} v^4 \arctan \bar{t}\Biggr] \nonumber \\
   &\quad+ O(b^{-3}),
   \end{align}
where 
\begin{equation}
\bar{t}=vt/b 
\end{equation}
is a rescaled time.

By substituting this into Eqs.~\eqref{radial_motion_wrt_xi} and~(\ref{eq:dphidchi}), we find
\begin{align}
r_p(\bar{t})&= b \sqrt{1+\bar{t}^2}-\frac{M}{v^2}+\frac{M\left(\bar{t}-3 \tb v^2\right) {\rm arcsinh}\,\bar{t}}{\sqrt{1+\bar{t}^2} v^2} \nonumber \\
&\quad + \dfrac{M^2}{b}\left[\frac{1-4 v^2}{2 \sqrt{1+\bar{t}^2} v^4}+\frac{\bar{t} \left(1-3 v^2\right)^2 {\rm arcsinh}\,\bar{t}}{\left(1+\bar{t}^2\right) v^4}\right.\nonumber \\
&\qquad\quad  \left.+\frac{\left(1-3 v^2\right)^2
   {\rm arcsinh}^2\,\bar{t}}{2 \left(1+\bar{t}^2\right)^{3/2} v^4}-\frac{15 \bar{t} \arctan\,\bar{t}}{2 \sqrt{1+\bar{t}^2}}\right] \nonumber\\
   &\quad + O(b^{-2}), \\
   \phi_p(\bar{t}) & = \arctan \bar{t} + \frac{M}{bv^2\sqrt{1+\bar{t}^2}}\Biggl[ \tb \, \left(1+v^2\right)\nonumber \\*
   &\qquad\qquad \qquad+\frac{\left(1-3 v^2\right) {\rm arcsinh}\,\bar{t}}{\sqrt{1+\bar{t}^2}}\Biggr] \nonumber \\*
   &\quad + \frac{M^2}{4b^2v^4
   \left(1+\bar{t}^2\right)}\Biggl\{3 \bar{t} v^2 \left(4+v^2\right)\nonumber \\*
   &\qquad  +3 v^2 \left[4-9 v^2+\bar{t}^2 \left(4+v^2\right)\right] \nonumber \\*
   &\qquad+ \frac{8 \left(1-4 v^2+3 v^4\right) {\rm arcsinh}\,\bar{t}}{\sqrt{1+\bar{t}^2}}\nonumber \\*
   &\qquad-\frac{4 \bar{t} \left(1-3 v^2\right)^2 {\rm arcsinh}^2\,\bar{t}}{1+\bar{t}^2}\Biggr\} + O(b^{-3}).
\end{align}
We can compare this expression for $\phi_p$ with the analogous result~\eqref{phip(u) PN}. We find that they are in agreement. 

Finally, using our result for $\phi_p$, we find the PM expansion of the geodesic scattering angle~\eqref{chigeo} reads
\begin{align} \nonumber
    \chi &= \dfrac{2 M}{bv^2} (1+v^2) + \dfrac{3M^2\pi}{4b^2v^4}(4+v^2) \\
    &\quad -\frac{2M^3}{3b^3v^6} \left[1-5 v^2
   \left(3+9 v^2+v^4\right)\right] + O(b^{-4}),
\end{align}
which can be shown to agree with Eq. \eqref{chigeoPN}.

\subsection{Self-forced motion}\label{Sec3partb}

A detailed discussion of how self-forces correct the geodesic scattering motion can be found in Ref.~\cite{Barack:2022pde}. Here we present an alternative formulation that will be particularly useful in our later calculations.

A point particle that scatters off the Schwarzschild black hole feels an acceleration due to its self-force, obeying the equation of motion
\begin{equation}
    u^\beta\nabla_\beta u_\alpha = F_\alpha.
\end{equation}

Here $u^\alpha$ is the particle's four-velocity, $F_\alpha$ is the self-force per unit mass, and $\nabla_\beta$ is the covariant derivative compatible with the Schwarzschild metric $g_{\alpha\beta}$.For the case of a scalar charged particle that we consider in later sections, the scalar self-force scales as $F_\alpha\propto q^2/(\mu M^2)$, where $q$ and $\mu$ are the particle's charge and mass, respectively. For a gravitating point mass, the gravitational self-force scales as $F_\alpha\propto \mu/M^2$. In this section we leave the force unspecified; we only assume it is small. Concretely, we write 
\begin{equation}
    F_\alpha = O(\e/M),
\end{equation}
with 
\begin{equation}
\label{eq:scalareps}
\e=q^2/(\mu M)\ll1 
\end{equation}
or $\e=\mu/M\ll1$, as appropriate. We then expand all quantities to linear order in $\e$. From Section \ref{sec6} onwards we will explicitly use the definition \eqref{eq:scalareps}.

When expanding for small $\e$, we must specify what quantities are being held fixed. We specifically expand for small $\e$ at fixed values of the initial energy $E_-$ and angular momentum $L_-$ that the particle has when it begins its orbit from infinite distance; we use a minus sign to indicate that these are the values in the infinite past. Put another way, we expand the perturbed orbit around a geodesic that has the same initial energy and angular momentum. The self-force correction to the scattering angle is then obtained as a function of $E_-$ and $L_-$. Equivalently, we can write the self-force correction as a function of $b$ and $v$, defining those quantities from $E_-$ and $L_-$ using Eq.~\eqref{relation_EL_bv}.

From the $t$ and $\phi$ components of the force, we can obtain this linear correction to the scattering angle via a formula of the form
\begin{equation}\label{scattering_angle_corrections}
    \delta\chi = \int_{-\infty}^{\infty} d\tau \left\{ \mathcal{G}_E(\tau) F_t(\tau) + \mathcal{G}_L(\tau) F_\phi(\tau) \right\} ,
\end{equation}
where $\mathcal{G}_{E/L}$ are two functions of geodesic quantities, derived in \cite{Barack:2022pde}. 
However, we will derive an alternative, equivalent expression that we find more convenient. In the following, we use hats to denote zeroth-order, background quantities and $\delta$ to denote corrections that are linear in $\e$.

We start by defining the accelerated orbit's energy and angular momentum as $E=-u_t$ and $L=u_\phi$, which evolve according to 
\begin{equation}\label{dEdt =-F}
    \frac{dE}{d\tau} = -F_t,\quad \frac{dL}{d\tau} = F_\phi.
\end{equation}
These $E$ and $L$ are not the conserved energy and angular momentum that would be defined in the conservative sector of the self-forced dynamics. However, they are convenient because (i) as mentioned below Eq.~\eqref{rdot}, they allow us to still make use of Eq.~\eqref{eq:geodeq} for $\dot x^\alpha_p$, and (ii) they can be immediately expressed in terms of the self-force. We express them as functions of $\tau$ as 
\begin{align}
\label{eq:deltaELdef}
E(\tau)
&=E_--\int_{-\infty}^0 F_t d\tau-\int_0^\tau F_t d\tau\nonumber\\*
&\equiv E_-+\delta E_0+\delta \tilde E(\tau)
\,,\nonumber\\
L(\tau)
&=L_-+\int_{-\infty}^0 F_\phi d\tau+\int_0^\tau F_\phi d\tau\nonumber\\*
&\equiv L_-+\delta L_0+\delta \tilde L(\tau)
\,.
\end{align}
The reason for splitting each correction into two terms will become clear below.

At this point, it is convenient to use $r$ as a parameter along the orbit.
In order to compute integrals along the perturbed orbit one has then to consider the incoming ($\hat u^r<0$) and outgoing ($\hat u^r>0$) branches separately, with $r$ decreasing from infinity ($\tau \to-\infty$) up to a minimum value $r_{\rm min}=\hat r_{\rm min}+\delta r_{\rm min}$ ($\tau=\tau_0=0$), then increasing again to infinity ($\tau \to\infty$).
Let us denote by a label $\pm$ the values of the various quantities corresponding to the incoming ($-$) and outgoing ($+$) branch, respectively, so that we formally have a slight redefinition of the scattering of Eq. \eqref{chigeo} as
\begin{equation}\label{chimodified}
    \chi=\sum_\pm\int_{r_{\rm min}}^\infty dr \left(\frac{d\phi_p}{dr}\right)^\pm-\pi\,.
\end{equation}
The lower limit of integration is expanded at fixed $(E_-,L_-)$, while  the integrand is expanded to linear order in $\e$ at fixed $r$ (as well as fixed $E_-$ and $L_-$), meaning $\phi_p(r,\e)=\hat\phi(r)+\delta\phi(r,\e)$. Explicitly, 
\begin{equation}\label{chi_full} 
    \chi =  \sum_\pm\int_{\hat r_{\rm min} + \delta r_{\rm{min}}}^\infty dr \left(\frac{d\hat\phi}{dr} +  \dfrac{d\delta\phi}{dr}\right)^\pm-\pi\, ,
\end{equation}
which can be written to linear order as
\begin{equation}\label{splitted_chi}
\chi=\hat\chi+\delta\hat\chi+\delta\tilde{\chi}\,,
\end{equation}
where  
\begin{align}
\delta\hat\chi &= \delta r_{\rm min}\frac{\partial}{\partial\hat r_{\rm min}}\sum_\pm\int^\infty_{\hat r_{\rm min}} dr \left(\frac{d\hat\phi}{dr}\right)^\pm \nonumber\\*
&= \delta r_{\rm min}\frac{\partial\hat\chi}{\partial\hat r_{\rm min}}\label{dchihat}
\end{align}
is the $O(\e)$ correction due to the shift $\delta r_{\rm min}$, and
\begin{equation}\label{dchitilde}
    \delta\tilde{\chi} = \sum_\pm\int_{\hat r_{\rm min}}^\infty dr \left(\dfrac{d\delta\phi}{dr}\right)^\pm.
\end{equation}

To evaluate Eq.~\eqref{dchihat}, we recall that Eq.~\eqref{rdot} remains valid for the self-forced motion. The closest approach of the self-forced orbit, which we have defined to occur at $\tau=0$, therefore satisfies
\begin{equation}\label{drmin}
    (E_-+\delta E_0)^2 = V(\hat r_{\rm min}+\delta r_{\rm min};L_-+\delta L_0);
\end{equation}
this was one reason for isolating the corrections to $E_-$ and $L_-$ at periastron in Eq.~\eqref{eq:deltaELdef}. 
Equation~\eqref{drmin} is identical to the equation for the closest approach of a geodesic with energy $E_-+\delta E_0$ and angular momentum $L_-+\delta L_0$. Therefore
\begin{equation}
    \delta r_{\rm min} = \frac{\partial \hat r_{\rm min}}{\partial E_-}\delta E_0 + \frac{\partial \hat r_{\rm min}}{\partial L_-}\delta L_0.
\end{equation}
Substituting this into Eq.~\eqref{dchihat} and appealing to the chain rule, we obtain the simple result
\begin{equation}
\delta\hat\chi=\frac{\partial\hat\chi}{\partial E_-}\delta E_0+\frac{\partial\hat\chi}{\partial L_-}\delta L_0\,.
\end{equation}

We next turn to the correction $\delta\tilde\chi$ in Eq.~\eqref{dchitilde}. To express the integrand in terms of energy and angular momentum, we write $\frac{d\phi_p}{dr}=\frac{d\phi_p/d\tau}{dr_p/d\tau}$. Expanded to linear order in $\e$ at fixed $r$, this becomes
\begin{align}\label{dphidr}
\frac{d\phi_p}{dr}&= \frac{\hat u^\phi+\delta u^\phi}{\hat u^r+\delta u^r}\nonumber\\*
&= \frac{\hat u^\phi}{\hat u^r}\left(1+\frac{\delta u^\phi}{\hat u^\phi}-\frac{\delta u^r} {\hat u^r}\right).  
\end{align}
Comparing to $\frac{d\phi_p}{dr}=\frac{d\hat \phi}{dr}+\frac{d\,\delta\phi}{dr}$, we see
\begin{equation}
   \frac{d\,\delta\phi}{dr}=\frac{\hat u^\phi}{\hat u^r}\left(\frac{\delta u^\phi}{\hat u^\phi}-\frac{\delta u^r} {\hat u^r}\right).\label{ddphi/dr} 
\end{equation}
We next express the right-hand side in terms of $\delta E$ and $\delta L$. All perturbations to the 4-velocity can be obtained from $\delta u_t =  - \delta E(r)$ and $\delta u_\phi = \delta L(r)$ together with the normalization $g_{\mu\nu}u^\mu u^\nu=-1$, where
\begin{equation}
\delta E = \delta E_0 + \delta\tilde E,
\end{equation}
and analogously for $\delta L$. Since we are linearizing in $\e$ at fixed $r$,  and since $g_{\mu\nu}$ is a function of $r$ only, the linear term in the normalization condition becomes $g_{\mu\nu}\delta u^\mu \hat{u}^\nu = 0$, which we use to obtain $\delta u_r$ in terms of $\delta E$ and $\delta L$.  Substituting these results into Eq.~\eqref{ddphi/dr}, we get
\begin{equation}
\label{eqdeltaphi}
\frac{d\,\delta\phi}{dr} =a_E(r) \delta E(r)+a_L(r) \delta L(r)\,,
\end{equation}
with
\begin{equation}
a_E(r)=-\frac{L_- E_-}{r^2(\hat u^r)^3 }
\,,\qquad
a_L(r)=\frac{E_-^2-f(r)}{r^2(\hat u^r)^3 }\,.
\end{equation}

Equation~\eqref{dchitilde} is then
\begin{equation}
\delta\tilde{\chi}=\sum_\pm\int_{\hat r_{\rm min}}^\infty dr \left[a_E^\pm(r)\delta E^\pm(r)+a_L^\pm(r)\delta L^\pm(r)\right]\,.
\end{equation}
By noting that 
\begin{equation}
a_E^-(r)=-a_E^+(r)\,,\qquad
a_L^-(r)=-a_L^+(r)\,,
\end{equation}
we see that $\delta E_0$ and $\delta L_0$ vanish from the integral; this was another reason for isolating these terms in Eq.~\eqref{eq:deltaELdef}. We are then left with
\begin{align}
\label{hatdeltaphifin}
\delta\tilde{\chi}&=\int_{\hat r_{\rm min}}^{\infty} dr \left\{a_E^+(r)\left[\delta \tilde E^+(r)-\delta \tilde E^-(r)\right]\right.\nonumber \\
& \left. \hspace{1.7cm} +a_L^+(r)\left[\delta \tilde L^+(r)-\delta \tilde L^-(r)\right]\right\}\nonumber\\
&=\int_{\hat r_{\rm min}}^{\infty} dr \left\{
-a_E^+(r)\int_{\hat r_{\rm min}}^r\frac{dr}{\hat u^r}\left[F_t^+(r)-F_t^-(r)\right]
\right. \nonumber \\
& \left. \hspace{1.7cm} +a_L^+(r)\int_{\hat r_{\rm min}}^r\frac{dr}{\hat u^r}\left[F_\phi^+(r)-F_\phi^-(r)\right]
\right\}\,,
\end{align}
having used
\begin{equation}
 \delta \tilde E^\pm(r)=-\int_{\hat r_{\rm min}}^r\frac{dr}{\hat u^r}F_t^\pm(r)\,,\quad 
\delta \tilde L^\pm(r)=\int_{\hat r_{\rm min}}^r\frac{dr}{\hat u^r}F_\phi^\pm(r)\,.
\end{equation}

At this point we can split both $\delta\hat{\chi}$ and $\delta\tilde{\chi}$ into their conservative and dissipative parts by firstly splitting the components of the forces as follows: 
\begin{align}
F_\alpha^{\rm cons\,+}(r)&=\frac12[F_\alpha^+(r)-F_\alpha^-(r)]
=-F_\alpha^{\rm cons\,-}(r)
\,,\nonumber\\
F_\alpha^{\rm diss\,+}(r)&=\frac12[F_\alpha^+(r)+F_\alpha^-(r)]
=F_\alpha^{\rm diss\,-}(r)\,,\label{Fdiss}
\end{align}
for $\alpha=t,\phi$, so that 
\begin{align}\nonumber
\delta\tilde\chi^{\rm cons}&=-2\int_{\hat r_{\rm min}}^{\infty} dr \left[
a_E^+(r)\int_{\hat r_{\rm min}}^r\frac{dr}{\hat u^r}F_t^{\rm cons\,+}
\right.\\*
&\left. \hspace{2.2cm} -a_L^+(r)\int_{\hat r_{\rm min}}^r\frac{dr}{\hat u^r}F_\phi^{\rm cons\,+}
\right].
\end{align}
Interestingly, $\delta\tilde\chi^{\rm diss}=0$ because of the symmetries of $F_{\alpha}^{\rm{diss}}$. As a consequence, the dissipative correction to the scattering angle is simply given by 
\begin{equation}\label{delta_chi_diss}
\delta\chi^{\rm diss}=\delta\hat\chi^{\rm diss}=-\frac12\left(\frac{\partial\hat\chi}{\partial E_-}E_{\rm rad}+\frac{\partial\hat\chi}{\partial L_-}L_{\rm rad}\right),
\end{equation}
(a reminder of the linear response formula, \cite{Damour:2020tta,Bini:2012ji}). Here $E_{\rm rad}$ and $L_{\rm rad}$ are the total radiated energy and angular momentum, which are related to the integrated dissipative self-force by
\begin{equation}\label{Erad and Lrad}
E_{\rm rad}=-2\delta E_0^{\rm diss},\quad L_{\rm rad}=-2\delta L_0^{\rm diss},
\end{equation}
respectively. Recall that our local definitions of orbital $E$ and $L$ are not the conserved energy and angular momentum of the perturbed orbit in the conservative sector, as emphasized below Eq.~\eqref{dEdt =-F}. As a consequence, the conservative self-force does cause $E$ and $L$ to change with time. However, this change averages to zero over the complete orbit, and only the dissipative self-force is related to the emitted fluxes of energy and angular momentum in scalar radiation.

It is also worthwhile to mention that we can equivalently compute  the scattering angle through a time parameterization in Eq. \eqref{chi_full}. 

\section{Field equations and PM expansion: general set up} \label{sec4}

We wish to describe our system in a weak-field (PM) and low-velocity (PN) limit. Specifically, we will impose 
\begin{align}
\label{eq:pmPNlimit}
b\gg M,\qquad v\ll1.
\end{align}
Our ultimate goal will be to calculate the scattering angle~\eqref{splitted_chi} in this limit. Doing so requires computing the self-force, which in turn requires knowledge of the field near the particle's worldline.  We thus introduce rescaled variables to parameterise both the worldline coordinates and field points as follows:
\begin{subequations}\label{eq:pmPNscaling}
\begin{align}
\bar{t}&=\frac{v t}{b}, \\
\bar{r}&=\frac{r}{b}, \\
\bar{r}_p(\bar{t})&=\frac{r_p(t(\bar{t}))}{b}.
\end{align}
\end{subequations}
Following the rescaling of the time variable, it is also natural to introduce a rescaled frequency variable for the Fourier transform of the field $\bar{\omega}=b \omega/v$. All of the barred variables will henceforth be formally treated as $O(1)$ quantities.

\subsection{PM+PN field modes: Generic treatment}

Perturbations to astrophysical black hole spacetimes are generically treated using the spin-$s$ Teukolsky equation, written schematically as ${}_s{\cal O}\,{}_s\psi = {}_s T$, where ${}_s{\cal O}$ is a second-order partial differential operator~\cite{Pound:2021qin}. For example, $s=0$ gives the massless Klein-Gordon (KG) equation with a scalar charge distribution source:
\begin{equation}\label{KG eqn}
    \Box\,{}_0\psi = -4\pi\rho,
\end{equation}
where $\Box=g^{\alpha\beta}\nabla_\alpha\nabla_\beta$ is the d'Alembertian and ${}_0T=-4\pi r^2 \rho$. In our case $\rho$ will be a point-particle (scalar) charge density, \begin{align}
\rho(x^\mu) &= q\int(-g)^{-1/2}\delta^4(x^\mu - x^\mu_p(\tau))d\tau\nonumber\\
 &= \frac{q}{u^t r^2}\delta(r-r_p(t))\delta(\theta-\pi/2)\delta(\phi-\phi_p(t)),\label{rho deltas}
\end{align}
where $g=-r^4\sin^2\theta$ is the determinant of the metric. In the gravitational case, one would instead solve the $|s|=2$ Teukolsky equation for the maximal spin-weight Weyl scalars. In that case, the source ${}_s T$ is constructed from a second-order linear operator acting on a point-mass stress-energy tensor $T^{\alpha\beta}$.

Our discussion in this section will remain agnostic about the spin weight, meaning it applies to both the scalar and the gravitational case. In either case, we seek solutions to the spin-$s$ Teukolsky equation in the PN-PM limit after the rescalings~\eqref{eq:pmPNscaling}. 

Restricting to Schwarzschild spacetime, we write the Teukolsky solution in the time domain as 
\begin{equation}\label{eq:Teukolsky solution}
    {}_s\psi(t,r,\theta,\phi) =\sum_{l=|s|}^\infty\sum_{m=-l}^{l}{}_s\psi_{\ell m}(t,r){}_s Y_{lm}(\theta,\phi),
\end{equation}
where the modes are expressed as an integral over a retarded Green function ${}_s G_{\ell m \omega}$:
\begin{align}\label{eq:Teukolsky solution}
    {}_s\psi_{\ell m}(t,r) &= \nonumber\\*
    &\frac{1}{2\pi}\int_{-\infty}^\infty d\omega\, e^{-i\omega t} \int dr'\, {}_s G_{\ell m \omega}(r,r')\, {}_s 
T_{\ell m\omega}(r').
\end{align}
This formula is standard but will be developed from scratch in the $s=0$ case in Sec.~\ref{sec5}. Using 
\begin{equation}
    {}_s T_{\ell m\omega}(r') = \int_{-\infty}^\infty dt' e^{i\omega t'}{}_s T_{\ell m}(t', r')
\end{equation}
we rearrange Eq.~\eqref{eq:Teukolsky solution} as
\begin{align}\label{eq:Teukolsky solution v2}
    {}_s\psi_{\ell m} &= \frac{1}{2\pi}\int dr'\int_{-\infty}^\infty d\omega\,  {}_sG_{\ell m \omega}(r,r') \nonumber \\*
    &\qquad \times\int_{-\infty}^\infty dt' e^{i\omega (t'-t)}{}_s T_{\ell m}(t', r').
\end{align}
Using our PM-PN scalings, we can rewrite this with all variables changed to their barred counterparts. 

In the next section we will explicitly give the form of the retarded Green function modes appearing here, but for now we will focus on its structure as a function of the Fourier frequency. In the PM-PN expansion,
the Green function has the structure
\begin{align}\label{GF_log_structure}
    {}_s G_{\ell m \omega}(\rb,\rb')&=\frac{1}{b}\sum_{j=0}^{\infty}\sum_{k=0}^1\,  {}_s G_{\ell m}^{(j,k)}(\rb,\rb')v^j\bar{\omega}^{j}\log^k(-i\bar{\omega}),
\end{align}
where the coefficients ${}_s G_{\ell m}^{(j,k)}(\rb,\rb')$ are dimensionless. As a consequence of our use of scaled coordinates, ${}_s G_{\ell m \omega}(\rb,\rb')$ depends on $b$ starting at 1PM, and the coefficients $ {}_s G_{\ell m}^{(j,k)}(\rb,\rb')$ here are expansions in $1/b$.
For the non-logarithmic contributions (i.e., $k=0$), ${}_s G_{\ell m}^{(j,0)}(\rb,\rb')$ starts at order $1/b^0$ for all values of $j$, $l$, and $m$. Meanwhile, for $k\neq0$, the first logarithm appears at $j=2$ for $s=l=0$, for example, and at $j=4$ for $s=l=1$. If we had not adopted scaled coordinates, this behavior would instead arise after integrating against the source and evaluating the field near the particle, such that the arguments of the Green function would inherit the behavior of the orbit.

In the pure PN expansion one finds the same structure in $\omega$ for the Green function (with no associated expansion in $1/b$). So we will omit the details here and continue the discussion for the remainder of this formalism section. We will return to the PN approach when explicitly solving the field equations in Sec.~\ref{sec5}.

In the following, we will refer to the non-logarithmic contributions as \textit{local terms} and the logarithmic terms as the \textit{non-local pieces}.   

\subsection{Local field contributions}

The non-logarithmic contribution to the field modes is given as
\begin{align}
    {}_s\psi^{\rm L}_{\ell m} &= \sum_j\frac{v^j}{2\pi}\int d\rb'\int_{-\infty}^\infty d\tb'\,  {}_sG^{(j,0)}_{\ell m }(\rb,\rb') {}_s T_{\ell m}(\tb', \rb')\nonumber \\
    &\quad  \times\int_{-\infty}^\infty d\bar{\omega} e^{i\bar{\omega} (\tb'-\tb)} \bar{\omega}^j.
\end{align}
Using the identity 
\begin{equation}
    \int_{-\infty}^\infty d\omega\, e^{i\omega y} \omega^j = \frac{2\pi}{i^j} \delta^{(j)}(y),
\end{equation}
we find
\begin{align}
\label{Eq:non-log Teuk solution}
    {}_s\psi^{\rm L}_{\ell m} &= \sum_j i^jv^j\frac{\partial^j}{\partial \tb^j}\int d\rb'\,  {}_sG^{(j,0)}_{\ell m }(\rb,\rb') {}_s T_{\ell m}(\tb, \rb').
\end{align}

Equation~\eqref{Eq:non-log Teuk solution} applies for any source that scales suitably with $b$ (in congruence with our use of scaled coordinates). But we have a final simplification because for a point-particle source, ${}_s T_{\ell m}(\bar t, \rb')$ is a sum of terms proportional to $\delta[\rb'-\rb_p(\bar t)]$ and its derivatives. These delta functions allow us to evaluate the integral over $r'$, reducing ${}_s\psi_{\ell m}(t,r)$ to a simple sum of analytically known terms. For example, for $s=0$ we can expand the source~\eqref{rho deltas} in spherical harmonics as 
\begin{multline}
\rho(x^\mu)  = \frac{q}{u^t \bar r^2 b^3}\delta(\bar r-\bar r_p(\bar t))\\
\times\sum_{lm}Y^*_{lm}(\pi/2,0)Y_{lm}(\theta,\phi)e^{-i m\phi_p(\bar t)},\label{rho lm expansion}
\end{multline}
where we used the identity
\begin{equation}
    \delta(\theta-\pi/2)\delta(\phi-\phi(t)) = \sum_{l,m}Y_{lm}(\theta,\phi)Y^*_{lm}(\pi/2,\phi(t)).
\end{equation}
Equation~\eqref{Eq:non-log Teuk solution} then becomes\footnote{The apparent conflict between the powers of $\bar r$ and $b$ in Eqs.~\eqref{rho lm expansion} and~\eqref{Eq:lowvspin0} comes from the fact that $_0T=-4\pi r^2\rho = -4\pi b^2\bar r^2\rho$.} 
\begin{align}
    {}_0\psi^{{\rm{L}}}_{\ell m} &= \frac{q}{b} \sum_{j\geq0}i^j v^j\frac{\partial^j}{\partial\tb^j}\left\{{}_0{ G}^{(j,0)}_{\ell m}[\bar r,\bar r_p(\bar t)]\frac{e^{-im\phi_p( \bar t)}}{u^t(\bar t)}\right\} \nonumber \\
    &\quad \times Y^*_{\ell m}(\pi/2,0).
    \label{Eq:lowvspin0}
\end{align}

\subsection{Non-local contribution}\label{nonlocal contributions generic}

The first logarithmic-in-frequency term we encounter is 
\begin{align} \nonumber
{}_s\psi^{\rm NL}_{\ell m}  &= \sum_j\frac{v^j}{2\pi}\int d\rb'\,G_{\ell m}^{(j,1)}(\rb,\rb') \\
&\times\int_{-\infty}^\infty dy {}_s T_{\ell m}(\tb+y, \rb')\int_{-\infty}^\infty d\bar{\omega}\, \bar{\omega}^j\log(-i\bar{\omega}) e^{i\bar{\omega} y},
\end{align}
where $y=\tb'-\tb$. The logarithmic inverse Fourier transform (IFT) here must be treated delicately. Firstly, care must be taken with the branch choice of the logarithm for positive and negative frequencies. Secondly, the IFT  must be handled as a distribution in order to obtain a finite value for the field after the integral in $y$; if treated as an ordinary function, the integral over $\bar\omega$ diverges. As a distribution, it is given by
\begin{align}
    \int_{-\infty}^\infty d\bar{\omega}\, e^{i\bar{\omega} y}&\log(-i\bar{\omega})= \nonumber\\*
    &-\pi\left[2\gammaE\delta(y)+\left(\frac{1}{|y|}\right)_1-{\rm p.v.}\!\left(\frac{1}{y}\right)\right],\label{Eq:logmiwFT}
\end{align}
where the latter two terms are defined distributionally by 
\begin{align}
    \int_{-\infty}^\infty dy \left(\frac{1}{|y|}\right)_1\phi(y)&=\int_{|y|\leq1}dy\frac{\phi(y)-\phi(0)}{|y|}\nonumber\\*
    &\quad+\int_{|y|\geq1}dy\frac{\phi(y)}{|y|},\label{(1/|y|)_1 def}\\
    \int_{-\infty}^\infty dy\; {\rm p.v.}\!\left(\frac{1}{y}\right)\phi(y)&=\int_{0}^{\infty} dy\frac{\phi(y)-\phi(-y)}{y}\label{pv(1/y) def}
\end{align}
for any test function $\phi$. We review the derivation of Eq.~\eqref{Eq:logmiwFT} in Appendix~\ref{appendix_B}. 

It is important to point out that the distributional definition of the IFT inherently yields finite results. An alternative approach, common in PN calculations, is to apply the Hadamard \emph{partie finie} (Pf) regularization, which is often computationally simpler than our distributional treatment. However, for the integrals in question, we find that the Pf operation introduces an arbitrary scale in the final result. 
In Appendix~\ref{appendix_A}, we provide a detailed discussion of these two procedures. 

 In brief, by using the distributional definition of the IFT we obtain a finite integral of the form
\begin{align}
   \hspace{-0.5cm} {}_s\psi^{\rm NL}_{\ell m}  &= 
   -\sum_{j\geq0} i^jv^j \nonumber\\*
   &\times\int d\rb'\,G_{\ell m}^{(j,1)}(\rb,\rb') 
    \left[\gammaE\,{}_sT_{lm}^{(j)}(\bar{t}, \rb')+ F_{lm}(\bar{t}, \rb')\right],
\end{align}
where
\begin{align}
    F_{lm}(\tb,\rb)=\int_0^{\infty}\log(y){}_sT_{lm}^{(j+1)}(\tb-y,\rb)dy.
\end{align}
We could use an equivalent expression for ${}_s\psi^{{\rm{NL}}}_{lm}$ by integrating first with respect to $\rb'$ and get
    \begin{align}\nonumber
    {}_s\psi^{{\rm{NL}}}_{lm} &=-\lim_{\epsilon\rightarrow0}  \sum_{j\geq0}i^jv^j\int d\rb'  \biggl\{ \left(\gamma_E + \log\epsilon\right) \\
    & \nonumber \times\dfrac{d^j}{d\bar{t}^j}\left[G^{(j,1)}_{lm}(\bar{r},\bar{r}'){}_sT_{lm}(\bar{t},\rb')\right] \\
    & + \int_{\epsilon}^{+\infty} \dfrac{dy}{y} \dfrac{d^j}{d\bar{t'}^j}\! \left[ G^{(j,1)}_{lm}(\bar{r},\bar{r}'){}_sT_{lm}(\bar{t}',\rb') \right]\biggr\vert_{\bar{t}' = \bar{t}-y} \biggr\} , \label{psiNL v2}
\end{align}
which is again finite, but now due to the $\log\epsilon$ counterterm that cancels the divergence arising from the integral. Using both methods independently we obtain the same results, providing internal consistency checks.

For $s=0$, we can write a direct analog of the local term~\eqref{Eq:lowvspin0}:
\begin{align}\label{psi_nonlocal}\nonumber
    {}_0\psi^{{\rm{NL}}}_{lm} &=-\frac{q}{b}\lim_{\epsilon\rightarrow0}  \sum_{j\geq0}i^jv^jY^*_{lm}(\pi/2,0)\Biggl\{ (\gamma_E+\log\epsilon)  \\
    &  \hspace{-1cm} 
     \times \dfrac{d^j}{d\bar{t}^j}\left[G^{(j,1)}_{lm}(\bar{r},\bar{r}_p(\bar{t}))\dfrac{e^{-im\phi(\bar{t})}}{u^t(\bar{t})}\right]\nonumber\\
    & \hspace{-1cm}+ \int_{\epsilon}^{+\infty} \dfrac{dy}{y} \dfrac{d^j}{d\bar{t'}^j} \left[ G^{(j,1)}_{lm}(\bar{r},\bar{r}_p(\bar{t'}))\dfrac{e^{-im\phi_p(\bar{t'})}}{u^t(\bar{t'})} \right]\bigg\vert_{\bar{t}' = \bar{t}-y} \Biggr\}.
\end{align}

\section{Scalar-field results} \label{sec5}

We now restrict our attention to the scalar case, with $s=0$, where the equation we have to solve is the massless KG equation~\eqref{KG eqn}. 
For the sake of brevity, from now on we will drop the subscript zero on the scalar field ${}_0\psi$, and we will write the spherical-harmonic expansion of the charge distribution, given in Eq.~\eqref{rho lm expansion}, as
\begin{align} \nonumber
    \rho(x^\mu) 
    &=\sum_{l,m} \rho_{lm}(t,r(t))Y_{lm}(\theta,\phi).
\end{align}

To express the retarded solution to the time-domain KG equation in the form~\eqref{eq:Teukolsky solution v2}, we first write it as 
\begin{equation}\label{eq:scalar_field}
\psi(t,r,\theta,\phi)=\sum_{l,m}\int \frac{d\omega}{2\pi} \psi_{lm\omega}(r) e^{-i\omega t} Y_{lm}(\theta,\phi),
\end{equation}
separating the time, radial, and angular dependence. The radial functions ${\psi}_{lm\omega}(r)$ then satisfy 
\begin{equation}
\label{eq:radialODE}
    \mathcal{L}_r (\psi_{lm\omega}(r)) = -\dfrac{4\pi}{f(r)} \rho_{lm\omega}(r),
\end{equation}
where the differential operator is
\begin{equation}
     \mathcal{L}_r  \equiv \dfrac{d^2}{dr^2} + \frac{2}{r} \dfrac{d}{dr} + \left[\dfrac{\omega^2}{ f^2(r)} - \dfrac{l(l+1)}{r^2 f(r)} \right],
\end{equation}
and the Fourier transform of the $l,m$ coefficients of the scalar charge density is
\begin{align} \nonumber
    \rho_{lm\omega}(r) & =   \int dt \hspace{0.1cm} e^{i\omega t} \rho_{lm}(t,r(t))  \\*
   & = \frac{q}{r^2} Y^*_{lm}(\pi/2,0)\int dt \hspace{0.1cm} \delta(r-r(t))  \dfrac{e^{i(\omega t-m\phi(t))}}{u^t(r)}.
\end{align}
The solution to Eq.~\eqref{eq:radialODE} can be obtained via the Green function method as
\begin{equation}
\psi_{lm\omega}(r)=-4\pi\int dr' G_{lm\omega}(r,r') r'^2 \rho_{lm\omega}(r'),
\end{equation}
which becomes 
\begin{align}\nonumber
{\psi}_{lm\omega}(r)&=-4\pi qY^*_{lm}(\pi/2,0)\\*
& \times\int dt\,\int dr' G_{lm\omega}(r,r')    \delta(r'-r(t))  \dfrac{e^{i(\omega t-m\phi(t))}}{u^t(r)},
\end{align}
leading to a trivial $r'$-integration. Inserting this in Eq.~\eqref{eq:scalar_field}, we find the more explicit form of Eq.~\eqref{eq:Teukolsky solution v2}:
\begin{align} \nonumber\label{integral_scalarfieldgen}
\psi(t,r,\theta,\phi)&= -4\pi q\sum_{l,m}Y^*_{lm}(\pi/2,0)Y_{lm}(\theta,\phi) \\*
&\times \int dt'\dfrac{e^{-im\phi(t')}}{u^t(t')}  \int \frac{d\omega}{2\pi} e^{i(t'-t)\omega} G_{lm\omega}(r,r(t')).
\end{align}

The physically relevant, retarded solution uses the retarded Green function, which is given by 
\begin{align}\nonumber
\label{eq:FD-GF}
    G_{lm\omega}(r,r') &= \dfrac{1}{r^2f(r)W_{lm\omega}}\left\{ R_{{\rm{in}}}^{lm\omega}(r)R_{{\rm{up}}}^{lm\omega}(r')H(r'-r)\right. \\*
    &\qquad\left.+ R_{{\rm{in}}}^{lm\omega}(r')R_{{\rm{up}}}^{lm\omega}(r)H(r-r')\right\}.
\end{align}
Here $W_{lm\omega}$ is the  Wronskian, $H(\cdot)$ is the Heaviside step function, and the two functions $R_{in}^{lm\omega}(r)$ and $R_{up}^{lm\omega}(r)$ are two independent solutions of the homogeneous radial equation which satisfy retarded boundary conditions at the horizon and radial infinity, respectively.

The retarded solution~\eqref{integral_scalarfieldgen} naturally diverges (as a Coulomb field) when we evaluate it at the particle's position, i.e. by sending $(r,\theta,\phi)\rightarrow (r(t),\pi/2,\phi(t))$. At the particle's position, the physically relevant field is instead the Detweiler-Whiting regular field $\psi^{\rm R}$; this is the field that exerts the self-force on the particle. 

We will compute the regular field at the particle using the standard method of mode-sum regularization \cite{Barack:1999wf,Barack:2009ux}. This method takes advantage of the fact that, while the full retarded field diverges at the particle, the individual spherical-harmonic modes do not. $\psi^{\rm R}$ can then calculated on the particle by subtracting from each harmonic mode a `regularization parameter'. We first define the $l$ modes of the retarded field at the particle as
\begin{equation}
    \psi^{\rm{ret}}_l(t) = \sum_{m=-l}^l \psi_{lm}(t,r)  Y_{lm}\left(\theta,\phi\right) \bigg\vert_{(r,\theta,\phi)\rightarrow (r(t),\pi/2,\phi(t))}.
\end{equation}
The regular field is then computed by subtracting an $l$-independent function $B(t)$ before summing over $l$:
\begin{equation}
    \psi^{\rm{R}}(t) = \sum_{l} [\psi^{\rm{ret}}_l(t)-B(t)].\label{psi mode sum reg}
\end{equation}
The regularization parameter  $B(\bar{t})$ has been computed in Ref.~\cite{Barack:2009ux,Barack:2022pde}\footnote{Note that here $L$ is the specific orbital angular momentum, while in Ref.~\cite{Barack:2009ux} it denotes $l+1/2$.} as
\begin{equation}\label{B_term_field}
    B(t)=\frac{2}{\pi}\frac{q}{L}k(t)K[k(t)]\,, \qquad
k(t)=\sqrt{\frac{L^2}{L^2+r_p(\bar{t})^2}}\,,
\end{equation}
where $K(k)$ denotes the complete elliptic integral of the first kind.

\subsection{Scalar Green function in the PN/PM limits}

Obtaining analytic expressions for the time-domain field using \eqref{integral_scalarfieldgen} (and correspondingly for the self-force, energy and angular momentum losses, and scattering angle) will require the spherical harmonic modes of the frequency-domain Green function \eqref{eq:FD-GF} for \emph{all} values of $l$ and $m$ and as a function of $\omega$. As in previous weak-field analytic self-force treatments in the bound-orbit case, e.g.~\cite{Bini:2013rfa,Kavanagh:2015lva,Munna:2022gio}, calculating the modes of the Green function can be split into two sectors: low-$l$ modes and large-$l$ modes. We will overview the strategy here.

For a small set of modes $l\leq l_{\rm max}$ we will use the exact expression for the retarded Green function using the homogeneous MST solutions~\cite{Mano:1996vt}, which fully satisfy the physical retarded boundary conditions. Here, the value of $l_{\rm max}$ will depend on the order in the expansion of interest: higher PN/PM order requires higher $l_{\rm max}$. 

For all values $l> l_{\rm max}$ we can 
treat our homogeneous solutions using an ansatz with both $l$ and $m$ as parameters. This ansatz will only approximately satisfy the retarded boundary conditions. However the violation of the boundary conditions by the ansatz will only appear beyond the order of interest when $l> l_{\rm max}$. This reflects the fact that in constructing weak-field solutions to the field equation, low multipoles  require matching to, e.g., far-field expansions, whereas higher multipoles do not require such matching. The MST solutions have this matching built in, which makes them ideal for low-$l$, but too cumbersome for large generic $l$ values.

We apply the above strategy to solve the field equations in two ways. Method one uses a standard PN scaling, which will later in the computation of the inhomogeneous field introduce the further PM expansion. Method two uses the PM-PN limit given in \eqref{eq:pmPNlimit} from the start. Ultimately, having two independent methods was useful for cross validation.

The generic-$l$ ansatz for the PN method corresponds to writing the solutions to the KG equation as a PN series
\begin{eqnarray}
\label{PNsols}
R_{\rm in(PN)}^{lm\omega}(r) &=& (r/M)^l \left[1 + \sum_{k=1}^{n}A_{2k}^{lm\omega}(r/M)\eta^{2k}\right]\,,\nonumber\\
R_{\rm up(PN)}^{lm\omega}(r) &=& R^{\rm in(PN)}_{-l-1m\omega}(r)\,,
\end{eqnarray}
 where we recall the PN rescalings of Eq. \eqref{PN_rescalings} together with the rescaling of $\omega \to \omega\eta$ and the coefficients $A_{2k}^{lm\omega}(r)$ have a polynomial structure in $\omega$, hence the Green functions constructed from these homogeneous solutions will inherit this property.  On the other hand, the MST solutions have the structure depicted in Eq. \eqref{GF_log_structure}.

In method two, the direct PM-PN approach, the radial equation  \eqref{eq:radialODE} is solved in the generic-$l$ case by applying the PN/PM scaling \eqref{eq:pmPNscaling} and  by employing the ansatz
\begin{align}
\label{eq:pmPNlargel}
    R_{\rm in(PM)}^{lm\omega}(r)&=\bar{r}^\nu\sum_{i=0}^{i_{\rm max}}\sum_{j=0}^{j_{\rm max}}B_{ij}\left(\frac{M}{b\bar{r}}\right)^i (\bar{r}\bar{\omega}v)^j,\\
    R_{\rm up(PM)}^{lm\omega}(r)&=\bar{r}^{-\nu-1}\sum_{i=0}^{i_{\rm max}}\sum_{j=0}^{j_{\rm max}}C_{ij}\left(\frac{M}{b\bar{r}}\right)^i (\bar{r}\bar{\omega}v)^j,
\end{align}
where $\nu$ is the `renormalized angular momentum' of MST~\cite{Sasaki:2003xr}, (see also the discussion in Sec.~IIB of~\cite{Kavanagh:2015lva}), $i_{\rm max}$ and $j_{\rm max}$ are the desired PM and PN orders, respectively, and the coefficients $B_{ij}, C_{ij}$ are purely functions of $l$. Once again, it is clear that for large values of $l$ the field will be a pure polynomial in the frequency, and will therefore only contribute to our local field as defined above.

The low-$l$ modes are again treated with the MST expressions for the homogeneous solutions, which are standardly written as infinite sums of hypergeometric functions of varying forms; see, e.g., Chapter 4 of~\cite{Sasaki:2003xr} for a detailed review. Their weak-field expansion follows standard methodology which we will omit from our discussion. The main point of note is for our 5PM, 4.5PN accuracy, we require the MST expressions for $l\leq4$. Ultimately the solutions follow a similar form to \eqref{eq:pmPNlargel} with the addition of the logarithmic-in-frequency dependence discussed previously.

In the following subsections we will provide the explicit expressions for the regular scalar field $\psi^R$. We will separate our results in two sectors: the local (non-logarithmic) and non-local (logarithmic) part. While the local terms can be obtained in two different parameterizations, as has been done for the solutions of the geodesic equations, the logarithmic contributions can be evaluated only with the time parameterization. The reason is that the first approach relies on a single PN expansion, while the second relies on a double PN and PM expansion.

\subsection{Local terms}

\subsubsection{PN regular scalar field}

We maintain the same conventions we used for the PN solution of the geodesic equation. We find for the regularization parameter
\begin{align}\nonumber 
B(t)&=\frac{qu_p(t)}{M}\left\{1-\frac{1}{4}[ju_p(t)]^2\eta^2+\frac{9}{64}[ju_p(t)]^4\eta^4\right. \\
& \left. -\frac{25}{256}[ju_p(t)]^6\eta^6+\frac{1225}{16384}[ju_p(t)]^8\eta^8
+O(\eta^{10})\right\}\,,
\end{align}
where we recall $u_p=M/r_p$. This corresponds to the PN expansion of Eq.~\eqref{B_term_field}.

Using this regularization parameter together with the retarded field modes, we find the local piece of the PN expansion of $\psi^R$ is given by
\begin{align}\nonumber
\psi^{\rm R,L}(t)&=-\frac{q\alpha^2j^2 u_p^2(t)}{M}\!\left\{
\frac{\epsilon_r\eta^3}{\alpha^3j^3}\sqrt{1+2 u_p(t) j^2\alpha^2-u_p^2(t) j^4\alpha^2} \right. \\
& \left.
+\left[2-3\alpha^2j^2u_p(t)(u_p(t)j^2-2)\right]\frac{\eta^4}{\alpha^4j^4}\right.\nonumber\\
&\left.
+\dfrac{\epsilon_r}{\sqrt{1+2 u_p(t) j^2\alpha^2-u_p^2(t) j^4\alpha^2}}\left[2+6\alpha^4 j^8 u_p^4(t)\right. \right.\nonumber\\
&\left. \left. -19\alpha^4 j^6 u_p^3(t)+8 j^4\alpha^2(-1+2\alpha^2)u_p^2(t)\right. \right. \nonumber \\
& \left. \left. +12 u_p(t) j^2\alpha^2\right]\frac{\eta^5}{\alpha^5j^5}
+O(\eta^6)
\right\}
\,.
\end{align}

\subsubsection{PN-PM regular scalar field}

As in the previous subsection, we firstly compute $B(\bar{t})$ and get
\begin{widetext}
    \begin{align} \nonumber
    &B(\bar{t}) = \dfrac{q}{b}\bigg\{ \frac{v^2}{\sqrt{1+\bar{t}^2}}-\frac{v^4}{4 \left(1+\bar{t}^2\right)^{3/2}}-\frac{v^6
   \left(7+16 \bar{t}^2\right)}{64 \left(1+\bar{t}^2\right)^{5/2}} + O(v^8)  \bigg\}   - \dfrac{q M}{b^2} \Biggl\{ \frac{v^2 \left({\rm{arcsinh}}\left(\bar{t}\right)
   \bar{t}-\sqrt{1+\bar{t}^2}\right)}{\left(1+\bar{t}^2\right)^{3/2}} \nonumber \\
   &\quad +\frac{3 v^4 \left[\sqrt{1+\bar{t}^2}-{\rm{arcsinh}}\left(\bar{t}\right) \bar{t} \left(5+4
   \bar{t}^2\right)\right]}{4 \left(1+\bar{t}^2\right)^{5/2}}  + \frac{3 v^6 \left[\sqrt{1+\bar{t}^2} \left(1+16 \bar{t}^2\right)+{\rm{arcsinh}}\left(\bar{t}\right) \bar{t} \left(47+32 \bar{t}^2\right)\right]}{64
   \left(1+\bar{t}^2\right)^{7/2}}  + O(v^8) \Biggr\} + O(b^{-3}), 
\end{align}
\end{widetext}
which corresponds to the PN-PM expansion of  Eq.~\eqref{B_term_field}.

From $B$ and the retarded field modes, we can finally obtain the local contribution to the regular scalar field in PN-PM expanded form:
\begin{align}\nonumber
    \psi^{\rm{R,L}}(\bar{t})&= -\frac{q v M}{b^2} \Biggl\{ \frac{\bar{t}}{\left(1+\bar{t}^2\right)^{3/2}}-\frac{v \left(1-2
   \bar{t}^2\right)}{\left(1+\bar{t}^2\right)^2}\\
   & \hspace{-1cm} -\frac{v^2 \bar{t} \left(7-5
   \bar{t}^2\right)}{2 \left(1+\bar{t}^2\right)^{5/2}} + O(v^3) \Biggr\} \nonumber \\
   & \hspace{-1cm} - \frac{3 q M^2}{b^3 v} \Biggl\{ \frac{{\rm{arcsinh}}\left(\bar{t}\right) \left(1-2 \bar{t}^2\right)}{3
   \left(1+\bar{t}^2\right)^{5/2}}+\frac{\bar{t}}{\left(1+\bar{t}^2\right)^2} \nonumber \\
   & \hspace{-1cm} +v \left[\frac{4 {\rm{arcsinh}}\left(\bar{t}\right) \bar{t} \left(2-\bar{t}^2\right)}{3
   \left(1+\bar{t}^2\right)^3}-\frac{2 \left(1-5 \bar{t}^2\right)}{3
   \left(1+\bar{t}^2\right)^{5/2}}\right] \nonumber \\
   & \hspace{-1cm} -v^2 \left[\frac{\bar{t} \left(31-29 \bar{t}^2\right)}{6
   \left(1+\bar{t}^2\right)^3}+\frac{{\rm{arcsinh}}\left(\bar{t}\right) \left(13-49
   \bar{t}^2-2 \bar{t}^4\right)}{6 \left(1+\bar{t}^2\right)^{7/2}}\right] \nonumber \\
   & \hspace{-1cm} + O(v^3)\Biggr\} + O(b^{-4}).
\end{align}

\subsection{Non-local terms}

The logarithmic contribution could be again computed both with a single PN or with the double PN-PM expansion. However, when the problem is approached with the first method one must compute integrals that, to the best of our knowledge, do not have a closed analytical form. For this reason, from now on the non-local sector will be approached only with a double PN-PM expansion. We recall Eq.~\eqref{psi_nonlocal}, and we divide in two sectors the logarithmic contributions to the field: one contribution does not require any integration and is proportional to $\gamma_E$, while the second needs both an integration and a distributional treatment. From the integrals we get the PolyLog structure previously observed in unbound-motion calculations using scattering amplitudes techniques~\cite{Klemm:2024wtd}. We will call these integrals \textit{tail integrals}.

\subsubsection{$\gamma_{\rm E}$}

For the first group of terms, we find 
\begin{widetext}
    \begin{align} \label{psi_NL1}\nonumber
    \psi^{{\rm{R,NL}},1}(\bar{t}) &= - \dfrac{q}{M}\gamma_E \Biggl\{ \dfrac{8v^2M^3}{3b^3}\left[ \frac{  \left(1-2 \bar{t}^2\right)}{ \left(1+\bar{t}^2\right)^{5/2}}  -\frac{ v^2
   \left(7-61 \bar{t}^2+22 \bar{t}^4\right)}{2 \left(1+\bar{t}^2\right)^{7/2}} + O(v^3)\right]   + \dfrac{M^4}{b^4}\left[ \dfrac{4}{3\left(1+\bar{t}^2\right)^{7/2}} \nonumber \right. \\
   & \left.\times\Bigl( \sqrt{1+\bar{t}^2} \left(7-23 \bar{t}^2\right)  -6\, {\rm{arcsinh}}\left(\bar{t}\right) \bar{t} \left(3-2 \bar{t}^2\right) \Bigr)+\frac{2 v^2}{\left(1+\bar{t}^2\right)^{9/2}} \Bigl( 10\, {\rm{arcsinh}}\left(\bar{t}\right) \bar{t} \left(15-25 \bar{t}^2+2 \bar{t}^4\right) \nonumber \right. \\
   & \left.-\sqrt{1+\bar{t}^2} \left(19-240 \bar{t}^2+161 \bar{t}^4\right) \Bigr)   + \frac{76 v^3 \bar{t} \left(3-2 \bar{t}^2\right)}{15
   \left(1+\bar{t}^2\right)^{7/2}} +   O(v^4)\right]  
   
   + O(b^{-5})\Biggr\}.
\end{align}

\subsubsection{ Tail integrals}

For the second group of terms, we find 
    \begin{align}
    \nonumber
    \psi^{{\rm{R,NL}},2}(\bar{t}) &= \dfrac{qM^2}{b^3} \bigg\{ \dfrac{4 v^2}{3  \left(1+\bar{t}^2\right)^{5/2}} \left[ -3+10 \bar{t}^2+6 \bar{t} \sqrt{1+\bar{t}^2}  +\left(2-4 \bar{t}^2\right)\left({\rm{arcsinh}}\left(\bar{t}\right) + \log 
   \left(1+\bar{t}^2\right) + \log2\right)  \right]  + O(v^4) \bigg\} \\
   & + \dfrac{qM^3}{b^4}\Biggl\{ \frac{2}{3 \left(1+\bar{t}^2\right)^{7/2}} \Biggl[ \left(26+3 \pi ^2\right) \bar{t}  +2 \left(11-\pi ^2\right) \bar{t}^3-4 \bar{t}^5 -2 \sqrt{1+\bar{t}^2} \left(7-38 \bar{t}^2+2 \bar{t}^4\right)  \nonumber \\
   &  \left. +6 \left(1-4 \bar{t}^2\right) \sqrt{1+\bar{t}^2} \log \left(1+\bar{t}^2\right)  -2 \log (2) \left(1+\bar{t}^2\right)^{3/2} -12 {\rm{arcsinh}}^2\left(\bar{t}\right) \bar{t} \left(3-2 \bar{t}^2\right) \right. \nonumber \\
   &  \left. +2\,{\rm{arcsinh}}(\bar{t})\left(  \left(7-23 \bar{t}^2\right) \sqrt{1+\bar{t}^2}  + \bar{t}(39-38\bar{t}^2) -6 \log (2) \bar{t} \left(3-2 \bar{t}^2\right) -6 \log \left(1+\bar{t}^2\right) \bar{t} \left(3-2 \bar{t}^2\right) \right) \right. \nonumber \\
   &   -\left(9 \bar{t}-6 \bar{t}^3\right)
   \text{Li}_2\left(\frac{\bar{t}+\sqrt{1+\bar{t}^2}}{\bar{t}-\sqrt{1+\bar{t}^2}}\right) + O(v^2) \Biggr]\Biggr\} + O(b^{-5}).
\end{align} %
\end{widetext}%
Here we have only presented the leading PN terms at each PM order due to the complexity of the complete result. However, the full expression for this group of terms is in the attached \texttt{Mathematica} ancillary file.

\section{Scalar self-force} \label{sec6}

We recall that the scalar self-force (per unit mass) acting on a particle with scalar charge $q$ and mass $\mu$  is given by~\cite{Poisson:2011nh}
\begin{equation}
  F_\alpha = \frac{q}{\mu} P_\alpha{}^\beta  \nabla_\beta \psi^{\rm R}, 
\end{equation}
where $\psi^{\rm R}$ is the Detweiler-Whiting regular field, extracted from the solution of the KG equation, and $P_\alpha{}^\beta =\delta_\alpha^\beta + u_\alpha u^\beta$ projects orthogonally to the particle's worldline. Because we are only interested in the leading-order, linear corrections due to the force, we can approximate $P_\alpha{}^\beta$ by $\hat P_\alpha{}^\beta =\delta_\alpha^\beta + \hat u_\alpha \hat u^\beta$. Similarly, although $\mu$ is not constant~\cite{Poisson:2011nh}, we can treat it as constant in our leading-order analysis.

As we are discussing equatorial scattering, the non-vanishing components of $F_\alpha$ are
\begin{align}
    F_t &= \frac{q}{\mu}\left(1-\hat{u}_t^2 f(r)\right)\partial_t \psi^{\rm R} \nonumber\\
    &\quad -\frac{q}{\mu}f(r)\hat{u}_t\left(  \hat{u}_r  \partial_r \psi^{\rm R} +\hat{u}_{\phi }\partial_\phi \psi^{\rm R} \right) , \label{Ft formula}\\
    F_r &= \frac{q}{\mu}(1+\hat{u}_r^2 f^{-1}(r))\partial_r\psi^{\rm R}\nonumber\\
    &\quad+\frac{q}{\mu}\hat{u}_r f^{-1}(r) \left( \hat{u}_t \partial_t \psi^{\rm R} + \hat{u}_\phi \partial_\phi \psi^{\rm R}  \right), \\
    F_{\phi} &= \frac{q}{\mu}(1+r^2 \hat{u}_\phi^2)\partial_\phi\psi^{\rm R}\nonumber\\
    &\quad +\frac{q}{\mu}\hat{u}_\phi r^2 \left( \hat{u}_t \partial_t\psi^{\rm R} + \hat{u}_r \partial_r\psi^{\rm R}\right) .\label{Fphi formula}
\end{align}
Practically speaking, only the $t$ and $\phi$ components will be used for the computation of the conservative and dissipative dynamics, as already discussed in Sec.~\ref{Sec3partb}. 

To compute $F_\alpha$ from the retarded field modes $\psi_{lm}$, we again use the method of mode-sum regularization, as explained above Eq.~\eqref{psi mode sum reg}. 
The derivatives of the retarded field possess a linear-in-$l$ divergence on the particle, and the mode-sum formula for the force is
\begin{equation}
    F_\alpha(t) = \sum_{l} \left\{F^{l\,\pm}_\alpha(t) - (l+1/2)A^{l\,\pm}_\alpha(t) - B_\alpha(t)\right\},\!\!
\end{equation}
where $F^{l\,\pm}_\alpha$ denotes the result of substituting a retarded-field mode $\psi_{lm}Y_{lm}$ into Eqs.~\eqref{Ft formula}--\eqref{Fphi formula}, evaluating on the particle, and summing over $m$. The $\pm$ indicates\footnote{It is important to remind that this $\pm$ is different from the one used in Section \ref{Sec3partb}.} whether the particle is approached from the right or left, i.e. $r^+_p(t)$ or $r^-_p(t)$, when we send $r\to r_p(t)$. It is well known~\cite{Barack:2009ux} that if we compute the average mode contribution from the two directions, as in
\begin{equation}
    F^l_\alpha(t) = \dfrac{1}{2}\left[F^{l\,+}_\alpha(t) + F^{l\,-}_\alpha(t) \right],
\end{equation}
then the contribution from $A^{l\,\pm}_\alpha(t)$ vanish and only the $B_\alpha(t)$ terms are required in the mode-sum formula. For this reason we will always compute the averaged components of the force.

These calculations require us to compute derivatives of the retarded field modes. The derivatives that come from the local sector are trivial to compute. However, the ones that comes from the non-local sector could be problematic, in particular the derivative with respect to $t$ when the field is written in the form of (superficially) divergent integrals. However, it is easy to realize that for an integral of the form
\begin{align}
    \dfrac{\partial I}{\partial t} = \dfrac{\partial }{\partial t} \int dt' f(r,t') \int \frac{d\omega}{2\pi}e^{i\omega (t'-t)} \omega^j \log\omega,
\end{align}
where $f(r,t')={}_0 G_{\ell m}^{(j,1)}(r,r(t'))e^{-im\phi(t')}$, we have
\begin{equation}
    \partial_t I = -i \int dt' f(r,t') \int \frac{d\omega}{2\pi}e^{i\omega (t'-t)} \omega^{j+1} \log\omega.
\end{equation}
Hence, for derivatives of the field we can employ the same distributional method we used for the field itself.

We will present here the expressions for $F_t$ and $F_\phi$ by splitting the contributions from the local and non-local sectors, as we did for the field. But we will directly focus on the results obtained with a double PN-PM expansion.

\subsection{Local terms}

We present the result in a similar fashion as we did for the field: we will present firstly the regularization parameter $B_\alpha(t)$ and then the components of the force.

In the case of the forces we have two different regularization parameters: $B_t$ and $B_{\phi}$ for the two components of the force we are interested in. Starting from the generic formulas in Ref.~\cite{Barack:2009ux,Barack:2022pde}, we find
\begin{align}\nonumber
    B_t &= \frac{\e v M}{2 b^2} \bigg\{ \frac{\bar{t}}{\left(1+\bar{t}^2\right)^{3/2}}+\frac{v^2 \bar{t} \left(7+4 \bar{t}^2\right)}{4
   \left(1+\bar{t}^2\right)^{5/2}} + O(v^4) \bigg\} \\
  & + \frac{\e M^2}{2 b^3 v} \bigg\{ \frac{3 \bar{t}}{\left(1+\bar{t}^2\right)^2} + \frac{{\rm{arcsinh}}\left(\bar{t}\right) \left(1-2 \bar{t}^2\right)}{\left(1+\bar{t}^2\right)^{5/2}} \nonumber \\
  & + v^2\left[ \frac{\bar{t} \left(23+8 \bar{t}^2\right)}{4 \left(1+\bar{t}^2\right)^3}-\frac{{\rm{arcsinh}}\left(\bar{t}\right) \left(5+4 \bar{t}^2-16 \bar{t}^4\right)}{4 \left(1+\bar{t}^2\right)^{7/2}} \right]\nonumber \\
  &+ O(v^4) \bigg\} + O(b^{-4}), 
\end{align}
and
\begin{align}\nonumber
    B_\phi &= -\frac{3 \e v^2 M}{4 b}\bigg\{ \frac{\bar{t}}{\left(1+\bar{t}^2\right)^{3/2}}+\frac{v^2 \bar{t} \left(3+8 \bar{t}^2\right)}{8
   \left(1+\bar{t}^2\right)^{5/2}}+O(v^4)\bigg\} \\
   & -\frac{9 \e M^2 }{4 b^2} \bigg\{ \frac{{\rm{arcsinh}}\left(\bar{t}\right) \left(1-2 \bar{t}^2\right)}{3
   \left(1+\bar{t}^2\right)^{5/2}}+\frac{\bar{t}}{\left(1+\bar{t}^2\right)^2} \nonumber \\
   & + v^2\left[ \frac{\bar{t} \left(-9+16 \bar{t}^2\right)}{24 \left(1+\bar{t}^2\right)^3}+\frac{{\rm{arcsinh}}\left(\bar{t}\right) \left(-21+36 \bar{t}^2+32 \bar{t}^4\right)}{24
   \left(1+\bar{t}^2\right)^{7/2}} \right] \nonumber \\
   & + O(v^4) \bigg\} + O(b^{-3}).
\end{align}

The local terms in the force, as computed with mode-sum regularization, are then
\begin{align}\nonumber
    F^{\rm{L}}_t (\bar{t})&= \frac{\e v^2 M^2}{3 b^3} \bigg\{ \frac{1-2 \bar{t}^2}{\left(1+\bar{t}^2\right)^{5/2}} +\frac{6 v \bar{t}}{\left(1+\bar{t}^2\right)^3} -\frac{v^2 \left(1-2 \bar{t}^2\right)}{2 \left(1+\bar{t}^2\right)^{5/2}} \\
    & + O(v^3)\bigg\} + \frac{5 \e M^3}{3 b^4} \bigg\{ \frac{1-2 \bar{t}^2}{\left(1+\bar{t}^2\right)^3}\nonumber\\
    &-\frac{3  \bar{t}\,{\rm{arcsinh}}\left(\bar{t}\right) \left(3-2 \bar{t}^2\right)}{5
   \left(1+\bar{t}^2\right)^{7/2}} \nonumber\\
   & +v\left[ \frac{36 \bar{t}}{5 \left(1+\bar{t}^2\right)^{7/2}}+\frac{6 {\rm{arcsinh}}\left(\bar{t}\right) \left(1-5 \bar{t}^2\right)}{5
   \left(1+\bar{t}^2\right)^4} \right] \nonumber \\
   & -v^2 \left[ \frac{3 \left(1-6 \bar{t}^2\right)}{2 \left(1+\bar{t}^2\right)^3}-\frac{21 \bar{t}\,{\rm{arcsinh}}\left(\bar{t}\right)  \left(3-2 \bar{t}^2\right)}{10
   \left(1+\bar{t}^2\right)^{7/2}} \right] \nonumber \\
   & + O(v^3)\bigg\} + O(b^{-5}),
\end{align}
and
\begin{align}\nonumber 
    F^{\rm{L}}_\phi (\bar{t})&=-\frac{\e v M^2}{3 b^2}\bigg\{ \frac{1}{\left(1+\bar{t}^2\right)^{3/2}} +\frac{6 v \bar{t}}{\left(1+\bar{t}^2\right)^2}+\frac{v^2 \left(-1+17
   \bar{t}^2\right)}{2 \left(1+\bar{t}^2\right)^{5/2}} \\*
   & + O(v^3) \bigg\} -\frac{\e M^3}{b^3 v} \bigg\{ \frac{1}{\left(1+\bar{t}^2\right)^2}-\frac{{\bar{t}\,\rm{arcsinh}}\left(\bar{t}\right)
   }{\left(1+\bar{t}^2\right)^{5/2}} \nonumber \\
   & + v\left[ \frac{8 \bar{t}}{\left(1+\bar{t}^2\right)^{5/2}}+\frac{{\rm{arcsinh}}\left(\bar{t}\right) \left(2-6 \bar{t}^2\right)}{\left(1+\bar{t}^2\right)^3}  \right] \nonumber \\
   & -v^2\left[ \frac{3 \left(1-9 \bar{t}^2\right)}{2 \left(1+\bar{t}^2\right)^3}-\frac{\bar{t} \,{\rm{arcsinh}}\left(\bar{t}\right) \left(19-11 \bar{t}^2\right)}{2
   \left(1+\bar{t}^2\right)^{7/2}} \right] \nonumber \\ 
   & + O(v^{3})\bigg\} + O(b^{-4}).
\end{align}

It is important to mention that these two components of the force have different leading orders in $1/b$.

\subsection{Non-local terms}

As we did for the regular field on the particle, we divide the non-local terms in  the force into $\gamma_E$ terms and tail-integral terms.

\subsubsection{$\gamma_E$}

We find the coefficients of $\gamma_E$ to be
\begin{widetext}
    \begin{align}\label{Ft_nonlocal1}
    F^{\rm{NL},1}_t (\bar{t}) &= \frac{4 \gamma_E \e \tb v^3 M^3}{b^4}\bigg\{\frac{2 \tb^2-3}{\left(\tb^2+1\right)^{7/2}}-\frac{\left(2 \tb^2-3\right) v^2}{2
   \left(\tb^2+1\right)^{7/2}}\nonumber + O(v^{3}) \bigg\}+\frac{4 \e \tb v M^4}{3 b^5}\bigg\{\frac{44 \tb^2-61}{\left(\tb^2+1\right)^4}\nonumber-\frac{3 \left(8 \tb^4-24 \tb^2+3\right) {\rm{arcsinh}}
   (\tb)}{\tb \left(t^2+1\right)^{9/2}} \\
   &\quad+ v^2\bigg[\frac{235-308 \tb^2}{2 \left(\tb^2+1\right)^4} +\frac{21 \left(8 \tb^4-24 \tb^2+3\right) {\rm{arcsinh}}
   (\tb)}{2 \tb \left(\tb^2+1\right)^{9/2}}\bigg]+O(v^{3}) \bigg\} + O(b^{-6}),
\end{align}
and
\begin{align}\label{Fphi_nonlocal1}
    F^{\rm{NL},1}_\phi (\bar{t}) &=  \frac{ 8 \gamma_E \e \tb v^2 M^3}{b^3}\bigg\{\frac{5 \left(3 \tb^2-1\right) v^2}{2
   \left(\tb^2+1\right)^{7/2}}+\frac{1}{\left(\tb^2+1\right)^{5/2}}+ O(v^{3}) \bigg\}+\frac{40 \e \tb M^4}{b^4}\bigg\{\frac{1}{\left(\tb^2+1\right)^3}+\frac{\left(1-4 \tb^2\right) {\rm{arcsinh}}(\tb)}{5 \tb
   \left(\tb^2+1\right)^{7/2}} \nonumber\\
   &\quad+v^2\bigg[\frac{299 \tb^2-121}{30 \left(\tb^2+1\right)^4}+\frac{\left(-36 \tb^4+93 \tb^2-11\right) {\rm{arcsinh}}
   (\tb)}{10 \tb \left(\tb^2+1\right)^{9/2}}\bigg]+O(v^{3}) \bigg\} + O(b^{-5}).
\end{align}
\end{widetext}

\subsubsection{Tail integrals}
The explicit expressions for the forces arising from the tail integrals are much lengthier than previous expressions. As such we only display the leading-order-in-velocity terms for each of the $t$ and $\phi$ components of the force: 
\begin{widetext}
    \begin{align}\label{Ft_nonlocal2}
    F^{\rm{NL},2}_t (\bar{t}) &= - \frac{8\e v^3 M^3}{3 \left(\bar{t}^2+1\right)^{7/2}b^4}\bigg\{\bar{t} \left(11 \bar{t}^2-12\right)+\frac{1}{2} \sqrt{\bar{t}^2+1} \left(11
   \bar{t}^2-4\right)+\left(\!\frac{9 \bar{t}}{2}-3 \bar{t}^3\!\right)\! \biggl[\log \left(2
   \left(\bar{t}^2+1\right)\right)+{\rm{arcsinh}} \left(\bar{t}\right)\biggr]\!+ O(v)\!\bigg\} \nonumber \\*
   &-\frac{4 \e v M^4}{3 \left(\bar{t}^2+1\right)^{5/2} b^5}\Bigg\{\frac{\big(2 \bar{t}^4+55 \bar{t}^2-89\big)
   \bar{t}}{\big(\bar{t}^2+1\big)^{3/2}}+2 \big(\bar{t}^2-2\big) +\frac{7 \log (2)
   \big(8 \bar{t}^2-7\big) \bar{t}}{\big(\bar{t}^2+1\big)^{3/2}}+\frac{6 \bar{t}
   }{\sqrt{\bar{t}^2+1}}\Bigl[\log \big(\bar{t}^2+1\big)+{\rm{arcsinh}}
   \big(\bar{t}\big)\Bigr] \nonumber \\*
   &-\frac{4 \big(25 \bar{t}^4-57
   \bar{t}^2+6\big) {\rm{arcsinh}} \big(\bar{t}\big)}{\big(\bar{t}^2+1\big)^2} + O(v)\Bigg\} + O(b^{-6}),
\end{align}
and
\begin{align}\label{Fphi_nonlocal2}
    F^{\rm{NL},2}_\phi (\bar{t})& = - \frac{4 \e v^2 M^3}{3 \left(\bar{t}^2+1\right)^{5/2}b^3}\bigg\{17 \bar{t}-2 \bar{t}^3-2 \left(\bar{t}^2-2\right) \sqrt{\bar{t}^2+1}   -6 \bar{t} \left[\log \left(2 \left(\bar{t}^2+1\right)\right)+{\rm{arcsinh}} \left(\bar{t}\right)\right] + O(v)\bigg\} \nonumber\\
   &   -\frac{2 \e M^4}{ 3\left(\bar{t}^2+1\right)^{7/2}b^4}\bigg\{-8+2 \bar{t} \sqrt{\bar{t}^2+1} \left(11 \bar{t}^2+28\right)   +\left[8 \bar{t}^4-172 \bar{t}^2+4 \left(11-4 \bar{t}^2\right) \sqrt{\bar{t}^2+1}
   \bar{t}+34\right] {\rm{arcsinh}}\left(\bar{t}\right) \nonumber \\
   &  +22 \bar{t}^4+14 \bar{t}^2 -6 \bar{t} \sqrt{\bar{t}^2+1} \left[4 \log (2) \left(\bar{t}^2-4\right)+\left(2
   \bar{t}^2-3\right) \log \left(\bar{t}^2+1\right)\right]  \nonumber\\
   &-3 \left(4 \bar{t}^2-1\right)\Biggl[\pi ^2-4 \log \left(2 \left(\bar{t}^2+1\right)\right)
   {\rm{arcsinh}}\left(\bar{t}\right)-4  {\rm{arcsinh}}\left(\bar{t}\right)^2-6 \text{Li}_2\left(\frac{\bar{t}+\sqrt{\bar{t}^2+1}}{\bar{t}-\sqrt{\bar{t}^2+1}}\right)\Biggr] + O(v) \Biggr\} + O(b^{-5}).
\end{align}
\end{widetext}

\section{Results and Discussions}\label{sec7}

Once the self-force is computed, we can investigate observables of interest, in particular the loss of energy and angular momentum and the self-force correction to the scattering angle. The first two quantities are purely dissipative effects, hence associated with radiation. The scattering angle contains also a contribution from the conservative sector, as discussed in Sec.~\ref{Sec3partb}.

Given our results for the self-force and the formal expressions derived in Sec.~\ref{Sec3partb}, the procedure for extracting the observables we are interested in is then conceptually easy, though it can be technically challenging when evaluating integrals from the non-local sector. 

We will present our results at 5PM-4.5PN firstly for the dissipative dynamics and then for the conservative sector. We have also restored the mass $M$ of the Schwarzschild black hole.

\subsection{Dissipative sector}

Firstly we show the  total radiated energy and angular momentum, as computed by integrating the dissipative piece of the self-force along the worldline according to Eq.~\eqref{Erad and Lrad} with Eqs.~\eqref{eq:deltaELdef} and~\eqref{Fdiss}:
\begin{align}\label{Erad_3pm}
    E^{\rm{rad}}_{1{\rm{PM}}}&= E^{\rm{rad}}_{2{\rm{PM}}} = 0, \\
    E^{\rm{rad}}_{3{\rm{PM}}} &= \dfrac{\pi \e M^3}{b^3v}\!\left[ \frac{1}{6}+\frac{11 v^2}{12}+\frac{25 v^4}{24}+\frac{311 v^6}{240} + O(v^7) \right]\!, \\
    E^{\rm{rad}}_{4{\rm{PM}}} &= \dfrac{\e M^4}{b^4}\dfrac{4 }{3 v^3} \biggl[ 1+\dfrac{23 v^2 }{3}-2 v^3 +\dfrac{28 v^4}{5}  \nonumber \\
    &\quad -\dfrac{16 v^5 }{3}+\dfrac{4324 v^6 }{525}  + O(v^7)  \biggr], \\
    E^{\rm{rad}}_{5{\rm{PM}}} &= \dfrac{\pi \e M^5}{b^5}\dfrac{1 }{2 v^5} \left\{1+14v^2-\dfrac{3\pi^2v^3}{4} +\dfrac{89 v^4}{8} \right. \nonumber \\
    &\quad +\left(-\frac{96}{5}-\frac{9 \pi ^2}{40}\right) v^5+ \biggl[\frac{20033}{600}+\frac{\pi ^2}{2} \nonumber \\
    &\quad -\frac{19 \log
   (2)}{2}-\frac{19 \log (b/M)}{5}+\frac{19 \log (v)}{10}\biggr]v^6\nonumber\\
   &\quad + O(v^7)  \biggr\}\,,
\end{align}
and 
\begin{align}
    &L^{\rm{rad}}_{1{\rm{PM}}}= 0, \\
    &L^{\rm{rad}}_{2{\rm{PM}}} = \dfrac{\e M^2}{b}\left[\frac{2}{3}+\frac{4 v^2}{3}+\frac{4 v^4}{3}+\frac{4 v^6}{3} + O(v^7) \right], \\
    &L^{\rm{rad}}_{3{\rm{PM}}} = \dfrac{ \pi \e M^3}{b^2v^2}\left[ \frac{1}{3}+\frac{5 v^2}{6}+\frac{v^4}{12}-\frac{31 v^6}{120} + O(v^7) \right], \\
    &L^{\rm{rad}}_{4{\rm{PM}}} = \dfrac{\e M^4}{b^3} \frac{2 }{3 v^4} \left[ 1+8 v^2 -4 v^3 -\frac{23 v^4 }{3}-\frac{16 v^5 }{5} \right. \nonumber \\
    & \left. \hspace{1.2cm} -\frac{1304
   v^6 }{75} + O(v^7) \right], \\
    &L^{\rm{rad}}_{5{\rm{PM}}} = \dfrac{\pi \e M^5}{b^4} \frac{5 }{2 v^4} \left\{ 1-\frac{\pi ^2 v}{10}-\frac{11 v^2}{10}-\left(\frac{16}{5}-\frac{\pi ^2}{100}\right) v^3 \right. \nonumber \\*
    & \left. \hspace{1.2cm} - \left[\frac{9883}{3000}-\frac{\pi ^2}{15}+\frac{152 \log (2)}{75}+\frac{19 \log (b/M)}{25}  \nonumber \right. \right. \\*
    & \left. \left. \hspace{1.2cm} -\frac{38 \log (v)}{75}\right]v^4 + O(v^5) \right\},
\end{align}
where we recall that $\e=q^2/(\mu M)\ll1$ as defined in \eqref{eq:scalareps}, and note the leading PN order is $1.5$ for both $\delta E^{\rm diss}$ and $\delta L^{\rm diss}$, except for the $5$PM radiated angular momentum, which has its first contribution at 2.5PN.  

In order to complete the radiative sector, we compute the dissipative scattering angle. From Eq.~\eqref{delta_chi_diss}, we find
\begin{align}
    &  \delta \chi^{\rm{diss}}_{1{\rm{PM}}}=\delta \chi^{\rm{diss}}_{2{\rm{PM}}} = 0, \\
    & \delta \chi^{\rm{diss}}_{3{\rm{PM}}}=\dfrac{\e M^3}{b^3 v^3} \left[ \frac{2}{3}+\frac{5 v^2}{3}+\frac{19 v^4}{12}+\frac{25 v^6}{24} + O(v^7) \right], \\
    & \delta \chi^{\rm{diss}}_{4{\rm{PM}}}=\dfrac{\pi \e M^4}{b^4 v^5} \left[ \frac{1}{2}+\frac{10 v^2}{3}+\frac{89 v^4}{48}+\frac{53 v^6}{80}  + O(v^7) \right], \\
    & \delta \chi^{\rm{diss}}_{5{\rm{PM}}}=\dfrac{4 \e M^5}{3  b^5 v^7} \left[ 1+\left(\frac{91}{6}+\frac{3 \pi ^2}{4}\right) v^2-4 v^3 \nonumber \right. \\ & \left. \hspace{1.2cm} +\left(\frac{1357}{120}+\frac{39 \pi
   ^2}{32}\right) v^4-\frac{14 v^5}{15} \right.\nonumber \\
   & \left. \hspace{1.2cm} + \left( \frac{16103}{560}-\frac{45 \pi ^2}{16}\right) v^6  + O(v^7) \right].
\end{align}

We see that the 3PM energy and angular momentum loss and the 4PM scattering angle present only (relative) integer PN orders (corresponding to even powers of $v$). This is because only the local sector contributes at this PM order. However, at the next PM order, we start to see half-integer PN terms (odd powers of $v$), which are usually related to tail effects in classical PN theory. This structure between local and non-local terms become apparent after the final integration over the orbit, and not, for instance in the self-force or in the regular field. Another smoking gun for tail contributions is the presence of $\log(b/M)$ and $\log v$, which are related to the physical scales of the problem.

It is also worth commenting on the absence of the Euler-Mascheroni constant  $\gamma_E$ in the final results. The presence of $\gamma_E$ is usually tied up, in PN theory, to the presence of logarithms in the observables. It is then surprising that, for the scattering motion, the PN expansions of these quantities do contain logarithms but not $\gamma_E$. It is possible, however, that at higher orders $\gamma_E$ will appear.   

Comparing our results against those obtained with amplitudes methods in Ref.~\cite{Barack:2023oqp}, we find a  mismatch beginning at 3PM-2.5PN,
\begin{equation}\label{diff_Erad}
    E^{\rm{rad}}_{3{\rm{PM}}}- E^{\text{rad,Ref.\cite{Barack:2023oqp}}}_{\rm{3PM}}=\dfrac{\e M^3}{b^3}\frac{\pi}{2}v\left[1+\dfrac{v^2}{2} + O(v^5)\right], 
\end{equation}
which propagates to the comparison between the scattering angle. This can be attributed to the fact that it is purely the loss of energy away from the binary system that is being computed in \cite{Barack:2023oqp}, whereas the results here, coming from the local dissipative force, also include energy loss into the black hole horizon. In this sense, we generalize their result by computing the full dissipative scattering angle at each PM order.

Nonetheless, we can still provide some verification of our results. In a recent work \cite{Jones:2023ugm}, the 3PM mass shift of the Schwarzschild black hole in the case of scalar absorption was computed. The horizon flux that they obtained is
\begin{equation}
    E_H = \dfrac{\e M^3}{b^3}\dfrac{\pi}{2}\dfrac{v}{\sqrt{1-v^2}},
\end{equation}
which is precisely the resummation of Eq. \eqref{diff_Erad}.

\subsection{Conservative sector}

The conservative dynamics affects the shape of the hyperbolic orbit, by changing the relationships between orbital parameters and, more importantly, the scattering angle. Recalling the results in Sec.~\ref{Sec3partb}, in the conservative sector we have that $\delta\chi_{\rm{cons}}$ takes the form
\begin{align}\label{delta_chi_cons}\nonumber
    \delta\chi_{\rm{cons}} & = \dfrac{\partial \bar{\chi}}{\partial E_-}\delta E_0^{\rm{cons}} + \dfrac{\partial \bar{\chi}}{\partial L_-}\delta L_0^{\rm{cons}} \\
    &\quad -2\int_{\bar r_{\rm min}}^{\infty} dr \left[
a_E^+(r)\int_{\bar r_{\rm min}}^r\frac{dr}{\bar u^r}F_t^{\rm cons\,+}
\right.\nonumber\\
&\left. \hspace{2cm} -a_L^+(r)\int_{\bar r_{\rm min}}^r\frac{dr}{\bar u^r}F_\phi^{\rm cons\,+}
\right]
\,,
\end{align}
where we used
\begin{equation}
    \delta E_0^{\rm{cons}} = -\int_{-\infty}^0 d\tau F^{\rm{cons}}_{t}, \hspace{0.5cm} \delta L_0^{\rm{cons}} = \int_{-\infty}^0 d\tau F^{\rm{cons}}_{\phi}, 
\end{equation}
and because we are dealing with conservative dynamics we have
\begin{equation}
    \int_{-\infty}^{+\infty} d\tau F^{\rm{cons}}_{t/\phi} = 0 .
\end{equation}

Evaluating these formulas, we find
\begin{align}\label{cons_chi_2pm_res}
    \delta \chi^{\rm{cons}}_{2{\rm{PM}}} &= -\frac{\pi  \e M^2 }{4 b^2} , \\
    \label{cons_chi_3pm_res}\delta \chi^{\rm{cons}}_{3{\rm{PM}}}&=-\dfrac{\e M^3}{b^3 v^2} \left[ 4+\frac{2 v^2}{3}+\frac{5 v^4}{6} + O(v^5) \right], \\ \nonumber
    \label{cons_chi_4pm_res} \delta \chi^{\rm{cons}}_{4{\rm{PM}}}&=-\dfrac{\pi \e M^4}{b^4 v^4} \biggl\{ \frac{9}{4}+v^2 \left[\frac{91}{24}+\frac{21 \pi ^2}{128}+\log (v/2)\right]  \\
    &\quad \left. +v^4 \left[\frac{493}{480}+\frac{4335 \pi ^2}{8192}-\frac{3 \log (b/M)}{2}+2 \log\Bigl(\frac{v}{2}\Bigr)\right]  \nonumber \right. \\
    &\quad  + O(v^5) \biggr\}, \\ \nonumber
    \label{cons_chi_5pm_res} \delta \chi^{\rm{cons}}_{5{\rm{PM}}}&=-\frac{16 \e M^5}{3  b^5v^6} \biggl\{ 1+v^2 \left[\frac{47}{6}+\frac{21 \pi ^2}{64}+2 \log (2v)\right]  \\*
    &\quad  -v^4\! \biggl[\frac{797}{72}-\frac{205 \pi ^2}{128}-\frac{43 \log (2)}{3}+4 \log \Bigl(\frac{b}{M}\Bigr) \nonumber  \\*&\qquad\quad -\frac{19 \log
   (v)}{3}\biggr] 
      +\frac{19 v^5}{15} + O(v^6) \biggr\},
\end{align}
where the leading term is at $2$PN.
Consistent with the dissipative sector, in the conservative case we do not see $\gamma_E$ appearing in these final results.

Another interesting feature is that in our results both the energy and the angular momentum contribute to the leading 2PM term of the conservative scattering angle, displayed in Eq.~\eqref{cons_chi_2pm_res}, and this leading term is totally determined by the local sector of the forces. 

The non-local sector starts to contribute at 4PM, in Eq.~\eqref{cons_chi_4pm_res}, introducing a transcendental structure to the result. In this case, we get first the logarithmic contributions and then at 5PM the odd powers of the velocity. 
However, by looking at our non-local expressions for the forces in Eqs.~\eqref{Ft_nonlocal1}-\eqref{Fphi_nonlocal2}, we see that the leading terms are at lower powers in $1/b$. It is a non-trivial statement that these leading terms in the non-local components of the force are integrated to zero once they are multiplied by $a^+_{E/L}$.

We now compare our results with the conservative scattering angle computed using amplitudes methods in Ref.~\cite{Barack:2023oqp}. In their framework, the scattering angle up to 4PM was obtained to \textit{all orders in $v$}, but with the presence of two free parameters, namely $c_1$ and $c_2(\bar\mu)$, that appear precisely at 4PM. Finding immediate agreement at 2PM and 3PM (to our precision in the velocity expansion), we now focus our attention on the 4PM scattering angle, which is \cite{Barack:2023oqp}
\begin{align}\nonumber
\label{eq:BarackBern4pmchi}
    \chi_{\rm{cons}}^{\rm{4PM}} &= \dfrac{3\pi \e M^4}{8b^4} \dfrac{1}{\sigma^2-1}\Biggl\{ - \Biggl[4\mathcal{M}^t_4\log\left( \dfrac{\sqrt{\sigma^2-1}}{2}\right) + \mathcal{M}_{4}^{\pi^2}\\
    &\quad- \mathcal{M}_4^{\rm{rem}}\Biggr] + (\sigma^2-5)c_1 + \biggl[c_2(\bar{\mu}) -\dfrac{31}{3}  \nonumber \\
    &\quad + 2\log\left(2b^2e^{2\gamma_E}\bar{\mu}^2\right) \biggr](\sigma^2 - 1) \Biggr\},
\end{align}
where $\sigma=(1-v^2)^{-1/2}$ and the functions $\mathcal{M}_4^t$, $\mathcal{M}_4^{\pi^2}$ and $\mathcal{M}_4^{\rm{rem}}$ are complicated functions of $\sigma$ given in Eqs.~(5.7)--(5.8) of Ref.~\cite{Barack:2023oqp}.

The low-velocity expansion of the 4PM scattering angle~\eqref{eq:BarackBern4pmchi} gives
\begin{align} \nonumber
    \chi_{\rm{cons}}^{\rm{4PM}} &= -\frac{9 \pi \e M^4 }{4 b^4 v^4}\Bigg\{ 1 + v^2 \left[ \frac{85}{54}+\frac{2 c_1}{3} +\frac{7 \pi ^2}{96}-\frac{4 \log2}{9} \right. \\
    &\quad \left. +\frac{4 \log v}{9} \right] + v^4 \left[ \frac{313}{1080}-\frac{5 c_1}{6}-\frac{c_2(\bar{\mu})}{6} -\frac{2 \gamma_E}{3} \right. \nonumber \\
    &\quad \left. +\frac{1445 \pi ^2}{6144}-\frac{8 \log2}{9} +\frac{8 \log v}{9}-\frac{1}{3} \log \left(2 b^2 \bar{\mu} ^2\right) \right] \nonumber \\
    &\quad +O(v^5)  \Bigg\}.
\end{align}
An important feature of this expression is that $c_1$ and $c_2(\bar{\mu})$ appear independently and at successive PN orders. It is thus clear from this expression that by matching the powers of $v$ with our 4PM scattering angle~\eqref{cons_chi_4pm_res}, we will have sufficient conditions to uniquely determine the two coefficients. In particular we find 
\begin{align}\label{c1_c2}
    c_1  = \dfrac{1}{6}, \hspace{0.2cm}
    c_2 (\bar{\mu})  = -\frac{11}{6}-4 \gamma_E -2 \log (2)-4 \log (\bar{\mu} M ).
\end{align}

It was suggested at the end of Sec.~3.2 of Ref.~\cite{Barack:2023oqp} that $c_1$ could be associated with a static Love number, which vanishes for a black hole in general relativity, whereas $c_2(\bar{\mu})$ arises from a regularization procedure. In Ref.~\cite{Barack:2023oqp} some constraints on these parameters were found by comparing the analytical PM scattering angle to fully relativistic, numerical self-force calculations, with the scale fixed to $\bar{\mu}=(2M)^{-1}$. Specifically they found for the first coefficient $c_1=0.31\pm0.38$, whereas for the second an interval was presented in the approximate range $(-25,-35)$. Clearly our value for $c_1$
is compatible with the numerical values they find inside their error bars. If $c_1$ is proportional to a static Love number it must be zero, but the case of a non-vanishing $c_1$ was already considered to be possible in Ref.~\cite{Barack:2023oqp} because of the field redefinitions used in the construction of the leading tidal action. For $c_2$ there is no particular theoretical value one might expect. Our $c_2$, with $\bar\mu=(2M)^{-1}$, has a numerical value $c_2\approx-2.76$, in disagreement with the numerically obtained range, but as we discuss further below, we do not believe this implies an inconsistency between our analytical calculation and the numerical calculation of Ref.~\cite{Barack:2023oqp}.

\section{Conclusion}

In this work we have given, for the first time, a methodology for obtaining analytical, high-order post-Minkowskian results for hyperbolic scattering of two black holes within the self-force framework. Working at linear order in the mass ratio, we followed the traditional, MST-based methods of analytical self-force but extended their validity to continuous Fourier spectra. Our framework, while valid also for gravitational perturbations, was explicitly implemented in a scalar self-force model, where we solved the massless Klein-Gordon equation and, by following the procedure first described in \cite{Barack:2022pde}, computed the self-force correction to the scattering angle at 4.5PN and 5PM accuracy.

Comparing with results in the literature obtained using scattering amplitudes methods, we were able to completely fix the two unknown parameters $c_1$ and $c_2(\bar{\mu})$ that appeared in the analytical results of Ref.~\cite{Barack:2023oqp}. Our result for $c_1$ is consistent with the range of possible values that Ref.~\cite{Barack:2023oqp} obtained by fitting to numerical self-force data. Our result for $c_2(\bar{\mu})$ disagrees with their best-fit value(s), but we emphasise that in Ref.~\cite{Barack:2023oqp}, the authors ``conclude that [they] have insufficient data to extract the two unknown coefficients $c_1$ and $c_2$ individually.'' We therefore do not have cause to doubt our values (which, we note, were confirmed using three independent codes). A valuable complementary approach would be to obtain values of these constants following Ref.~\cite{Ivanov:2024sds}. In the meantime, we highlight that our values have recently been used directly in Ref.~\cite{Long:2024ltn}.

Although our calculations are limited to a linear scalar problem, our methods should open a new path to self-force calculations for the unbound problem more generally, complementing the numerical approach in Refs.~\cite{Long:2021ufh,Barack:2022pde,Whittall:2023xjp,Barack:2023oqp}. In particular, we have shown how  to generalize analytical self-force theory to the case of a continuous Fourier spectrum.

In a subsequent paper we will calculate the gravitational fluxes produced by hyperbolic scattering of a massive particle. Future calculations will be devoted to reconstructing the perturbed metric and hence calculate the gravitational self-force correction to the scattering angle. Ultimately, we expect analytical self-force calculations such as these to be a starting point for the types of resummations explored in Ref.~\cite{Long:2024ltn}, extending the accuracy of the analytical results into the regime of strong fields and large velocities.  

\section*{Acknowledgments}

C.K. and A.P. particularly thank Rafael Porto for motivating this work. We also thank Leor Barack and Olly Long for many helpful discussions.  
D.B. acknowledges the sponsorship of the Italian Gruppo Nazionale per la
Fisica Matematica (GNFM) of the Istituto Nazionale
di Alta Matematica (INDAM), as well as the hospitality and the highly stimulating environment of the Institut des Hautes Etudes Scientifiques where this project was started.
D.U. acknowledges Prof. S. Capozziello for continuous encouragement at all stages during the development of this project and the Scuola Superiore Meridionale's sponsorship for providing the possibility to visit UCD School of Mathematics and Statistics. A.P. acknowledges the support of a Royal Society University Research Fellowship and the UKRI Frontier Research Grant GWModels, as selected by the ERC and funded by UKRI under the Horizon Europe Guarantee scheme [grant number EP/Y008251/1]. CK acknowledges support from Science Foundation Ireland under Grant number 21/PATH-S/9610.

\appendix

\section{Distributional Fourier transform of a logarithm}\label{appendix_B}

In this appendix we review the derivation of Eq.~\eqref{Eq:logmiwFT}. For simplicity, we consider the Fourier integral 
\begin{equation} \nonumber
     \hat{g}(s) = \int_{-\infty}^{+\infty} \dfrac{d\omega}{2\pi} e^{is\omega}\log\omega,
\end{equation}
with $\omega,s \in \mathbb{R}$. The integral in Eq.~\eqref{Eq:logmiwFT} is obtained straightforwardly from this one.

We define the integral as a distribution in the usual way. We denote $g(\omega)=\log\omega$ and $\hat{g}$ its Fourier transform, written as
\begin{equation}
    \hat{g}(s) = \mathcal{F}\left\{g\right\}(s).
\end{equation}
For any test function $\phi \in \mathbb{S}$, the Fourier transform of a distribution $g$ is defined to act as
\begin{equation}\label{Fourier def}
    \langle\mathcal{F}\left\{g\right\},\phi\rangle = \langle g,\mathcal{F}\left\{\phi\right\}\rangle,
\end{equation}
where $\mathbb{S}$ is the Schwartz space,
\begin{multline}
    \hspace{-0.3cm} \mathbb{S} = \Bigl\{ \phi: \mathbb{R} \rightarrow \mathbb{C} \hspace{0.05cm} \Big|\\ \hspace{0.05cm} \phi \hspace{0.05cm} {\rm{is}} \hspace{0.05cm} \mathbb{C}^\infty , \sup_{x\in\mathbb{R}}|x^n \phi^{(m)}(x)| < \infty , \forall \hspace{0.05cm} n,m \in \mathbb{N}  \Bigr\}.
\end{multline}

We now define
\begin{equation}
    \hat{g}_\epsilon(s) = \int_{-\infty}^{+\infty} \dfrac{d\omega}{2\pi} e^{-\epsilon|\omega|} e^{is\omega}\log\omega.
\end{equation}
It is easy to see from the definition~\eqref{Fourier def} that 
\begin{equation}
    \hat{g}(s) = \lim_{\epsilon\rightarrow0^+}\hat{g}_\epsilon(s).
\end{equation}
Here the limit is defined in the distributional sense, $\langle \lim_{\epsilon\rightarrow0^+}\hat{g}_\epsilon,\phi\rangle=\lim_{\epsilon\rightarrow0^+}\langle \hat{g}_\epsilon,\phi\rangle$.

$\hat{g}_\epsilon$ is an ordinary integral which can be readily evaluated. Explicitly splitting the result into its real and imaginary parts gives
\begin{align} 
    \hat{g}_\epsilon 
    & = \dfrac{1}{2\pi}\left(\hat{g}^{(1)}_\epsilon(s)+\hat{g}^{(2)}_\epsilon(s)+\hat{g}^{(3)}_\epsilon(s)+\hat{g}^{(4)}_\epsilon(s)\right),
\end{align}
with
\begin{align} \nonumber
& \hat{g}^{(1)}_\epsilon(s)=\dfrac{\pi s}{s^2+\epsilon^2}, \hspace{0.15cm}\hat{g}^{(2)}_\epsilon(s)= (i\pi - 2\gamma_E)\dfrac{\epsilon}{s^2+\epsilon^2}, \\ 
& \hat{g}^{(3)}_\epsilon(s)= - \dfrac{2s\arctan{s/\epsilon}}{s^2+\epsilon^2}, \hspace{0.15cm}\hat{g}^{(4)}_\epsilon(s)=- \dfrac{\epsilon\log(s^2+\epsilon^2)}{s^2+\epsilon^2}.
\end{align}
Here $\gamma_E$ is the Euler constant. We next study the action of each $\hat{g}^{(i)}_\epsilon$ on a test function $\phi$. 

\subsection{$\hat{g}^{(1)}_\epsilon(s)$}
We have to compute
\begin{align}\label{eq:first_Fourier} \nonumber
    &\lim_{\epsilon\rightarrow0^+}\langle \hat{g}^{(1)}_\epsilon,\phi\rangle=\pi\lim_{\epsilon\rightarrow0^+} \int_{-\infty}^{+\infty} ds \dfrac{ s}{s^2+\epsilon^2} \phi(s) = \\  
    & = \pi\;  {\rm p.v.}\!\int_{-\infty}^{+\infty} ds \dfrac{\phi(s)}{s} = \pi\int_{0}^{+\infty} ds \dfrac{\phi(s)-\phi(-s)}{s},
\end{align}
where p.v. denotes the principal value. Hence  in the distributional sense we have
\begin{equation}
   \hat{g}^{(1)}(s) = \lim_{\epsilon\to0^+}\hat{g}^{(1)}_\epsilon(s) = \pi\, {\rm p.v.}\!\left(\dfrac{1}{s}\right)
\end{equation}
where we define ${\rm p.v.}\!\left(\dfrac{1}{s}\right)$ through its action on $\phi$ as in Eq.~\eqref{eq:first_Fourier}.

\subsection{$\hat{g}^{(2)}_\epsilon(s)$}
We continue by computing
\begin{align}\label{eq:second_Fourier} \nonumber
    \lim_{\epsilon\rightarrow0^+}\langle \hat{g}^{(2)}_\epsilon,\phi\rangle &= (i\pi - 2\gamma_E)\lim_{\epsilon\rightarrow0^+} \int_{-\infty}^{+\infty} ds \dfrac{\epsilon}{s^2+\epsilon^2} \phi(s) \\*
    & = (i\pi - 2\gamma_E) \lim_{\epsilon\rightarrow0^+} \int_{-\infty}^{+\infty} dx \dfrac{1}{1+x^2}\phi(\epsilon x) \nonumber\\*
    &= (i\pi-2\gamma_E)\pi \phi(0),
\end{align}
and in the distributional sense we have
\begin{equation}
   \hat{g}^{(2)}(s) =\lim_{\epsilon\to0^+}\hat{g}^{(2)}_\epsilon(s) = (i\pi - 2\gamma_E)\pi \delta(s).
\end{equation}

\subsection{$\hat{g}^{(3)}_\epsilon(s)$}
The action of $\hat{g}^{(3)}_\epsilon(s)$ on a test function is  the most complicated because of the $\arctan(s/\epsilon)$ and must be handled carefully. This term does not have a well-defined limit at $\epsilon\to0^+$, implying we must work at finite (small) $\epsilon$. We find
\begin{widetext}
\begin{align}\label{eq:third_Fourier} \nonumber
    \langle \hat{g}^{(3)}_\epsilon,\phi\rangle &=  -2\left\{ \int_{|s|\leq1} ds  \dfrac{s\arctan{s/\epsilon}}{s^2+\epsilon^2} \phi(s) + \int_{|s|\geq1} ds  \dfrac{s\arctan{s/\epsilon}}{s^2+\epsilon^2} \phi(s) \right\}  \\ \nonumber
    & = -2\left\{\int_{|s|\leq1} ds  \dfrac{s\arctan{s/\epsilon}}{s^2+\epsilon^2} \phi(0)+ \int_{|s|\leq1} ds  \dfrac{s\arctan{s/\epsilon}}{s^2+\epsilon^2} \left(\phi(s)-\phi(0)\right) + \int_{|s|\geq1} ds  \dfrac{s\arctan{s/\epsilon}}{s^2+\epsilon^2} \phi(s) \right\}  \nonumber \\
    & = -2\left\{\phi(0)\int_{|s|\leq1} ds  \dfrac{s\arctan{s/\epsilon}}{s^2+\epsilon^2} + \frac{\pi}{2} \int_{|s|\leq1} ds  \dfrac{\phi(s)-\phi(0)}{|s|}+  \frac{\pi}{2} \int_{|s|\geq1} ds  \dfrac{\phi(s)}{|s|} \right\} \nonumber \\
    & = (2\pi\log\epsilon+2\pi\log2)\phi(0) + O(\epsilon) - \pi \left\{ \int_{|s|\leq1} ds  \dfrac{\phi(s)-\phi(0)}{|s|}+  \int_{|s|\geq1} ds  \dfrac{\phi(s)}{|s|} \right\},
\end{align}    
\end{widetext}
where we have immediately split the integration domain between $|s|\leq 1$ and $|s|\geq1$: the first one contains a divergence at $s=0$, while the second is finite. In order to isolate the divergence we add and remove the same integral with the test function $\phi(s)$ evaluated at $s=0$.

Hence, for small $\epsilon$, $\hat{g}^{(3)}_\epsilon$ can be written in the distributional sense as
\begin{equation}
   \hat{g}^{(3)}_\epsilon(s) = 2\pi\left\{(\log\epsilon+\log2) + O(\epsilon)\right\} \delta(s) - \pi \left( \dfrac{1}{|s|} \right)_{1},
\end{equation}
where we have defined
\begin{equation}
    \left( \dfrac{1}{|s|} \right)_{1} = \int_{|s|\leq1} ds  \dfrac{\phi(s)-\phi(0)}{|s|}+  \int_{|s|\geq1} ds  \dfrac{\phi(s)}{|s|}.
\end{equation}

\subsection{$\hat{g}^{(4)}_\epsilon(s)$}
The last piece is again non-trivial and can only be evaluated for finite $\epsilon$:
    \begin{align}\label{eq:fourth_Fourier} \nonumber
    \langle \hat{g}^{(4)},\phi\rangle&=-\int_{-\infty}^{+\infty} ds  \dfrac{\epsilon\log(s^2+\epsilon^2)}{s^2+\epsilon^2} \phi(s) \\ \nonumber
    &= -\left\{ 2\log\epsilon \int_{-\infty}^\infty dx \dfrac{\phi(\epsilon x)}{1+x^2} \right.\nonumber\\
    &\qquad \left.+ \int_{-\infty}^\infty dx \dfrac{\log(1+x^2)}{1+x^2}\phi(\epsilon x)\right\} \nonumber\\
    &=  -(2\pi\log\epsilon+2\pi\log2)\phi(0) + O(\epsilon),
\end{align}

Hence $\hat{g}^{(4)}_\epsilon$ can be written in the distributional sense as
\begin{equation}
   \hat{g}^{(4)}_\epsilon (s) = -\left\{(2\pi\log\epsilon+2\pi\log2) + O(\epsilon)\right\} \delta(s) .
\end{equation}

\subsection{Final result}
At this point we can combine the above results to obtain $\mathcal{F}\left\{ \log\omega \right\}(s) = \hat{g}(s)$:
\begin{equation}
\mathcal{F}\left\{ \log\omega \right\}= \dfrac{1}{2\pi}\lim_{\epsilon\to0}\left\{\hat{g}^{(1)}_\epsilon+\hat{g}^{(2)}_\epsilon+\hat{g}^{(3)}_\epsilon+\hat{g}^{(4)}_\epsilon\right\},
\end{equation}
which evaluates to
\begin{multline}
   \mathcal{F}\left\{ \log\omega \right\}(s) \\={\rm p.v.}\!\left(\dfrac{1}{2s}\right)
   - \left(\dfrac{1}{2|s|}\right)_1 -\left(\gamma_E  - \frac{i\pi}{2}\right) \delta(s).\label{IFT log omega}
\end{multline}
We see that the terms involving $\log\epsilon$ have cancelled, as they must given the definition~\eqref{Fourier def}. By following the same steps, it is easy to prove that
\begin{align}
    \mathcal{F}\left\{ \log(-i\omega) \right\}(s) &= {\rm p.v.}\!\left(\dfrac{1}{2s}\right)
    - \left(\dfrac{1}{2|s|}\right)_1 -  \gamma_E  \delta(s).
\end{align}

As a final remark we emphasize that we split the integration domain between $|s|\leq1$ and $|s|\geq1$ for simplicity. One could redo all the computations with a generic partition of the integration domain, for instance
\begin{equation}
    \left( \dfrac{1}{|s|} \right)_{\alpha} = \int_{|s|\leq\alpha} ds  \dfrac{\phi(s)-\phi(0)}{|s|}+  \int_{|s|\geq\alpha} ds  \dfrac{\phi(s)}{|s|}.\label{(1/|s|)_alpha def}
\end{equation}
Different choices of $\alpha$ will move terms between these contributions and the Dirac delta contributions, leaving the total distribution $\mathcal{F}\left\{ \log(-i\omega) \right\}$ unaltered. Concretely,
\begin{multline}    
    \mathcal{F}\left\{ \log(-i\omega) \right\}(s) \\= {\rm p.v.}\!\left(\dfrac{1}{2s}\right)
    - \left(\dfrac{1}{2|s|}\right)_\alpha - (\gamma_E+\log\alpha)  \delta(s).
\end{multline}

\section{Derivation of the logarithmic terms in the scalar field} \label{appendix_A}

Here we give the full derivation of the logarithmic contributions to the field described in Sec.~\ref{nonlocal contributions generic}, specialising to $s=0$. The discussion will be for a generic $(l,m)$ mode and $k=1$. 

We have to compute integrals of the form
\begin{multline}
    I = \int_{-\infty}^\infty dt' {}_0 G_{\ell m}^{(j,1)}(r,r(t'))e^{-im\phi(t')}\\
    \times\int_{-\infty}^\infty \frac{d\omega}{2\pi}e^{i\omega (t'-t)} \omega^j \log\omega,
\end{multline}
where $t$, $t'$, and $\omega$ are coordinate times and frequency, i.e. not yet rescaled with $b,v$.

There are several ways to deal with this integral. 
Here we work with the convolution theorem, which states
\begin{align}
    \int \frac{d\omega }{2\pi}e^{i\omega \xi} f(\omega) g(\omega)&=\int dy\, \hat{g}(y)\hat{f}(\xi-y)\nonumber\\
&=\int dy \hat{f}(y)\hat{g}(\xi-y)\,,
\end{align}
where hatted quantities denote inverse Fourier transforms (IFTs), and we have $ f(\omega)=\omega^j$, $ g(\omega)=\log\omega$. 
Knowing that $\hat f(t) = i^{-j}\delta^{(j)}(t)$, we can write
\begin{align} \label{I_2}
    I &= \int dt' {}_0 G_{\ell m}^{(j,1)}(r,r(t'))e^{-im\phi(t')} \int \frac{d\omega}{2\pi}e^{i\omega (t'-t)} \omega^j \log\omega \nonumber \\
    &=
i^{-j}\int dy\,\hat{g}(y)\nonumber\\
&\quad\times\int dt'\, {}_0 G_{\ell m}^{(j,1)}(r,r(t'))e^{-im\phi(t')}\delta^{(j)}(t'-t-y)\nonumber\\
&=i^{-j}(-1)^j\nonumber\\
&\quad\times\int dy\,\hat{g}(y)\frac{d^j}{dt'^j}\left[ {}_0 G_{\ell m}^{(j,1)}(r,r(t'))e^{-im\phi(t')} \right]\bigg\vert_{t'=t+y}\nonumber\\
&=i^{j}\int_{-\infty}^\infty dy\,c_{lm}(r,t+y)\hat{g}(y)
\,,
\end{align}
with
\begin{equation}\label{def:derivatives_GF}
    c_{lm}(r,t+y)=\frac{d^j}{dt^j}\left[{}_0 G_{\ell m}^{(j,1)}(r,r(t+y))e^{-im\phi(t+y)}\right].\!
\end{equation}
The remaining integral can be handled using the distributional definition of $\hat g$ (the IFT of $\log\omega$), as calculated in Appendix~\ref{appendix_B}. We describe this approach in the first subsection below.

An alternative approach is to use the Hadamard \emph{partie finie} procedure to evaluate both the IFT and the remaining integral~\eqref{I_2}. We describe that approach at the end of this appendix. While the partie finie usually leads to simpler results in PN theory, we show that here it leads to a result involving an arbitrary length scale. We fix the length scale by comparing to the distributional definition. 

\subsection{Distributional inverse Fourier transform}


The distributional IFT of $\log\omega$, denoted $\hat g$ in Eq.~\eqref{I_2}, is computed in detail in Appendix~\ref{appendix_B}. It is given by Eq.~\eqref{IFT log omega} with the definitions Eq.~\eqref{(1/|y|)_1 def} and~\eqref{pv(1/y) def}.
We now plug this expression into Eq. \eqref{I_2}, and we find
\begin{widetext}
\begin{align} \label{I_2_distr}\nonumber
    I 
   &= i^j \left\{ -\int dy \, c_{lm}(r,t+y)\left(\gamma_E  - \frac{i\pi}{2}\right)\delta(y)  + \dfrac{1}{2}  \int dy \left[{\rm p.v.}\left( \dfrac{1}{y} \right) - \left( \dfrac{1}{|y|} \right)_1 \right] c_{lm}(r,t+y)   \right\} \nonumber \\
    &  = -i^j\left\{ \left(\gamma_E - \dfrac{i\pi}{2} \right)c_{lm}(r,t)  + \left[ \int_0^1 dy \frac{c_{lm}(r,t-y) -c_{lm}(r,t)}{y}+\int_1^\infty dy \frac{c_{lm}(r,t-y)}{y}\right] \right\}\,.
\end{align}
We observe that the terms in square brackets can equivalently be written in the form we use in Eq.~\eqref{psiNL v2}, as $\displaystyle\lim_{\epsilon\to0}\left[c_{lm}(r,t)\log \epsilon + \int_\epsilon^\infty dy \frac{c_{lm}(r,t-y)}{y}\right]$.


By following the same steps it is possible to compute the analogous result with $\hat g$ now the IFT of $\log(-i\omega)$. In this case we find
\begin{align}
   I&=-i^j \bigg\{\gamma_E \,c_{lm}(r,t) +\int_{1}^{\infty}\, \frac{c_{lm}(r,t-y)}{y}dy  + \int_{0}^{1} \, \frac{c_{lm}(r,t-y)-c_{lm}(r,t)}{y} dy \bigg\}.
\end{align}

\end{widetext}

\subsection{Hadamard \emph{partie finie} regularization}

We now consider the alternative of adopting a partie finie (Pf) regularization of the IFT and of Eq.~\eqref{I_2}. The Pf of a divergent integral can be defined in several equivalent ways. We write it as 
\begin{equation}
    \Pf_{y_0}\int f(y)dy = \FP_{B\to0}\int (y/y_0)^B f(y) dy,
\end{equation}
where $y_0$ is an arbitrary length scale, $B$ is a complex number, and the `finite part' operation $\FP_{B\to0}$ picks out the coefficient of $B^0$ in a Laurent series around $B=0$.

In the case of the IFT of $\log \omega$, we can write
\begin{multline}
    \Pf_{\omega_0}\int_{-\infty}^\infty e^{i\omega t}\log\omega\, d\omega \\= \FP_{B\to0}\FP_{\epsilon\to0}\int_{-\infty}^\infty e^{i\omega t}(\omega/\omega_0)^B \frac{\omega^\epsilon}{\epsilon} \, d\omega,\label{Pf IFT log}
\end{multline}
where we have used $\FP_{\epsilon\to0}\displaystyle\frac{\omega^\epsilon}{\epsilon}=\log\omega$ together with the fact that Pf applied to a convergent integral returns the integral itself, and we work in the region $-1<{\rm Re}\,B<0$.\footnote{Note any arbitrary scale $\omega_1$ inserted into this second finite part cannot contribute to the IFT because $\Pf_{\omega_1}\int_{-\infty}^\infty e^{i\omega t} d\omega=0$ for all $\omega_1$.} Since the integrand no longer involves logarithms, it can be straightforwardly evaluated to find
\begin{equation}
     \hat g_{\Pf} \equiv \frac{1}{2\pi}\Pf_{\omega_0}\int_{-\infty}^\infty e^{i\omega t}\log\omega\, d\omega = \frac{1}{2y}-\frac{1}{2|y|}, 
\end{equation}
which is to be compared to Eq.~\eqref{IFT log omega}. This result is independent of the arbitrary scale $\omega_0$.

To obtain the Pf-regularized Eq.~\eqref{I_2}, we first consider a generic version of the integral,
\begin{multline}
    \Pf_{y_0}\int_{-\infty}^\infty \phi(y)\hat g_{\rm Pf}(y)dy \\*
    = \FP_{B\to0}\int^\infty_{-\infty}\frac{y^{B}}{2y_0^B}\left(\frac{1}{y}-\frac{1}{|y|}\right)\phi(y)dy  \label{Pf ghatPf def}
\end{multline}
for an arbitrary test function $\phi$. The first term immediately yields
\begin{equation}
    \FP_{B\to0}\int^\infty_{-\infty}\frac{y^{B-1}}{2y_0^B}\phi(y)dy = {\rm p.v.}\int^\infty_{-\infty}\frac{\phi(y)}{2y}dy,\label{Pf1/y}
\end{equation}
where p.v. denotes the principal value, defined in Eq.~\eqref{pv(1/y) def}. The second term in Eq.~\eqref{Pf ghatPf def} yields
\begin{align}
     \Pf_{y_0}&\int_{-\infty}^\infty \frac{\phi(y)}{2|y|}dy \nonumber\\*
     &= \FP_{B\to0}\int^\infty_0\frac{(y/y_0)^{B}\phi(y)+(-y/y_0)^{B}\phi(-y)}{2y}dy\nonumber\\*
     &= \FP_{B\to0}\int^\alpha_0\frac{(y/y_0)^{B}\phi(y)+(-y/y_0)^{B}\phi(-y)}{2y}dy \nonumber\\*
     &\quad + \int_\alpha^\infty\frac{\phi(s)+\phi(-s)}{2y}dy,\label{Pf1/|y| def}
\end{align}
where $\alpha>0$ is arbitrary and we have immediately set $B=0$ in the integral from $\alpha$ to $\infty$ because the integral is convergent. We now subtract and add $\phi(0)\FP_{B\to0}\int_0^\alpha \frac{(y/y_0)^B+(-y/y_0)^B}{2y}dy$ to obtain 
\begin{align}
     \Pf_{y_0}&\int_{-\infty}^\infty \frac{\phi(y)}{2|y|}dy \nonumber\\
     &= \int^\alpha_0\frac{\phi(y)+\phi(-y)-2\phi(0)}{2y}dy \nonumber\\
     &\quad + \phi(0)\FP_{B\to0}\int^\alpha_0\frac{(y/y_0)^{B}+(-y/y_0)^{B}}{2y}dy \nonumber\\
     &\quad + \int_\alpha^\infty\frac{\phi(s)+\phi(-s)}{2y}dy.
\end{align}
Here we have removed the FP operation on the first term because it is now convergent. The remaining Pf integral immediately evaluates to
\begin{multline}
    \phi(0)\FP_{B\to0}\int^\alpha_0\frac{(y/y_0)^{B}+(-y/y_0)^{B}}{2y}dy \\
    = \left[\frac{i\pi}{2}+\log(\alpha/y_0)\right]\phi(0).\label{Pf1/|y|}
\end{multline}
Given these results, we can rewrite Eq.~\eqref{Pf1/|y| def} as the action of a distribution $\Pf_{y_0}\left(\frac{1}{2|y|}\right)$ on a test function $\phi$, where
\begin{equation}
    \Pf_{y_0}\left(\frac{1}{2|y|}\right) = \left(\frac{1}{2|y|}\right)_{\alpha} + \left[\frac{i\pi}{2}+\log(\alpha/y_0)\right]\delta(y)\label{Pf1/|y| distribution}
\end{equation}
with $\left(\frac{1}{|y|}\right)_{\alpha}$ defined in Eq.~\eqref{(1/|s|)_alpha def}. Equation~\eqref{Pf1/|y| distribution} is independent of $\alpha$, for the reason explained below Eq.~\eqref{(1/|s|)_alpha def}. However, it depends on the arbitrary regularization scale~$y_0$.

Combining Eqs.~\eqref{Pf1/y} and \eqref{Pf1/|y|}, we can write the partie finie integral \eqref{Pf ghatPf def} as the action of a distribution, 
\begin{multline}
    \Pf_{y_0}\hat g_{\Pf}(y) \\*
    ={\rm p.v.}\!\left(\dfrac{1}{2y}\right)
   - \left(\dfrac{1}{2|y|}\right)_1 +\left(\log y_0  - \frac{i\pi}{2}\right) \delta(y),\label{Pf gPf distribution}
\end{multline}
on $\phi$. Here we have  chosen $\alpha=1$ (without loss of generality).

From this calculation, we observe that the partie finie introduces an arbitrary constant $y_0$ into the calculation. We can fix this constant by comparing to the unambiguous result~\eqref{IFT log omega}. The result is\footnote{The fact that $y_0$ is negative might be counter-intuitive. However, note that our original integrand $\frac{1}{2y}-\frac{1}{2|y|}$ is only nonzero for negative values of $y$.}
\begin{equation}
    y_0 = - e^{-\gamma_E}.
\end{equation}
An alternative method of determining $y_0$ would be via a matching calculation in which an observable is calculated in our PM-PN limit and compared to a calculation of the same observable from a separate regime that does not suffer from the same divergent integrals. However, we make two comments: First, $y_0$ does not depend on any physical scale in the problem. Second, the solution for our scalar field is manifestly finite in the PM-PN limit when formulated as an integral of a time-domain retarded Green function against the time-domain charge density; any superficial divergence that appears is then a consequence of the treatment of the forward and inverse Fourier transforms, and our distributional treatment of the inverse Fourier transform should automatically ensure we recover the same finite result we would if we worked entirely in the time domain. We therefore conclude that the appearance of an arbitrary constant must be a failure of the partie finie method rather than a breakdown of our perturbative expansions.

\bibliography{scalarpaperbib}

\begin{thebibliography}{79}%
\makeatletter
\providecommand \@ifxundefined [1]{%
 \@ifx{#1\undefined}
}%
\providecommand \@ifnum [1]{%
 \ifnum #1\expandafter \@firstoftwo
 \else \expandafter \@secondoftwo
 \fi
}%
\providecommand \@ifx [1]{%
 \ifx #1\expandafter \@firstoftwo
 \else \expandafter \@secondoftwo
 \fi
}%
\providecommand \natexlab [1]{#1}%
\providecommand \enquote  [1]{``#1''}%
\providecommand \bibnamefont  [1]{#1}%
\providecommand \bibfnamefont [1]{#1}%
\providecommand \citenamefont [1]{#1}%
\providecommand \href@noop [0]{\@secondoftwo}%
\providecommand \href [0]{\begingroup \@sanitize@url \@href}%
\providecommand \@href[1]{\@@startlink{#1}\@@href}%
\providecommand \@@href[1]{\endgroup#1\@@endlink}%
\providecommand \@sanitize@url [0]{\catcode `\\12\catcode `\$12\catcode
  `\&12\catcode `\#12\catcode `\^12\catcode `\_12\catcode `\%12\relax}%
\providecommand \@@startlink[1]{}%
\providecommand \@@endlink[0]{}%
\providecommand \url  [0]{\begingroup\@sanitize@url \@url }%
\providecommand \@url [1]{\endgroup\@href {#1}{\urlprefix }}%
\providecommand \urlprefix  [0]{URL }%
\providecommand \Eprint [0]{\href }%
\providecommand \doibase [0]{https://doi.org/}%
\providecommand \selectlanguage [0]{\@gobble}%
\providecommand \bibinfo  [0]{\@secondoftwo}%
\providecommand \bibfield  [0]{\@secondoftwo}%
\providecommand \translation [1]{[#1]}%
\providecommand \BibitemOpen [0]{}%
\providecommand \bibitemStop [0]{}%
\providecommand \bibitemNoStop [0]{.\EOS\space}%
\providecommand \EOS [0]{\spacefactor3000\relax}%
\providecommand \BibitemShut  [1]{\csname bibitem#1\endcsname}%
\let\auto@bib@innerbib\@empty
\bibitem [{\citenamefont {Abbott}\ \emph {et~al.}(2016)\citenamefont {Abbott}
  \emph {et~al.}}]{LIGOScientific:2016aoc}%
  \BibitemOpen
  \bibfield  {author} {\bibinfo {author} {\bibfnamefont {B.~P.}\ \bibnamefont
  {Abbott}} \emph {et~al.} (\bibinfo {collaboration} {LIGO Scientific,
  Virgo}),\ }\bibfield  {title} {\bibinfo {title} {{Observation of
  Gravitational Waves from a Binary Black Hole Merger}},\ }\href
  {https://doi.org/10.1103/PhysRevLett.116.061102} {\bibfield  {journal}
  {\bibinfo  {journal} {Phys. Rev. Lett.}\ }\textbf {\bibinfo {volume} {116}},\
  \bibinfo {pages} {061102} (\bibinfo {year} {2016})},\ \Eprint
  {https://arxiv.org/abs/1602.03837} {arXiv:1602.03837 [gr-qc]} \BibitemShut
  {NoStop}%
\bibitem [{\citenamefont {Abbott}\ \emph {et~al.}(2017)\citenamefont {Abbott}
  \emph {et~al.}}]{LIGOScientific:2017vwq}%
  \BibitemOpen
  \bibfield  {author} {\bibinfo {author} {\bibfnamefont {B.~P.}\ \bibnamefont
  {Abbott}} \emph {et~al.} (\bibinfo {collaboration} {LIGO Scientific,
  Virgo}),\ }\bibfield  {title} {\bibinfo {title} {{GW170817: Observation of
  Gravitational Waves from a Binary Neutron Star Inspiral}},\ }\href
  {https://doi.org/10.1103/PhysRevLett.119.161101} {\bibfield  {journal}
  {\bibinfo  {journal} {Phys. Rev. Lett.}\ }\textbf {\bibinfo {volume} {119}},\
  \bibinfo {pages} {161101} (\bibinfo {year} {2017})},\ \Eprint
  {https://arxiv.org/abs/1710.05832} {arXiv:1710.05832 [gr-qc]} \BibitemShut
  {NoStop}%
\bibitem [{\citenamefont {Amaro-Seoane}\ \emph {et~al.}(2017)\citenamefont
  {Amaro-Seoane} \emph {et~al.}}]{LISA:2017pwj}%
  \BibitemOpen
  \bibfield  {author} {\bibinfo {author} {\bibfnamefont {P.}~\bibnamefont
  {Amaro-Seoane}} \emph {et~al.} (\bibinfo {collaboration} {LISA}),\ }\bibfield
   {title} {\bibinfo {title} {{Laser Interferometer Space Antenna}},\
  }\href@noop {} {\  (\bibinfo {year} {2017})},\ \Eprint
  {https://arxiv.org/abs/1702.00786} {arXiv:1702.00786 [astro-ph.IM]}
  \BibitemShut {NoStop}%
\bibitem [{\citenamefont {Punturo}\ \emph {et~al.}(2010)\citenamefont {Punturo}
  \emph {et~al.}}]{Punturo:2010zz}%
  \BibitemOpen
  \bibfield  {author} {\bibinfo {author} {\bibfnamefont {M.}~\bibnamefont
  {Punturo}} \emph {et~al.},\ }\bibfield  {title} {\bibinfo {title} {{The
  Einstein Telescope: A third-generation gravitational wave observatory}},\
  }\href {https://doi.org/10.1088/0264-9381/27/19/194002} {\bibfield  {journal}
  {\bibinfo  {journal} {Class. Quant. Grav.}\ }\textbf {\bibinfo {volume}
  {27}},\ \bibinfo {pages} {194002} (\bibinfo {year} {2010})}\BibitemShut
  {NoStop}%
\bibitem [{\citenamefont {Pretorius}(2005)}]{Pretorius:2005gq}%
  \BibitemOpen
  \bibfield  {author} {\bibinfo {author} {\bibfnamefont {F.}~\bibnamefont
  {Pretorius}},\ }\bibfield  {title} {\bibinfo {title} {{Evolution of binary
  black hole spacetimes}},\ }\href
  {https://doi.org/10.1103/PhysRevLett.95.121101} {\bibfield  {journal}
  {\bibinfo  {journal} {Phys. Rev. Lett.}\ }\textbf {\bibinfo {volume} {95}},\
  \bibinfo {pages} {121101} (\bibinfo {year} {2005})},\ \Eprint
  {https://arxiv.org/abs/gr-qc/0507014} {arXiv:gr-qc/0507014} \BibitemShut
  {NoStop}%
\bibitem [{\citenamefont {Damour}\ \emph
  {et~al.}(2014{\natexlab{a}})\citenamefont {Damour}, \citenamefont
  {Guercilena}, \citenamefont {Hinder}, \citenamefont {Hopper}, \citenamefont
  {Nagar},\ and\ \citenamefont {Rezzolla}}]{Damour:2014afa}%
  \BibitemOpen
  \bibfield  {author} {\bibinfo {author} {\bibfnamefont {T.}~\bibnamefont
  {Damour}}, \bibinfo {author} {\bibfnamefont {F.}~\bibnamefont {Guercilena}},
  \bibinfo {author} {\bibfnamefont {I.}~\bibnamefont {Hinder}}, \bibinfo
  {author} {\bibfnamefont {S.}~\bibnamefont {Hopper}}, \bibinfo {author}
  {\bibfnamefont {A.}~\bibnamefont {Nagar}},\ and\ \bibinfo {author}
  {\bibfnamefont {L.}~\bibnamefont {Rezzolla}},\ }\bibfield  {title} {\bibinfo
  {title} {{Strong-Field Scattering of Two Black Holes: Numerics Versus
  Analytics}},\ }\href {https://doi.org/10.1103/PhysRevD.89.081503} {\bibfield
  {journal} {\bibinfo  {journal} {Phys. Rev. D}\ }\textbf {\bibinfo {volume}
  {89}},\ \bibinfo {pages} {081503} (\bibinfo {year} {2014}{\natexlab{a}})},\
  \Eprint {https://arxiv.org/abs/1402.7307} {arXiv:1402.7307 [gr-qc]}
  \BibitemShut {NoStop}%
\bibitem [{\citenamefont {Hopper}\ \emph {et~al.}(2023)\citenamefont {Hopper},
  \citenamefont {Nagar},\ and\ \citenamefont {Rettegno}}]{Hopper:2022rwo}%
  \BibitemOpen
  \bibfield  {author} {\bibinfo {author} {\bibfnamefont {S.}~\bibnamefont
  {Hopper}}, \bibinfo {author} {\bibfnamefont {A.}~\bibnamefont {Nagar}},\ and\
  \bibinfo {author} {\bibfnamefont {P.}~\bibnamefont {Rettegno}},\ }\bibfield
  {title} {\bibinfo {title} {{Strong-field scattering of two spinning black
  holes: Numerics versus analytics}},\ }\href
  {https://doi.org/10.1103/PhysRevD.107.124034} {\bibfield  {journal} {\bibinfo
   {journal} {Phys. Rev. D}\ }\textbf {\bibinfo {volume} {107}},\ \bibinfo
  {pages} {124034} (\bibinfo {year} {2023})},\ \Eprint
  {https://arxiv.org/abs/2204.10299} {arXiv:2204.10299 [gr-qc]} \BibitemShut
  {NoStop}%
\bibitem [{\citenamefont {Mino}\ \emph {et~al.}(1997)\citenamefont {Mino},
  \citenamefont {Sasaki},\ and\ \citenamefont {Tanaka}}]{Mino:1996nk}%
  \BibitemOpen
  \bibfield  {author} {\bibinfo {author} {\bibfnamefont {Y.}~\bibnamefont
  {Mino}}, \bibinfo {author} {\bibfnamefont {M.}~\bibnamefont {Sasaki}},\ and\
  \bibinfo {author} {\bibfnamefont {T.}~\bibnamefont {Tanaka}},\ }\bibfield
  {title} {\bibinfo {title} {{Gravitational radiation reaction to a particle
  motion}},\ }\href {https://doi.org/10.1103/PhysRevD.55.3457} {\bibfield
  {journal} {\bibinfo  {journal} {Phys. Rev. D}\ }\textbf {\bibinfo {volume}
  {55}},\ \bibinfo {pages} {3457} (\bibinfo {year} {1997})},\ \Eprint
  {https://arxiv.org/abs/gr-qc/9606018} {arXiv:gr-qc/9606018} \BibitemShut
  {NoStop}%
\bibitem [{\citenamefont {Poisson}\ \emph {et~al.}(2011)\citenamefont
  {Poisson}, \citenamefont {Pound},\ and\ \citenamefont
  {Vega}}]{Poisson:2011nh}%
  \BibitemOpen
  \bibfield  {author} {\bibinfo {author} {\bibfnamefont {E.}~\bibnamefont
  {Poisson}}, \bibinfo {author} {\bibfnamefont {A.}~\bibnamefont {Pound}},\
  and\ \bibinfo {author} {\bibfnamefont {I.}~\bibnamefont {Vega}},\ }\bibfield
  {title} {\bibinfo {title} {{The Motion of point particles in curved
  spacetime}},\ }\href {https://doi.org/10.12942/lrr-2011-7} {\bibfield
  {journal} {\bibinfo  {journal} {Living Rev. Rel.}\ }\textbf {\bibinfo
  {volume} {14}},\ \bibinfo {pages} {7} (\bibinfo {year} {2011})},\ \Eprint
  {https://arxiv.org/abs/1102.0529} {arXiv:1102.0529 [gr-qc]} \BibitemShut
  {NoStop}%
\bibitem [{\citenamefont {Pound}\ and\ \citenamefont
  {Wardell}(2021)}]{Pound:2021qin}%
  \BibitemOpen
  \bibfield  {author} {\bibinfo {author} {\bibfnamefont {A.}~\bibnamefont
  {Pound}}\ and\ \bibinfo {author} {\bibfnamefont {B.}~\bibnamefont
  {Wardell}},\ }\bibfield  {title} {\bibinfo {title} {{Black hole perturbation
  theory and gravitational self-force}}\ }\href
  {https://doi.org/10.1007/978-981-15-4702-7-38-1}
  {10.1007/978-981-15-4702-7-38-1} (\bibinfo {year} {2021}),\ \Eprint
  {https://arxiv.org/abs/2101.04592} {arXiv:2101.04592 [gr-qc]} \BibitemShut
  {NoStop}%
\bibitem [{\citenamefont {Blanchet}(2014)}]{Blanchet:2013haa}%
  \BibitemOpen
  \bibfield  {author} {\bibinfo {author} {\bibfnamefont {L.}~\bibnamefont
  {Blanchet}},\ }\bibfield  {title} {\bibinfo {title} {{Gravitational Radiation
  from Post-Newtonian Sources and Inspiralling Compact Binaries}},\ }\href
  {https://doi.org/10.12942/lrr-2014-2} {\bibfield  {journal} {\bibinfo
  {journal} {Living Rev. Rel.}\ }\textbf {\bibinfo {volume} {17}},\ \bibinfo
  {pages} {2} (\bibinfo {year} {2014})},\ \Eprint
  {https://arxiv.org/abs/1310.1528} {arXiv:1310.1528 [gr-qc]} \BibitemShut
  {NoStop}%
\bibitem [{\citenamefont {Damour}(2018)}]{Damour:2017zjx}%
  \BibitemOpen
  \bibfield  {author} {\bibinfo {author} {\bibfnamefont {T.}~\bibnamefont
  {Damour}},\ }\bibfield  {title} {\bibinfo {title} {{High-energy gravitational
  scattering and the general relativistic two-body problem}},\ }\href
  {https://doi.org/10.1103/PhysRevD.97.044038} {\bibfield  {journal} {\bibinfo
  {journal} {Phys. Rev. D}\ }\textbf {\bibinfo {volume} {97}},\ \bibinfo
  {pages} {044038} (\bibinfo {year} {2018})},\ \Eprint
  {https://arxiv.org/abs/1710.10599} {arXiv:1710.10599 [gr-qc]} \BibitemShut
  {NoStop}%
\bibitem [{\citenamefont {Damour}(2016)}]{Damour:2016gwp}%
  \BibitemOpen
  \bibfield  {author} {\bibinfo {author} {\bibfnamefont {T.}~\bibnamefont
  {Damour}},\ }\bibfield  {title} {\bibinfo {title} {{Gravitational scattering,
  post-Minkowskian approximation and Effective One-Body theory}},\ }\href
  {https://doi.org/10.1103/PhysRevD.94.104015} {\bibfield  {journal} {\bibinfo
  {journal} {Phys. Rev. D}\ }\textbf {\bibinfo {volume} {94}},\ \bibinfo
  {pages} {104015} (\bibinfo {year} {2016})},\ \Eprint
  {https://arxiv.org/abs/1609.00354} {arXiv:1609.00354 [gr-qc]} \BibitemShut
  {NoStop}%
\bibitem [{\citenamefont {Buonanno}\ and\ \citenamefont
  {Damour}(1999)}]{Buonanno:1998gg}%
  \BibitemOpen
  \bibfield  {author} {\bibinfo {author} {\bibfnamefont {A.}~\bibnamefont
  {Buonanno}}\ and\ \bibinfo {author} {\bibfnamefont {T.}~\bibnamefont
  {Damour}},\ }\bibfield  {title} {\bibinfo {title} {{Effective one-body
  approach to general relativistic two-body dynamics}},\ }\href
  {https://doi.org/10.1103/PhysRevD.59.084006} {\bibfield  {journal} {\bibinfo
  {journal} {Phys. Rev. D}\ }\textbf {\bibinfo {volume} {59}},\ \bibinfo
  {pages} {084006} (\bibinfo {year} {1999})},\ \Eprint
  {https://arxiv.org/abs/gr-qc/9811091} {arXiv:gr-qc/9811091} \BibitemShut
  {NoStop}%
\bibitem [{\citenamefont {Buonanno}\ and\ \citenamefont
  {Damour}(2000)}]{Buonanno:2000ef}%
  \BibitemOpen
  \bibfield  {author} {\bibinfo {author} {\bibfnamefont {A.}~\bibnamefont
  {Buonanno}}\ and\ \bibinfo {author} {\bibfnamefont {T.}~\bibnamefont
  {Damour}},\ }\bibfield  {title} {\bibinfo {title} {{Transition from inspiral
  to plunge in binary black hole coalescences}},\ }\href
  {https://doi.org/10.1103/PhysRevD.62.064015} {\bibfield  {journal} {\bibinfo
  {journal} {Phys. Rev. D}\ }\textbf {\bibinfo {volume} {62}},\ \bibinfo
  {pages} {064015} (\bibinfo {year} {2000})},\ \Eprint
  {https://arxiv.org/abs/gr-qc/0001013} {arXiv:gr-qc/0001013} \BibitemShut
  {NoStop}%
\bibitem [{\citenamefont {Ajith}\ \emph {et~al.}(2008)\citenamefont {Ajith}
  \emph {et~al.}}]{Ajith:2007kx}%
  \BibitemOpen
  \bibfield  {author} {\bibinfo {author} {\bibfnamefont {P.}~\bibnamefont
  {Ajith}} \emph {et~al.},\ }\bibfield  {title} {\bibinfo {title} {{A Template
  bank for gravitational waveforms from coalescing binary black holes. I.
  Non-spinning binaries}},\ }\href {https://doi.org/10.1103/PhysRevD.77.104017}
  {\bibfield  {journal} {\bibinfo  {journal} {Phys. Rev. D}\ }\textbf {\bibinfo
  {volume} {77}},\ \bibinfo {pages} {104017} (\bibinfo {year} {2008})},\
  \bibinfo {note} {[Erratum: Phys.Rev.D 79, 129901 (2009)]},\ \Eprint
  {https://arxiv.org/abs/0710.2335} {arXiv:0710.2335 [gr-qc]} \BibitemShut
  {NoStop}%
\bibitem [{\citenamefont {Santamaria}\ \emph {et~al.}(2010)\citenamefont
  {Santamaria} \emph {et~al.}}]{Santamaria:2010yb}%
  \BibitemOpen
  \bibfield  {author} {\bibinfo {author} {\bibfnamefont {L.}~\bibnamefont
  {Santamaria}} \emph {et~al.},\ }\bibfield  {title} {\bibinfo {title}
  {{Matching post-Newtonian and numerical relativity waveforms: systematic
  errors and a new phenomenological model for non-precessing black hole
  binaries}},\ }\href {https://doi.org/10.1103/PhysRevD.82.064016} {\bibfield
  {journal} {\bibinfo  {journal} {Phys. Rev. D}\ }\textbf {\bibinfo {volume}
  {82}},\ \bibinfo {pages} {064016} (\bibinfo {year} {2010})},\ \Eprint
  {https://arxiv.org/abs/1005.3306} {arXiv:1005.3306 [gr-qc]} \BibitemShut
  {NoStop}%
\bibitem [{\citenamefont {Kocsis}\ \emph {et~al.}(2006)\citenamefont {Kocsis},
  \citenamefont {Gaspar},\ and\ \citenamefont {Marka}}]{Kocsis:2006hq}%
  \BibitemOpen
  \bibfield  {author} {\bibinfo {author} {\bibfnamefont {B.}~\bibnamefont
  {Kocsis}}, \bibinfo {author} {\bibfnamefont {M.~E.}\ \bibnamefont {Gaspar}},\
  and\ \bibinfo {author} {\bibfnamefont {S.}~\bibnamefont {Marka}},\ }\bibfield
   {title} {\bibinfo {title} {{Detection rate estimates of gravity-waves
  emitted during parabolic encounters of stellar black holes in globular
  clusters}},\ }\href {https://doi.org/10.1086/505641} {\bibfield  {journal}
  {\bibinfo  {journal} {Astrophys. J.}\ }\textbf {\bibinfo {volume} {648}},\
  \bibinfo {pages} {411} (\bibinfo {year} {2006})},\ \Eprint
  {https://arxiv.org/abs/astro-ph/0603441} {arXiv:astro-ph/0603441}
  \BibitemShut {NoStop}%
\bibitem [{\citenamefont {Codazzo}\ \emph {et~al.}(2023)\citenamefont
  {Codazzo}, \citenamefont {Di~Giovanni}, \citenamefont {Harms}, \citenamefont
  {Dall'Amico},\ and\ \citenamefont {Mapelli}}]{Codazzo:2022aqj}%
  \BibitemOpen
  \bibfield  {author} {\bibinfo {author} {\bibfnamefont {E.}~\bibnamefont
  {Codazzo}}, \bibinfo {author} {\bibfnamefont {M.}~\bibnamefont
  {Di~Giovanni}}, \bibinfo {author} {\bibfnamefont {J.}~\bibnamefont {Harms}},
  \bibinfo {author} {\bibfnamefont {M.}~\bibnamefont {Dall'Amico}},\ and\
  \bibinfo {author} {\bibfnamefont {M.}~\bibnamefont {Mapelli}},\ }\bibfield
  {title} {\bibinfo {title} {{Study on the detectability of gravitational
  radiation from single-binary encounters between black holes in nuclear star
  clusters: The case of hyperbolic flybys}},\ }\href
  {https://doi.org/10.1103/PhysRevD.107.023023} {\bibfield  {journal} {\bibinfo
   {journal} {Phys. Rev. D}\ }\textbf {\bibinfo {volume} {107}},\ \bibinfo
  {pages} {023023} (\bibinfo {year} {2023})},\ \Eprint
  {https://arxiv.org/abs/2207.01326} {arXiv:2207.01326 [astro-ph.HE]}
  \BibitemShut {NoStop}%
\bibitem [{\citenamefont {Mukherjee}\ \emph {et~al.}(2021)\citenamefont
  {Mukherjee}, \citenamefont {Mitra},\ and\ \citenamefont
  {Chatterjee}}]{Mukherjee:2020hnm}%
  \BibitemOpen
  \bibfield  {author} {\bibinfo {author} {\bibfnamefont {S.}~\bibnamefont
  {Mukherjee}}, \bibinfo {author} {\bibfnamefont {S.}~\bibnamefont {Mitra}},\
  and\ \bibinfo {author} {\bibfnamefont {S.}~\bibnamefont {Chatterjee}},\
  }\bibfield  {title} {\bibinfo {title} {{Gravitational wave observatories may
  be able to detect hyperbolic encounters of black holes}},\ }\href
  {https://doi.org/10.1093/mnras/stab2721} {\bibfield  {journal} {\bibinfo
  {journal} {Mon. Not. Roy. Astron. Soc.}\ }\textbf {\bibinfo {volume} {508}},\
  \bibinfo {pages} {5064} (\bibinfo {year} {2021})},\ \Eprint
  {https://arxiv.org/abs/2010.00916} {arXiv:2010.00916 [gr-qc]} \BibitemShut
  {NoStop}%
\bibitem [{\citenamefont {Bini}\ \emph {et~al.}(2019)\citenamefont {Bini},
  \citenamefont {Damour},\ and\ \citenamefont {Geralico}}]{Bini:2019nra}%
  \BibitemOpen
  \bibfield  {author} {\bibinfo {author} {\bibfnamefont {D.}~\bibnamefont
  {Bini}}, \bibinfo {author} {\bibfnamefont {T.}~\bibnamefont {Damour}},\ and\
  \bibinfo {author} {\bibfnamefont {A.}~\bibnamefont {Geralico}},\ }\bibfield
  {title} {\bibinfo {title} {{Novel approach to binary dynamics: application to
  the fifth post-Newtonian level}},\ }\href
  {https://doi.org/10.1103/PhysRevLett.123.231104} {\bibfield  {journal}
  {\bibinfo  {journal} {Phys. Rev. Lett.}\ }\textbf {\bibinfo {volume} {123}},\
  \bibinfo {pages} {231104} (\bibinfo {year} {2019})},\ \Eprint
  {https://arxiv.org/abs/1909.02375} {arXiv:1909.02375 [gr-qc]} \BibitemShut
  {NoStop}%
\bibitem [{\citenamefont {Bini}\ \emph
  {et~al.}(2020{\natexlab{a}})\citenamefont {Bini}, \citenamefont {Damour},\
  and\ \citenamefont {Geralico}}]{Bini:2020wpo}%
  \BibitemOpen
  \bibfield  {author} {\bibinfo {author} {\bibfnamefont {D.}~\bibnamefont
  {Bini}}, \bibinfo {author} {\bibfnamefont {T.}~\bibnamefont {Damour}},\ and\
  \bibinfo {author} {\bibfnamefont {A.}~\bibnamefont {Geralico}},\ }\bibfield
  {title} {\bibinfo {title} {{Binary dynamics at the fifth and fifth-and-a-half
  post-Newtonian orders}},\ }\href
  {https://doi.org/10.1103/PhysRevD.102.024062} {\bibfield  {journal} {\bibinfo
   {journal} {Phys. Rev. D}\ }\textbf {\bibinfo {volume} {102}},\ \bibinfo
  {pages} {024062} (\bibinfo {year} {2020}{\natexlab{a}})},\ \Eprint
  {https://arxiv.org/abs/2003.11891} {arXiv:2003.11891 [gr-qc]} \BibitemShut
  {NoStop}%
\bibitem [{\citenamefont {Bini}\ \emph
  {et~al.}(2020{\natexlab{b}})\citenamefont {Bini}, \citenamefont {Damour},\
  and\ \citenamefont {Geralico}}]{Bini:2020nsb}%
  \BibitemOpen
  \bibfield  {author} {\bibinfo {author} {\bibfnamefont {D.}~\bibnamefont
  {Bini}}, \bibinfo {author} {\bibfnamefont {T.}~\bibnamefont {Damour}},\ and\
  \bibinfo {author} {\bibfnamefont {A.}~\bibnamefont {Geralico}},\ }\bibfield
  {title} {\bibinfo {title} {{Sixth post-Newtonian local-in-time dynamics of
  binary systems}},\ }\href {https://doi.org/10.1103/PhysRevD.102.024061}
  {\bibfield  {journal} {\bibinfo  {journal} {Phys. Rev. D}\ }\textbf {\bibinfo
  {volume} {102}},\ \bibinfo {pages} {024061} (\bibinfo {year}
  {2020}{\natexlab{b}})},\ \Eprint {https://arxiv.org/abs/2004.05407}
  {arXiv:2004.05407 [gr-qc]} \BibitemShut {NoStop}%
\bibitem [{\citenamefont {Antonelli}\ \emph
  {et~al.}(2020{\natexlab{a}})\citenamefont {Antonelli}, \citenamefont
  {Kavanagh}, \citenamefont {Khalil}, \citenamefont {Steinhoff},\ and\
  \citenamefont {Vines}}]{Antonelli:2020aeb}%
  \BibitemOpen
  \bibfield  {author} {\bibinfo {author} {\bibfnamefont {A.}~\bibnamefont
  {Antonelli}}, \bibinfo {author} {\bibfnamefont {C.}~\bibnamefont {Kavanagh}},
  \bibinfo {author} {\bibfnamefont {M.}~\bibnamefont {Khalil}}, \bibinfo
  {author} {\bibfnamefont {J.}~\bibnamefont {Steinhoff}},\ and\ \bibinfo
  {author} {\bibfnamefont {J.}~\bibnamefont {Vines}},\ }\bibfield  {title}
  {\bibinfo {title} {{Gravitational spin-orbit coupling through
  third-subleading post-Newtonian order: from first-order self-force to
  arbitrary mass ratios}},\ }\href
  {https://doi.org/10.1103/PhysRevLett.125.011103} {\bibfield  {journal}
  {\bibinfo  {journal} {Phys. Rev. Lett.}\ }\textbf {\bibinfo {volume} {125}},\
  \bibinfo {pages} {011103} (\bibinfo {year} {2020}{\natexlab{a}})},\ \Eprint
  {https://arxiv.org/abs/2003.11391} {arXiv:2003.11391 [gr-qc]} \BibitemShut
  {NoStop}%
\bibitem [{\citenamefont {Antonelli}\ \emph
  {et~al.}(2020{\natexlab{b}})\citenamefont {Antonelli}, \citenamefont
  {Kavanagh}, \citenamefont {Khalil}, \citenamefont {Steinhoff},\ and\
  \citenamefont {Vines}}]{Antonelli:2020ybz}%
  \BibitemOpen
  \bibfield  {author} {\bibinfo {author} {\bibfnamefont {A.}~\bibnamefont
  {Antonelli}}, \bibinfo {author} {\bibfnamefont {C.}~\bibnamefont {Kavanagh}},
  \bibinfo {author} {\bibfnamefont {M.}~\bibnamefont {Khalil}}, \bibinfo
  {author} {\bibfnamefont {J.}~\bibnamefont {Steinhoff}},\ and\ \bibinfo
  {author} {\bibfnamefont {J.}~\bibnamefont {Vines}},\ }\bibfield  {title}
  {\bibinfo {title} {{Gravitational spin-orbit and aligned spin$_1$-spin$_2$
  couplings through third-subleading post-Newtonian orders}},\ }\href
  {https://doi.org/10.1103/PhysRevD.102.124024} {\bibfield  {journal} {\bibinfo
   {journal} {Phys. Rev. D}\ }\textbf {\bibinfo {volume} {102}},\ \bibinfo
  {pages} {124024} (\bibinfo {year} {2020}{\natexlab{b}})},\ \Eprint
  {https://arxiv.org/abs/2010.02018} {arXiv:2010.02018 [gr-qc]} \BibitemShut
  {NoStop}%
\bibitem [{\citenamefont {Siemonsen}\ and\ \citenamefont
  {Vines}(2020)}]{Siemonsen:2019dsu}%
  \BibitemOpen
  \bibfield  {author} {\bibinfo {author} {\bibfnamefont {N.}~\bibnamefont
  {Siemonsen}}\ and\ \bibinfo {author} {\bibfnamefont {J.}~\bibnamefont
  {Vines}},\ }\bibfield  {title} {\bibinfo {title} {{Test black holes,
  scattering amplitudes and perturbations of Kerr spacetime}},\ }\href
  {https://doi.org/10.1103/PhysRevD.101.064066} {\bibfield  {journal} {\bibinfo
   {journal} {Phys. Rev. D}\ }\textbf {\bibinfo {volume} {101}},\ \bibinfo
  {pages} {064066} (\bibinfo {year} {2020})},\ \Eprint
  {https://arxiv.org/abs/1909.07361} {arXiv:1909.07361 [gr-qc]} \BibitemShut
  {NoStop}%
\bibitem [{\citenamefont {K\"alin}\ and\ \citenamefont
  {Porto}(2020{\natexlab{a}})}]{Kalin:2019rwq}%
  \BibitemOpen
  \bibfield  {author} {\bibinfo {author} {\bibfnamefont {G.}~\bibnamefont
  {K\"alin}}\ and\ \bibinfo {author} {\bibfnamefont {R.~A.}\ \bibnamefont
  {Porto}},\ }\bibfield  {title} {\bibinfo {title} {{From Boundary Data to
  Bound States}},\ }\href {https://doi.org/10.1007/JHEP01(2020)072} {\bibfield
  {journal} {\bibinfo  {journal} {JHEP}\ }\textbf {\bibinfo {volume} {01}},\
  \bibinfo {pages} {072}},\ \Eprint {https://arxiv.org/abs/1910.03008}
  {arXiv:1910.03008 [hep-th]} \BibitemShut {NoStop}%
\bibitem [{\citenamefont {K\"alin}\ and\ \citenamefont
  {Porto}(2020{\natexlab{b}})}]{Kalin:2019inp}%
  \BibitemOpen
  \bibfield  {author} {\bibinfo {author} {\bibfnamefont {G.}~\bibnamefont
  {K\"alin}}\ and\ \bibinfo {author} {\bibfnamefont {R.~A.}\ \bibnamefont
  {Porto}},\ }\bibfield  {title} {\bibinfo {title} {{From boundary data to
  bound states. Part II. Scattering angle to dynamical invariants (with
  twist)}},\ }\href {https://doi.org/10.1007/JHEP02(2020)120} {\bibfield
  {journal} {\bibinfo  {journal} {JHEP}\ }\textbf {\bibinfo {volume} {02}},\
  \bibinfo {pages} {120}},\ \Eprint {https://arxiv.org/abs/1911.09130}
  {arXiv:1911.09130 [hep-th]} \BibitemShut {NoStop}%
\bibitem [{\citenamefont {Cho}\ \emph {et~al.}(2022)\citenamefont {Cho},
  \citenamefont {K\"alin},\ and\ \citenamefont {Porto}}]{Cho:2021arx}%
  \BibitemOpen
  \bibfield  {author} {\bibinfo {author} {\bibfnamefont {G.}~\bibnamefont
  {Cho}}, \bibinfo {author} {\bibfnamefont {G.}~\bibnamefont {K\"alin}},\ and\
  \bibinfo {author} {\bibfnamefont {R.~A.}\ \bibnamefont {Porto}},\ }\bibfield
  {title} {\bibinfo {title} {{From boundary data to bound states. Part III.
  Radiative effects}},\ }\href {https://doi.org/10.1007/JHEP04(2022)154}
  {\bibfield  {journal} {\bibinfo  {journal} {JHEP}\ }\textbf {\bibinfo
  {volume} {04}},\ \bibinfo {pages} {154}},\ \bibinfo {note} {[Erratum: JHEP
  07, 002 (2022)]},\ \Eprint {https://arxiv.org/abs/2112.03976}
  {arXiv:2112.03976 [hep-th]} \BibitemShut {NoStop}%
\bibitem [{\citenamefont {Adamo}\ \emph {et~al.}(2024)\citenamefont {Adamo},
  \citenamefont {Gonzo},\ and\ \citenamefont {Ilderton}}]{Adamo:2024oxy}%
  \BibitemOpen
  \bibfield  {author} {\bibinfo {author} {\bibfnamefont {T.}~\bibnamefont
  {Adamo}}, \bibinfo {author} {\bibfnamefont {R.}~\bibnamefont {Gonzo}},\ and\
  \bibinfo {author} {\bibfnamefont {A.}~\bibnamefont {Ilderton}},\ }\bibfield
  {title} {\bibinfo {title} {{Gravitational bound waveforms from amplitudes}},\
  }\href {https://doi.org/10.1007/JHEP05(2024)034} {\bibfield  {journal}
  {\bibinfo  {journal} {JHEP}\ }\textbf {\bibinfo {volume} {05}},\ \bibinfo
  {pages} {034}},\ \Eprint {https://arxiv.org/abs/2402.00124} {arXiv:2402.00124
  [hep-th]} \BibitemShut {NoStop}%
\bibitem [{\citenamefont {K\"alin}\ and\ \citenamefont
  {Porto}(2020{\natexlab{c}})}]{Kalin:2020mvi}%
  \BibitemOpen
  \bibfield  {author} {\bibinfo {author} {\bibfnamefont {G.}~\bibnamefont
  {K\"alin}}\ and\ \bibinfo {author} {\bibfnamefont {R.~A.}\ \bibnamefont
  {Porto}},\ }\bibfield  {title} {\bibinfo {title} {{Post-Minkowskian Effective
  Field Theory for Conservative Binary Dynamics}},\ }\href
  {https://doi.org/10.1007/JHEP11(2020)106} {\bibfield  {journal} {\bibinfo
  {journal} {JHEP}\ }\textbf {\bibinfo {volume} {11}},\ \bibinfo {pages}
  {106}},\ \Eprint {https://arxiv.org/abs/2006.01184} {arXiv:2006.01184
  [hep-th]} \BibitemShut {NoStop}%
\bibitem [{\citenamefont {K\"alin}\ \emph {et~al.}(2020)\citenamefont
  {K\"alin}, \citenamefont {Liu},\ and\ \citenamefont {Porto}}]{Kalin:2020fhe}%
  \BibitemOpen
  \bibfield  {author} {\bibinfo {author} {\bibfnamefont {G.}~\bibnamefont
  {K\"alin}}, \bibinfo {author} {\bibfnamefont {Z.}~\bibnamefont {Liu}},\ and\
  \bibinfo {author} {\bibfnamefont {R.~A.}\ \bibnamefont {Porto}},\ }\bibfield
  {title} {\bibinfo {title} {{Conservative Dynamics of Binary Systems to Third
  Post-Minkowskian Order from the Effective Field Theory Approach}},\ }\href
  {https://doi.org/10.1103/PhysRevLett.125.261103} {\bibfield  {journal}
  {\bibinfo  {journal} {Phys. Rev. Lett.}\ }\textbf {\bibinfo {volume} {125}},\
  \bibinfo {pages} {261103} (\bibinfo {year} {2020})},\ \Eprint
  {https://arxiv.org/abs/2007.04977} {arXiv:2007.04977 [hep-th]} \BibitemShut
  {NoStop}%
\bibitem [{\citenamefont {K\"alin}\ \emph {et~al.}(2023)\citenamefont
  {K\"alin}, \citenamefont {Neef},\ and\ \citenamefont
  {Porto}}]{Kalin:2022hph}%
  \BibitemOpen
  \bibfield  {author} {\bibinfo {author} {\bibfnamefont {G.}~\bibnamefont
  {K\"alin}}, \bibinfo {author} {\bibfnamefont {J.}~\bibnamefont {Neef}},\ and\
  \bibinfo {author} {\bibfnamefont {R.~A.}\ \bibnamefont {Porto}},\ }\bibfield
  {title} {\bibinfo {title} {{Radiation-reaction in the Effective Field Theory
  approach to Post-Minkowskian dynamics}},\ }\href
  {https://doi.org/10.1007/JHEP01(2023)140} {\bibfield  {journal} {\bibinfo
  {journal} {JHEP}\ }\textbf {\bibinfo {volume} {01}},\ \bibinfo {pages}
  {140}},\ \Eprint {https://arxiv.org/abs/2207.00580} {arXiv:2207.00580
  [hep-th]} \BibitemShut {NoStop}%
\bibitem [{\citenamefont {Dlapa}\ \emph
  {et~al.}(2022{\natexlab{a}})\citenamefont {Dlapa}, \citenamefont {K\"alin},
  \citenamefont {Liu},\ and\ \citenamefont {Porto}}]{Dlapa:2021npj}%
  \BibitemOpen
  \bibfield  {author} {\bibinfo {author} {\bibfnamefont {C.}~\bibnamefont
  {Dlapa}}, \bibinfo {author} {\bibfnamefont {G.}~\bibnamefont {K\"alin}},
  \bibinfo {author} {\bibfnamefont {Z.}~\bibnamefont {Liu}},\ and\ \bibinfo
  {author} {\bibfnamefont {R.~A.}\ \bibnamefont {Porto}},\ }\bibfield  {title}
  {\bibinfo {title} {{Dynamics of binary systems to fourth Post-Minkowskian
  order from the effective field theory approach}},\ }\href
  {https://doi.org/10.1016/j.physletb.2022.137203} {\bibfield  {journal}
  {\bibinfo  {journal} {Phys. Lett. B}\ }\textbf {\bibinfo {volume} {831}},\
  \bibinfo {pages} {137203} (\bibinfo {year} {2022}{\natexlab{a}})},\ \Eprint
  {https://arxiv.org/abs/2106.08276} {arXiv:2106.08276 [hep-th]} \BibitemShut
  {NoStop}%
\bibitem [{\citenamefont {Dlapa}\ \emph
  {et~al.}(2022{\natexlab{b}})\citenamefont {Dlapa}, \citenamefont {K\"alin},
  \citenamefont {Liu},\ and\ \citenamefont {Porto}}]{Dlapa:2021vgp}%
  \BibitemOpen
  \bibfield  {author} {\bibinfo {author} {\bibfnamefont {C.}~\bibnamefont
  {Dlapa}}, \bibinfo {author} {\bibfnamefont {G.}~\bibnamefont {K\"alin}},
  \bibinfo {author} {\bibfnamefont {Z.}~\bibnamefont {Liu}},\ and\ \bibinfo
  {author} {\bibfnamefont {R.~A.}\ \bibnamefont {Porto}},\ }\bibfield  {title}
  {\bibinfo {title} {{Conservative Dynamics of Binary Systems at Fourth
  Post-Minkowskian Order in the Large-Eccentricity Expansion}},\ }\href
  {https://doi.org/10.1103/PhysRevLett.128.161104} {\bibfield  {journal}
  {\bibinfo  {journal} {Phys. Rev. Lett.}\ }\textbf {\bibinfo {volume} {128}},\
  \bibinfo {pages} {161104} (\bibinfo {year} {2022}{\natexlab{b}})},\ \Eprint
  {https://arxiv.org/abs/2112.11296} {arXiv:2112.11296 [hep-th]} \BibitemShut
  {NoStop}%
\bibitem [{\citenamefont {Kosower}\ \emph {et~al.}(2019)\citenamefont
  {Kosower}, \citenamefont {Maybee},\ and\ \citenamefont
  {O'Connell}}]{Kosower:2018adc}%
  \BibitemOpen
  \bibfield  {author} {\bibinfo {author} {\bibfnamefont {D.~A.}\ \bibnamefont
  {Kosower}}, \bibinfo {author} {\bibfnamefont {B.}~\bibnamefont {Maybee}},\
  and\ \bibinfo {author} {\bibfnamefont {D.}~\bibnamefont {O'Connell}},\
  }\bibfield  {title} {\bibinfo {title} {{Amplitudes, Observables, and
  Classical Scattering}},\ }\href {https://doi.org/10.1007/JHEP02(2019)137}
  {\bibfield  {journal} {\bibinfo  {journal} {JHEP}\ }\textbf {\bibinfo
  {volume} {02}},\ \bibinfo {pages} {137}},\ \Eprint
  {https://arxiv.org/abs/1811.10950} {arXiv:1811.10950 [hep-th]} \BibitemShut
  {NoStop}%
\bibitem [{\citenamefont {Bern}\ \emph
  {et~al.}(2019{\natexlab{a}})\citenamefont {Bern}, \citenamefont {Cheung},
  \citenamefont {Roiban}, \citenamefont {Shen}, \citenamefont {Solon},\ and\
  \citenamefont {Zeng}}]{Bern:2019nnu}%
  \BibitemOpen
  \bibfield  {author} {\bibinfo {author} {\bibfnamefont {Z.}~\bibnamefont
  {Bern}}, \bibinfo {author} {\bibfnamefont {C.}~\bibnamefont {Cheung}},
  \bibinfo {author} {\bibfnamefont {R.}~\bibnamefont {Roiban}}, \bibinfo
  {author} {\bibfnamefont {C.-H.}\ \bibnamefont {Shen}}, \bibinfo {author}
  {\bibfnamefont {M.~P.}\ \bibnamefont {Solon}},\ and\ \bibinfo {author}
  {\bibfnamefont {M.}~\bibnamefont {Zeng}},\ }\bibfield  {title} {\bibinfo
  {title} {{Scattering Amplitudes and the Conservative Hamiltonian for Binary
  Systems at Third Post-Minkowskian Order}},\ }\href
  {https://doi.org/10.1103/PhysRevLett.122.201603} {\bibfield  {journal}
  {\bibinfo  {journal} {Phys. Rev. Lett.}\ }\textbf {\bibinfo {volume} {122}},\
  \bibinfo {pages} {201603} (\bibinfo {year} {2019}{\natexlab{a}})},\ \Eprint
  {https://arxiv.org/abs/1901.04424} {arXiv:1901.04424 [hep-th]} \BibitemShut
  {NoStop}%
\bibitem [{\citenamefont {Bern}\ \emph
  {et~al.}(2019{\natexlab{b}})\citenamefont {Bern}, \citenamefont {Cheung},
  \citenamefont {Roiban}, \citenamefont {Shen}, \citenamefont {Solon},\ and\
  \citenamefont {Zeng}}]{Bern:2019crd}%
  \BibitemOpen
  \bibfield  {author} {\bibinfo {author} {\bibfnamefont {Z.}~\bibnamefont
  {Bern}}, \bibinfo {author} {\bibfnamefont {C.}~\bibnamefont {Cheung}},
  \bibinfo {author} {\bibfnamefont {R.}~\bibnamefont {Roiban}}, \bibinfo
  {author} {\bibfnamefont {C.-H.}\ \bibnamefont {Shen}}, \bibinfo {author}
  {\bibfnamefont {M.~P.}\ \bibnamefont {Solon}},\ and\ \bibinfo {author}
  {\bibfnamefont {M.}~\bibnamefont {Zeng}},\ }\bibfield  {title} {\bibinfo
  {title} {{Black Hole Binary Dynamics from the Double Copy and Effective
  Theory}},\ }\href {https://doi.org/10.1007/JHEP10(2019)206} {\bibfield
  {journal} {\bibinfo  {journal} {JHEP}\ }\textbf {\bibinfo {volume} {10}},\
  \bibinfo {pages} {206}},\ \Eprint {https://arxiv.org/abs/1908.01493}
  {arXiv:1908.01493 [hep-th]} \BibitemShut {NoStop}%
\bibitem [{\citenamefont {Mogull}\ \emph {et~al.}(2021)\citenamefont {Mogull},
  \citenamefont {Plefka},\ and\ \citenamefont {Steinhoff}}]{Mogull:2020sak}%
  \BibitemOpen
  \bibfield  {author} {\bibinfo {author} {\bibfnamefont {G.}~\bibnamefont
  {Mogull}}, \bibinfo {author} {\bibfnamefont {J.}~\bibnamefont {Plefka}},\
  and\ \bibinfo {author} {\bibfnamefont {J.}~\bibnamefont {Steinhoff}},\
  }\bibfield  {title} {\bibinfo {title} {{Classical black hole scattering from
  a worldline quantum field theory}},\ }\href
  {https://doi.org/10.1007/JHEP02(2021)048} {\bibfield  {journal} {\bibinfo
  {journal} {JHEP}\ }\textbf {\bibinfo {volume} {02}},\ \bibinfo {pages}
  {048}},\ \Eprint {https://arxiv.org/abs/2010.02865} {arXiv:2010.02865
  [hep-th]} \BibitemShut {NoStop}%
\bibitem [{\citenamefont {Bern}\ \emph {et~al.}(2022)\citenamefont {Bern},
  \citenamefont {Parra-Martinez}, \citenamefont {Roiban}, \citenamefont {Ruf},
  \citenamefont {Shen}, \citenamefont {Solon},\ and\ \citenamefont
  {Zeng}}]{Bern:2021yeh}%
  \BibitemOpen
  \bibfield  {author} {\bibinfo {author} {\bibfnamefont {Z.}~\bibnamefont
  {Bern}}, \bibinfo {author} {\bibfnamefont {J.}~\bibnamefont
  {Parra-Martinez}}, \bibinfo {author} {\bibfnamefont {R.}~\bibnamefont
  {Roiban}}, \bibinfo {author} {\bibfnamefont {M.~S.}\ \bibnamefont {Ruf}},
  \bibinfo {author} {\bibfnamefont {C.-H.}\ \bibnamefont {Shen}}, \bibinfo
  {author} {\bibfnamefont {M.~P.}\ \bibnamefont {Solon}},\ and\ \bibinfo
  {author} {\bibfnamefont {M.}~\bibnamefont {Zeng}},\ }\bibfield  {title}
  {\bibinfo {title} {{Scattering Amplitudes, the Tail Effect, and Conservative
  Binary Dynamics at O(G4)}},\ }\href
  {https://doi.org/10.1103/PhysRevLett.128.161103} {\bibfield  {journal}
  {\bibinfo  {journal} {Phys. Rev. Lett.}\ }\textbf {\bibinfo {volume} {128}},\
  \bibinfo {pages} {161103} (\bibinfo {year} {2022})},\ \Eprint
  {https://arxiv.org/abs/2112.10750} {arXiv:2112.10750 [hep-th]} \BibitemShut
  {NoStop}%
\bibitem [{\citenamefont {Driesse}\ \emph {et~al.}(2024)\citenamefont
  {Driesse}, \citenamefont {Jakobsen}, \citenamefont {Mogull}, \citenamefont
  {Plefka}, \citenamefont {Sauer},\ and\ \citenamefont
  {Usovitsch}}]{Driesse:2024xad}%
  \BibitemOpen
  \bibfield  {author} {\bibinfo {author} {\bibfnamefont {M.}~\bibnamefont
  {Driesse}}, \bibinfo {author} {\bibfnamefont {G.~U.}\ \bibnamefont
  {Jakobsen}}, \bibinfo {author} {\bibfnamefont {G.}~\bibnamefont {Mogull}},
  \bibinfo {author} {\bibfnamefont {J.}~\bibnamefont {Plefka}}, \bibinfo
  {author} {\bibfnamefont {B.}~\bibnamefont {Sauer}},\ and\ \bibinfo {author}
  {\bibfnamefont {J.}~\bibnamefont {Usovitsch}},\ }\bibfield  {title} {\bibinfo
  {title} {{Conservative Black Hole Scattering at Fifth Post-Minkowskian and
  First Self-Force Order}},\ }\href@noop {} {\  (\bibinfo {year} {2024})},\
  \Eprint {https://arxiv.org/abs/2403.07781} {arXiv:2403.07781 [hep-th]}
  \BibitemShut {NoStop}%
\bibitem [{\citenamefont {Bini}\ \emph {et~al.}(2021)\citenamefont {Bini},
  \citenamefont {Damour},\ and\ \citenamefont {Geralico}}]{Bini:2021gat}%
  \BibitemOpen
  \bibfield  {author} {\bibinfo {author} {\bibfnamefont {D.}~\bibnamefont
  {Bini}}, \bibinfo {author} {\bibfnamefont {T.}~\bibnamefont {Damour}},\ and\
  \bibinfo {author} {\bibfnamefont {A.}~\bibnamefont {Geralico}},\ }\bibfield
  {title} {\bibinfo {title} {{Radiative contributions to gravitational
  scattering}},\ }\href {https://doi.org/10.1103/PhysRevD.104.084031}
  {\bibfield  {journal} {\bibinfo  {journal} {Phys. Rev. D}\ }\textbf {\bibinfo
  {volume} {104}},\ \bibinfo {pages} {084031} (\bibinfo {year} {2021})},\
  \Eprint {https://arxiv.org/abs/2107.08896} {arXiv:2107.08896 [gr-qc]}
  \BibitemShut {NoStop}%
\bibitem [{\citenamefont {Bini}\ \emph {et~al.}(2024)\citenamefont {Bini},
  \citenamefont {Damour}, \citenamefont {De~Angelis}, \citenamefont {Geralico},
  \citenamefont {Herderschee}, \citenamefont {Roiban},\ and\ \citenamefont
  {Teng}}]{Bini:2024rsy}%
  \BibitemOpen
  \bibfield  {author} {\bibinfo {author} {\bibfnamefont {D.}~\bibnamefont
  {Bini}}, \bibinfo {author} {\bibfnamefont {T.}~\bibnamefont {Damour}},
  \bibinfo {author} {\bibfnamefont {S.}~\bibnamefont {De~Angelis}}, \bibinfo
  {author} {\bibfnamefont {A.}~\bibnamefont {Geralico}}, \bibinfo {author}
  {\bibfnamefont {A.}~\bibnamefont {Herderschee}}, \bibinfo {author}
  {\bibfnamefont {R.}~\bibnamefont {Roiban}},\ and\ \bibinfo {author}
  {\bibfnamefont {F.}~\bibnamefont {Teng}},\ }\bibfield  {title} {\bibinfo
  {title} {{Gravitational Waveform: A Tale of Two Formalisms}},\ }\href@noop {}
  {\  (\bibinfo {year} {2024})},\ \Eprint {https://arxiv.org/abs/2402.06604}
  {arXiv:2402.06604 [hep-th]} \BibitemShut {NoStop}%
\bibitem [{\citenamefont {Bautista}\ and\ \citenamefont
  {Siemonsen}(2022)}]{Bautista:2021inx}%
  \BibitemOpen
  \bibfield  {author} {\bibinfo {author} {\bibfnamefont {Y.~F.}\ \bibnamefont
  {Bautista}}\ and\ \bibinfo {author} {\bibfnamefont {N.}~\bibnamefont
  {Siemonsen}},\ }\bibfield  {title} {\bibinfo {title} {{Post-Newtonian
  waveforms from spinning scattering amplitudes}},\ }\href
  {https://doi.org/10.1007/JHEP01(2022)006} {\bibfield  {journal} {\bibinfo
  {journal} {JHEP}\ }\textbf {\bibinfo {volume} {01}},\ \bibinfo {pages}
  {006}},\ \Eprint {https://arxiv.org/abs/2110.12537} {arXiv:2110.12537
  [hep-th]} \BibitemShut {NoStop}%
\bibitem [{\citenamefont {Afshordi}\ \emph {et~al.}(2023)\citenamefont
  {Afshordi} \emph {et~al.}}]{LISAConsortiumWaveformWorkingGroup:2023arg}%
  \BibitemOpen
  \bibfield  {author} {\bibinfo {author} {\bibfnamefont {N.}~\bibnamefont
  {Afshordi}} \emph {et~al.} (\bibinfo {collaboration} {LISA Consortium
  Waveform Working Group}),\ }\bibfield  {title} {\bibinfo {title} {{Waveform
  Modelling for the Laser Interferometer Space Antenna}},\ }\href@noop {} {\
  (\bibinfo {year} {2023})},\ \Eprint {https://arxiv.org/abs/2311.01300}
  {arXiv:2311.01300 [gr-qc]} \BibitemShut {NoStop}%
\bibitem [{\citenamefont {Wardell}\ \emph {et~al.}(2023)\citenamefont
  {Wardell}, \citenamefont {Pound}, \citenamefont {Warburton}, \citenamefont
  {Miller}, \citenamefont {Durkan},\ and\ \citenamefont
  {Le~Tiec}}]{Wardell:2021fyy}%
  \BibitemOpen
  \bibfield  {author} {\bibinfo {author} {\bibfnamefont {B.}~\bibnamefont
  {Wardell}}, \bibinfo {author} {\bibfnamefont {A.}~\bibnamefont {Pound}},
  \bibinfo {author} {\bibfnamefont {N.}~\bibnamefont {Warburton}}, \bibinfo
  {author} {\bibfnamefont {J.}~\bibnamefont {Miller}}, \bibinfo {author}
  {\bibfnamefont {L.}~\bibnamefont {Durkan}},\ and\ \bibinfo {author}
  {\bibfnamefont {A.}~\bibnamefont {Le~Tiec}},\ }\bibfield  {title} {\bibinfo
  {title} {{Gravitational Waveforms for Compact Binaries from Second-Order
  Self-Force Theory}},\ }\href {https://doi.org/10.1103/PhysRevLett.130.241402}
  {\bibfield  {journal} {\bibinfo  {journal} {Phys. Rev. Lett.}\ }\textbf
  {\bibinfo {volume} {130}},\ \bibinfo {pages} {241402} (\bibinfo {year}
  {2023})},\ \Eprint {https://arxiv.org/abs/2112.12265} {arXiv:2112.12265
  [gr-qc]} \BibitemShut {NoStop}%
\bibitem [{\citenamefont {Albertini}\ \emph {et~al.}(2022)\citenamefont
  {Albertini}, \citenamefont {Nagar}, \citenamefont {Pound}, \citenamefont
  {Warburton}, \citenamefont {Wardell}, \citenamefont {Durkan},\ and\
  \citenamefont {Miller}}]{Albertini:2022rfe}%
  \BibitemOpen
  \bibfield  {author} {\bibinfo {author} {\bibfnamefont {A.}~\bibnamefont
  {Albertini}}, \bibinfo {author} {\bibfnamefont {A.}~\bibnamefont {Nagar}},
  \bibinfo {author} {\bibfnamefont {A.}~\bibnamefont {Pound}}, \bibinfo
  {author} {\bibfnamefont {N.}~\bibnamefont {Warburton}}, \bibinfo {author}
  {\bibfnamefont {B.}~\bibnamefont {Wardell}}, \bibinfo {author} {\bibfnamefont
  {L.}~\bibnamefont {Durkan}},\ and\ \bibinfo {author} {\bibfnamefont
  {J.}~\bibnamefont {Miller}},\ }\bibfield  {title} {\bibinfo {title}
  {{Comparing second-order gravitational self-force, numerical relativity, and
  effective one body waveforms from inspiralling, quasicircular, and
  nonspinning black hole binaries}},\ }\href
  {https://doi.org/10.1103/PhysRevD.106.084061} {\bibfield  {journal} {\bibinfo
   {journal} {Phys. Rev. D}\ }\textbf {\bibinfo {volume} {106}},\ \bibinfo
  {pages} {084061} (\bibinfo {year} {2022})},\ \Eprint
  {https://arxiv.org/abs/2208.01049} {arXiv:2208.01049 [gr-qc]} \BibitemShut
  {NoStop}%
\bibitem [{\citenamefont {van~de Meent}\ \emph {et~al.}(2023)\citenamefont
  {van~de Meent}, \citenamefont {Buonanno}, \citenamefont {Mihaylov},
  \citenamefont {Ossokine}, \citenamefont {Pompili}, \citenamefont {Warburton},
  \citenamefont {Pound}, \citenamefont {Wardell}, \citenamefont {Durkan},\ and\
  \citenamefont {Miller}}]{vandeMeent:2023ols}%
  \BibitemOpen
  \bibfield  {author} {\bibinfo {author} {\bibfnamefont {M.}~\bibnamefont
  {van~de Meent}}, \bibinfo {author} {\bibfnamefont {A.}~\bibnamefont
  {Buonanno}}, \bibinfo {author} {\bibfnamefont {D.~P.}\ \bibnamefont
  {Mihaylov}}, \bibinfo {author} {\bibfnamefont {S.}~\bibnamefont {Ossokine}},
  \bibinfo {author} {\bibfnamefont {L.}~\bibnamefont {Pompili}}, \bibinfo
  {author} {\bibfnamefont {N.}~\bibnamefont {Warburton}}, \bibinfo {author}
  {\bibfnamefont {A.}~\bibnamefont {Pound}}, \bibinfo {author} {\bibfnamefont
  {B.}~\bibnamefont {Wardell}}, \bibinfo {author} {\bibfnamefont
  {L.}~\bibnamefont {Durkan}},\ and\ \bibinfo {author} {\bibfnamefont
  {J.}~\bibnamefont {Miller}},\ }\bibfield  {title} {\bibinfo {title}
  {{Enhancing the SEOBNRv5 effective-one-body waveform model with second-order
  gravitational self-force fluxes}},\ }\href
  {https://doi.org/10.1103/PhysRevD.108.124038} {\bibfield  {journal} {\bibinfo
   {journal} {Phys. Rev. D}\ }\textbf {\bibinfo {volume} {108}},\ \bibinfo
  {pages} {124038} (\bibinfo {year} {2023})},\ \Eprint
  {https://arxiv.org/abs/2303.18026} {arXiv:2303.18026 [gr-qc]} \BibitemShut
  {NoStop}%
\bibitem [{\citenamefont {Long}\ and\ \citenamefont
  {Barack}(2021)}]{Long:2021ufh}%
  \BibitemOpen
  \bibfield  {author} {\bibinfo {author} {\bibfnamefont {O.}~\bibnamefont
  {Long}}\ and\ \bibinfo {author} {\bibfnamefont {L.}~\bibnamefont {Barack}},\
  }\bibfield  {title} {\bibinfo {title} {{Time-domain metric reconstruction for
  hyperbolic scattering}},\ }\href
  {https://doi.org/10.1103/PhysRevD.104.024014} {\bibfield  {journal} {\bibinfo
   {journal} {Phys. Rev. D}\ }\textbf {\bibinfo {volume} {104}},\ \bibinfo
  {pages} {024014} (\bibinfo {year} {2021})},\ \Eprint
  {https://arxiv.org/abs/2105.05630} {arXiv:2105.05630 [gr-qc]} \BibitemShut
  {NoStop}%
\bibitem [{\citenamefont {Mano}\ \emph {et~al.}(1996)\citenamefont {Mano},
  \citenamefont {Suzuki},\ and\ \citenamefont {Takasugi}}]{Mano:1996vt}%
  \BibitemOpen
  \bibfield  {author} {\bibinfo {author} {\bibfnamefont {S.}~\bibnamefont
  {Mano}}, \bibinfo {author} {\bibfnamefont {H.}~\bibnamefont {Suzuki}},\ and\
  \bibinfo {author} {\bibfnamefont {E.}~\bibnamefont {Takasugi}},\ }\bibfield
  {title} {\bibinfo {title} {{Analytic solutions of the Teukolsky equation and
  their low frequency expansions}},\ }\href
  {https://doi.org/10.1143/PTP.95.1079} {\bibfield  {journal} {\bibinfo
  {journal} {Prog. Theor. Phys.}\ }\textbf {\bibinfo {volume} {95}},\ \bibinfo
  {pages} {1079} (\bibinfo {year} {1996})},\ \Eprint
  {https://arxiv.org/abs/gr-qc/9603020} {arXiv:gr-qc/9603020} \BibitemShut
  {NoStop}%
\bibitem [{\citenamefont {Fujita}(2012)}]{Fujita:2012cm}%
  \BibitemOpen
  \bibfield  {author} {\bibinfo {author} {\bibfnamefont {R.}~\bibnamefont
  {Fujita}},\ }\bibfield  {title} {\bibinfo {title} {{Gravitational Waves from
  a Particle in Circular Orbits around a Schwarzschild Black Hole to the 22nd
  Post-Newtonian Order}},\ }\href {https://doi.org/10.1143/PTP.128.971}
  {\bibfield  {journal} {\bibinfo  {journal} {Prog. Theor. Phys.}\ }\textbf
  {\bibinfo {volume} {128}},\ \bibinfo {pages} {971} (\bibinfo {year}
  {2012})},\ \Eprint {https://arxiv.org/abs/1211.5535} {arXiv:1211.5535
  [gr-qc]} \BibitemShut {NoStop}%
\bibitem [{\citenamefont {Bini}\ and\ \citenamefont
  {Damour}(2014)}]{Bini:2013rfa}%
  \BibitemOpen
  \bibfield  {author} {\bibinfo {author} {\bibfnamefont {D.}~\bibnamefont
  {Bini}}\ and\ \bibinfo {author} {\bibfnamefont {T.}~\bibnamefont {Damour}},\
  }\bibfield  {title} {\bibinfo {title} {{High-order post-Newtonian
  contributions to the two-body gravitational interaction potential from
  analytical gravitational self-force calculations}},\ }\href
  {https://doi.org/10.1103/PhysRevD.89.064063} {\bibfield  {journal} {\bibinfo
  {journal} {Phys. Rev. D}\ }\textbf {\bibinfo {volume} {89}},\ \bibinfo
  {pages} {064063} (\bibinfo {year} {2014})},\ \Eprint
  {https://arxiv.org/abs/1312.2503} {arXiv:1312.2503 [gr-qc]} \BibitemShut
  {NoStop}%
\bibitem [{\citenamefont {Shah}\ \emph {et~al.}(2014)\citenamefont {Shah},
  \citenamefont {Friedman},\ and\ \citenamefont {Whiting}}]{Shah:2013uya}%
  \BibitemOpen
  \bibfield  {author} {\bibinfo {author} {\bibfnamefont {A.~G.}\ \bibnamefont
  {Shah}}, \bibinfo {author} {\bibfnamefont {J.~L.}\ \bibnamefont {Friedman}},\
  and\ \bibinfo {author} {\bibfnamefont {B.~F.}\ \bibnamefont {Whiting}},\
  }\bibfield  {title} {\bibinfo {title} {{Finding high-order analytic
  post-Newtonian parameters from a high-precision numerical self-force
  calculation}},\ }\href {https://doi.org/10.1103/PhysRevD.89.064042}
  {\bibfield  {journal} {\bibinfo  {journal} {Phys. Rev. D}\ }\textbf {\bibinfo
  {volume} {89}},\ \bibinfo {pages} {064042} (\bibinfo {year} {2014})},\
  \Eprint {https://arxiv.org/abs/1312.1952} {arXiv:1312.1952 [gr-qc]}
  \BibitemShut {NoStop}%
\bibitem [{\citenamefont {Kavanagh}\ \emph {et~al.}(2015)\citenamefont
  {Kavanagh}, \citenamefont {Ottewill},\ and\ \citenamefont
  {Wardell}}]{Kavanagh:2015lva}%
  \BibitemOpen
  \bibfield  {author} {\bibinfo {author} {\bibfnamefont {C.}~\bibnamefont
  {Kavanagh}}, \bibinfo {author} {\bibfnamefont {A.~C.}\ \bibnamefont
  {Ottewill}},\ and\ \bibinfo {author} {\bibfnamefont {B.}~\bibnamefont
  {Wardell}},\ }\bibfield  {title} {\bibinfo {title} {{Analytical high-order
  post-Newtonian expansions for extreme mass ratio binaries}},\ }\href
  {https://doi.org/10.1103/PhysRevD.92.084025} {\bibfield  {journal} {\bibinfo
  {journal} {Phys. Rev. D}\ }\textbf {\bibinfo {volume} {92}},\ \bibinfo
  {pages} {084025} (\bibinfo {year} {2015})},\ \Eprint
  {https://arxiv.org/abs/1503.02334} {arXiv:1503.02334 [gr-qc]} \BibitemShut
  {NoStop}%
\bibitem [{\citenamefont {Forseth}\ \emph {et~al.}(2016)\citenamefont
  {Forseth}, \citenamefont {Evans},\ and\ \citenamefont
  {Hopper}}]{Forseth:2015oua}%
  \BibitemOpen
  \bibfield  {author} {\bibinfo {author} {\bibfnamefont {E.}~\bibnamefont
  {Forseth}}, \bibinfo {author} {\bibfnamefont {C.~R.}\ \bibnamefont {Evans}},\
  and\ \bibinfo {author} {\bibfnamefont {S.}~\bibnamefont {Hopper}},\
  }\bibfield  {title} {\bibinfo {title} {{Eccentric-orbit extreme-mass-ratio
  inspiral gravitational wave energy fluxes to 7PN order}},\ }\href
  {https://doi.org/10.1103/PhysRevD.93.064058} {\bibfield  {journal} {\bibinfo
  {journal} {Phys. Rev. D}\ }\textbf {\bibinfo {volume} {93}},\ \bibinfo
  {pages} {064058} (\bibinfo {year} {2016})},\ \Eprint
  {https://arxiv.org/abs/1512.03051} {arXiv:1512.03051 [gr-qc]} \BibitemShut
  {NoStop}%
\bibitem [{\citenamefont {Isoyama}\ \emph {et~al.}(2022)\citenamefont
  {Isoyama}, \citenamefont {Fujita}, \citenamefont {Chua}, \citenamefont
  {Nakano}, \citenamefont {Pound},\ and\ \citenamefont
  {Sago}}]{Isoyama:2021jjd}%
  \BibitemOpen
  \bibfield  {author} {\bibinfo {author} {\bibfnamefont {S.}~\bibnamefont
  {Isoyama}}, \bibinfo {author} {\bibfnamefont {R.}~\bibnamefont {Fujita}},
  \bibinfo {author} {\bibfnamefont {A.~J.~K.}\ \bibnamefont {Chua}}, \bibinfo
  {author} {\bibfnamefont {H.}~\bibnamefont {Nakano}}, \bibinfo {author}
  {\bibfnamefont {A.}~\bibnamefont {Pound}},\ and\ \bibinfo {author}
  {\bibfnamefont {N.}~\bibnamefont {Sago}},\ }\bibfield  {title} {\bibinfo
  {title} {{Adiabatic Waveforms from Extreme-Mass-Ratio Inspirals: An
  Analytical Approach}},\ }\href
  {https://doi.org/10.1103/PhysRevLett.128.231101} {\bibfield  {journal}
  {\bibinfo  {journal} {Phys. Rev. Lett.}\ }\textbf {\bibinfo {volume} {128}},\
  \bibinfo {pages} {231101} (\bibinfo {year} {2022})},\ \Eprint
  {https://arxiv.org/abs/2111.05288} {arXiv:2111.05288 [gr-qc]} \BibitemShut
  {NoStop}%
\bibitem [{\citenamefont {Damour}\ \emph
  {et~al.}(2014{\natexlab{b}})\citenamefont {Damour}, \citenamefont
  {Jaranowski},\ and\ \citenamefont {Sch\"afer}}]{Damour:2014jta}%
  \BibitemOpen
  \bibfield  {author} {\bibinfo {author} {\bibfnamefont {T.}~\bibnamefont
  {Damour}}, \bibinfo {author} {\bibfnamefont {P.}~\bibnamefont {Jaranowski}},\
  and\ \bibinfo {author} {\bibfnamefont {G.}~\bibnamefont {Sch\"afer}},\
  }\bibfield  {title} {\bibinfo {title} {{Nonlocal-in-time action for the
  fourth post-Newtonian conservative dynamics of two-body systems}},\ }\href
  {https://doi.org/10.1103/PhysRevD.89.064058} {\bibfield  {journal} {\bibinfo
  {journal} {Phys. Rev. D}\ }\textbf {\bibinfo {volume} {89}},\ \bibinfo
  {pages} {064058} (\bibinfo {year} {2014}{\natexlab{b}})},\ \Eprint
  {https://arxiv.org/abs/1401.4548} {arXiv:1401.4548 [gr-qc]} \BibitemShut
  {NoStop}%
\bibitem [{\citenamefont {Damour}\ \emph {et~al.}(2016)\citenamefont {Damour},
  \citenamefont {Jaranowski},\ and\ \citenamefont
  {Sch\"afer}}]{Damour:2016abl}%
  \BibitemOpen
  \bibfield  {author} {\bibinfo {author} {\bibfnamefont {T.}~\bibnamefont
  {Damour}}, \bibinfo {author} {\bibfnamefont {P.}~\bibnamefont {Jaranowski}},\
  and\ \bibinfo {author} {\bibfnamefont {G.}~\bibnamefont {Sch\"afer}},\
  }\bibfield  {title} {\bibinfo {title} {{Conservative dynamics of two-body
  systems at the fourth post-Newtonian approximation of general relativity}},\
  }\href {https://doi.org/10.1103/PhysRevD.93.084014} {\bibfield  {journal}
  {\bibinfo  {journal} {Phys. Rev. D}\ }\textbf {\bibinfo {volume} {93}},\
  \bibinfo {pages} {084014} (\bibinfo {year} {2016})},\ \Eprint
  {https://arxiv.org/abs/1601.01283} {arXiv:1601.01283 [gr-qc]} \BibitemShut
  {NoStop}%
\bibitem [{\citenamefont {Damour}(2020{\natexlab{a}})}]{Damour:2019lcq}%
  \BibitemOpen
  \bibfield  {author} {\bibinfo {author} {\bibfnamefont {T.}~\bibnamefont
  {Damour}},\ }\bibfield  {title} {\bibinfo {title} {{Classical and quantum
  scattering in post-Minkowskian gravity}},\ }\href
  {https://doi.org/10.1103/PhysRevD.102.024060} {\bibfield  {journal} {\bibinfo
   {journal} {Phys. Rev. D}\ }\textbf {\bibinfo {volume} {102}},\ \bibinfo
  {pages} {024060} (\bibinfo {year} {2020}{\natexlab{a}})},\ \Eprint
  {https://arxiv.org/abs/1912.02139} {arXiv:1912.02139 [gr-qc]} \BibitemShut
  {NoStop}%
\bibitem [{\citenamefont {Cheung}\ \emph
  {et~al.}(2024{\natexlab{a}})\citenamefont {Cheung}, \citenamefont
  {Parra-Martinez}, \citenamefont {Rothstein}, \citenamefont {Shah},\ and\
  \citenamefont {Wilson-Gerow}}]{Cheung:2023lnj}%
  \BibitemOpen
  \bibfield  {author} {\bibinfo {author} {\bibfnamefont {C.}~\bibnamefont
  {Cheung}}, \bibinfo {author} {\bibfnamefont {J.}~\bibnamefont
  {Parra-Martinez}}, \bibinfo {author} {\bibfnamefont {I.~Z.}\ \bibnamefont
  {Rothstein}}, \bibinfo {author} {\bibfnamefont {N.}~\bibnamefont {Shah}},\
  and\ \bibinfo {author} {\bibfnamefont {J.}~\bibnamefont {Wilson-Gerow}},\
  }\bibfield  {title} {\bibinfo {title} {{Effective Field Theory for Extreme
  Mass Ratio Binaries}},\ }\href
  {https://doi.org/10.1103/PhysRevLett.132.091402} {\bibfield  {journal}
  {\bibinfo  {journal} {Phys. Rev. Lett.}\ }\textbf {\bibinfo {volume} {132}},\
  \bibinfo {pages} {091402} (\bibinfo {year} {2024}{\natexlab{a}})},\ \Eprint
  {https://arxiv.org/abs/2308.14832} {arXiv:2308.14832 [hep-th]} \BibitemShut
  {NoStop}%
\bibitem [{\citenamefont {Cheung}\ \emph
  {et~al.}(2024{\natexlab{b}})\citenamefont {Cheung}, \citenamefont
  {Parra-Martinez}, \citenamefont {Rothstein}, \citenamefont {Shah},\ and\
  \citenamefont {Wilson-Gerow}}]{Cheung:2024jpo}%
  \BibitemOpen
  \bibfield  {author} {\bibinfo {author} {\bibfnamefont {C.}~\bibnamefont
  {Cheung}}, \bibinfo {author} {\bibfnamefont {J.}~\bibnamefont
  {Parra-Martinez}}, \bibinfo {author} {\bibfnamefont {I.~Z.}\ \bibnamefont
  {Rothstein}}, \bibinfo {author} {\bibfnamefont {N.}~\bibnamefont {Shah}},\
  and\ \bibinfo {author} {\bibfnamefont {J.}~\bibnamefont {Wilson-Gerow}},\
  }\bibfield  {title} {\bibinfo {title} {{Gravitational Scattering and Beyond
  from Extreme Mass Ratio Effective Field Theory}},\ }\href@noop {} {\
  (\bibinfo {year} {2024}{\natexlab{b}})},\ \Eprint
  {https://arxiv.org/abs/2406.14770} {arXiv:2406.14770 [hep-th]} \BibitemShut
  {NoStop}%
\bibitem [{\citenamefont {Kosmopoulos}\ and\ \citenamefont
  {Solon}(2024)}]{Kosmopoulos:2023bwc}%
  \BibitemOpen
  \bibfield  {author} {\bibinfo {author} {\bibfnamefont {D.}~\bibnamefont
  {Kosmopoulos}}\ and\ \bibinfo {author} {\bibfnamefont {M.~P.}\ \bibnamefont
  {Solon}},\ }\bibfield  {title} {\bibinfo {title} {{Gravitational self force
  from scattering amplitudes in curved space}},\ }\href
  {https://doi.org/10.1007/JHEP03(2024)125} {\bibfield  {journal} {\bibinfo
  {journal} {JHEP}\ }\textbf {\bibinfo {volume} {03}},\ \bibinfo {pages}
  {125}},\ \Eprint {https://arxiv.org/abs/2308.15304} {arXiv:2308.15304
  [hep-th]} \BibitemShut {NoStop}%
\bibitem [{\citenamefont {Hopper}\ and\ \citenamefont
  {Cardoso}(2018)}]{Hopper:2017qus}%
  \BibitemOpen
  \bibfield  {author} {\bibinfo {author} {\bibfnamefont {S.}~\bibnamefont
  {Hopper}}\ and\ \bibinfo {author} {\bibfnamefont {V.}~\bibnamefont
  {Cardoso}},\ }\bibfield  {title} {\bibinfo {title} {{Scattering of point
  particles by black holes: gravitational radiation}},\ }\href
  {https://doi.org/10.1103/PhysRevD.97.044031} {\bibfield  {journal} {\bibinfo
  {journal} {Phys. Rev. D}\ }\textbf {\bibinfo {volume} {97}},\ \bibinfo
  {pages} {044031} (\bibinfo {year} {2018})},\ \Eprint
  {https://arxiv.org/abs/1706.02791} {arXiv:1706.02791 [gr-qc]} \BibitemShut
  {NoStop}%
\bibitem [{\citenamefont {Hopper}(2018)}]{Hopper:2017iyq}%
  \BibitemOpen
  \bibfield  {author} {\bibinfo {author} {\bibfnamefont {S.}~\bibnamefont
  {Hopper}},\ }\bibfield  {title} {\bibinfo {title} {{Unbound motion on a
  Schwarzschild background: Practical approaches to frequency domain
  computations}},\ }\href {https://doi.org/10.1103/PhysRevD.97.064007}
  {\bibfield  {journal} {\bibinfo  {journal} {Phys. Rev. D}\ }\textbf {\bibinfo
  {volume} {97}},\ \bibinfo {pages} {064007} (\bibinfo {year} {2018})},\
  \Eprint {https://arxiv.org/abs/1706.05455} {arXiv:1706.05455 [gr-qc]}
  \BibitemShut {NoStop}%
\bibitem [{\citenamefont {Gralla}\ and\ \citenamefont
  {Lobo}(2022)}]{Gralla:2021qaf}%
  \BibitemOpen
  \bibfield  {author} {\bibinfo {author} {\bibfnamefont {S.~E.}\ \bibnamefont
  {Gralla}}\ and\ \bibinfo {author} {\bibfnamefont {K.}~\bibnamefont {Lobo}},\
  }\bibfield  {title} {\bibinfo {title} {{Self-force effects in
  post-Minkowskian scattering}},\ }\href
  {https://doi.org/10.1088/1361-6382/ac5d88} {\bibfield  {journal} {\bibinfo
  {journal} {Class. Quant. Grav.}\ }\textbf {\bibinfo {volume} {39}},\ \bibinfo
  {pages} {095001} (\bibinfo {year} {2022})},\ \Eprint
  {https://arxiv.org/abs/2110.08681} {arXiv:2110.08681 [gr-qc]} \BibitemShut
  {NoStop}%
\bibitem [{\citenamefont {Barack}\ and\ \citenamefont
  {Long}(2022)}]{Barack:2022pde}%
  \BibitemOpen
  \bibfield  {author} {\bibinfo {author} {\bibfnamefont {L.}~\bibnamefont
  {Barack}}\ and\ \bibinfo {author} {\bibfnamefont {O.}~\bibnamefont {Long}},\
  }\bibfield  {title} {\bibinfo {title} {{Self-force correction to the
  deflection angle in black-hole scattering: A scalar charge toy model}},\
  }\href {https://doi.org/10.1103/PhysRevD.106.104031} {\bibfield  {journal}
  {\bibinfo  {journal} {Phys. Rev. D}\ }\textbf {\bibinfo {volume} {106}},\
  \bibinfo {pages} {104031} (\bibinfo {year} {2022})},\ \Eprint
  {https://arxiv.org/abs/2209.03740} {arXiv:2209.03740 [gr-qc]} \BibitemShut
  {NoStop}%
\bibitem [{\citenamefont {Whittall}\ and\ \citenamefont
  {Barack}(2023)}]{Whittall:2023xjp}%
  \BibitemOpen
  \bibfield  {author} {\bibinfo {author} {\bibfnamefont {C.}~\bibnamefont
  {Whittall}}\ and\ \bibinfo {author} {\bibfnamefont {L.}~\bibnamefont
  {Barack}},\ }\bibfield  {title} {\bibinfo {title} {{Frequency-domain approach
  to self-force in hyperbolic scattering}},\ }\href
  {https://doi.org/10.1103/PhysRevD.108.064017} {\bibfield  {journal} {\bibinfo
   {journal} {Phys. Rev. D}\ }\textbf {\bibinfo {volume} {108}},\ \bibinfo
  {pages} {064017} (\bibinfo {year} {2023})},\ \Eprint
  {https://arxiv.org/abs/2305.09724} {arXiv:2305.09724 [gr-qc]} \BibitemShut
  {NoStop}%
\bibitem [{\citenamefont {Long}\ \emph {et~al.}(2024)\citenamefont {Long},
  \citenamefont {Whittall},\ and\ \citenamefont {Barack}}]{Long:2024ltn}%
  \BibitemOpen
  \bibfield  {author} {\bibinfo {author} {\bibfnamefont {O.}~\bibnamefont
  {Long}}, \bibinfo {author} {\bibfnamefont {C.}~\bibnamefont {Whittall}},\
  and\ \bibinfo {author} {\bibfnamefont {L.}~\bibnamefont {Barack}},\
  }\bibfield  {title} {\bibinfo {title} {{Black hole scattering near the
  transition to plunge: Self-force and resummation of post-Minkowskian
  theory}},\ }\href@noop {} {\  (\bibinfo {year} {2024})},\ \Eprint
  {https://arxiv.org/abs/2406.08363} {arXiv:2406.08363 [gr-qc]} \BibitemShut
  {NoStop}%
\bibitem [{\citenamefont {Barack}\ \emph {et~al.}(2023)\citenamefont {Barack}
  \emph {et~al.}}]{Barack:2023oqp}%
  \BibitemOpen
  \bibfield  {author} {\bibinfo {author} {\bibfnamefont {L.}~\bibnamefont
  {Barack}} \emph {et~al.},\ }\bibfield  {title} {\bibinfo {title} {{Comparison
  of post-Minkowskian and self-force expansions: Scattering in a scalar charge
  toy model}},\ }\href {https://doi.org/10.1103/PhysRevD.108.024025} {\bibfield
   {journal} {\bibinfo  {journal} {Phys. Rev. D}\ }\textbf {\bibinfo {volume}
  {108}},\ \bibinfo {pages} {024025} (\bibinfo {year} {2023})},\ \Eprint
  {https://arxiv.org/abs/2304.09200} {arXiv:2304.09200 [hep-th]} \BibitemShut
  {NoStop}%
\bibitem [{\citenamefont {Jones}\ and\ \citenamefont
  {Ruf}(2024)}]{Jones:2023ugm}%
  \BibitemOpen
  \bibfield  {author} {\bibinfo {author} {\bibfnamefont {C.~R.~T.}\
  \bibnamefont {Jones}}\ and\ \bibinfo {author} {\bibfnamefont {M.~S.}\
  \bibnamefont {Ruf}},\ }\bibfield  {title} {\bibinfo {title} {{Absorptive
  effects and classical black hole scattering}},\ }\href
  {https://doi.org/10.1007/JHEP03(2024)015} {\bibfield  {journal} {\bibinfo
  {journal} {JHEP}\ }\textbf {\bibinfo {volume} {03}},\ \bibinfo {pages}
  {015}},\ \Eprint {https://arxiv.org/abs/2310.00069} {arXiv:2310.00069
  [hep-th]} \BibitemShut {NoStop}%
\bibitem [{\citenamefont {Detweiler}\ and\ \citenamefont
  {Whiting}(2003)}]{Detweiler:2002mi}%
  \BibitemOpen
  \bibfield  {author} {\bibinfo {author} {\bibfnamefont {S.~L.}\ \bibnamefont
  {Detweiler}}\ and\ \bibinfo {author} {\bibfnamefont {B.~F.}\ \bibnamefont
  {Whiting}},\ }\bibfield  {title} {\bibinfo {title} {{Selfforce via a Green's
  function decomposition}},\ }\href
  {https://doi.org/10.1103/PhysRevD.67.024025} {\bibfield  {journal} {\bibinfo
  {journal} {Phys. Rev. D}\ }\textbf {\bibinfo {volume} {67}},\ \bibinfo
  {pages} {024025} (\bibinfo {year} {2003})},\ \Eprint
  {https://arxiv.org/abs/gr-qc/0202086} {arXiv:gr-qc/0202086} \BibitemShut
  {NoStop}%
\bibitem [{\citenamefont {Damour}(2020{\natexlab{b}})}]{Damour:2020tta}%
  \BibitemOpen
  \bibfield  {author} {\bibinfo {author} {\bibfnamefont {T.}~\bibnamefont
  {Damour}},\ }\bibfield  {title} {\bibinfo {title} {{Radiative contribution to
  classical gravitational scattering at the third order in $G$}},\ }\href
  {https://doi.org/10.1103/PhysRevD.102.124008} {\bibfield  {journal} {\bibinfo
   {journal} {Phys. Rev. D}\ }\textbf {\bibinfo {volume} {102}},\ \bibinfo
  {pages} {124008} (\bibinfo {year} {2020}{\natexlab{b}})},\ \Eprint
  {https://arxiv.org/abs/2010.01641} {arXiv:2010.01641 [gr-qc]} \BibitemShut
  {NoStop}%
\bibitem [{\citenamefont {Bini}\ and\ \citenamefont
  {Damour}(2012)}]{Bini:2012ji}%
  \BibitemOpen
  \bibfield  {author} {\bibinfo {author} {\bibfnamefont {D.}~\bibnamefont
  {Bini}}\ and\ \bibinfo {author} {\bibfnamefont {T.}~\bibnamefont {Damour}},\
  }\bibfield  {title} {\bibinfo {title} {{Gravitational radiation reaction
  along general orbits in the effective one-body formalism}},\ }\href
  {https://doi.org/10.1103/PhysRevD.86.124012} {\bibfield  {journal} {\bibinfo
  {journal} {Phys. Rev. D}\ }\textbf {\bibinfo {volume} {86}},\ \bibinfo
  {pages} {124012} (\bibinfo {year} {2012})},\ \Eprint
  {https://arxiv.org/abs/1210.2834} {arXiv:1210.2834 [gr-qc]} \BibitemShut
  {NoStop}%
\bibitem [{\citenamefont {Barack}\ and\ \citenamefont
  {Ori}(2000)}]{Barack:1999wf}%
  \BibitemOpen
  \bibfield  {author} {\bibinfo {author} {\bibfnamefont {L.}~\bibnamefont
  {Barack}}\ and\ \bibinfo {author} {\bibfnamefont {A.}~\bibnamefont {Ori}},\
  }\bibfield  {title} {\bibinfo {title} {{Mode sum regularization approach for
  the selfforce in black hole space-time}},\ }\href
  {https://doi.org/10.1103/PhysRevD.61.061502} {\bibfield  {journal} {\bibinfo
  {journal} {Phys. Rev. D}\ }\textbf {\bibinfo {volume} {61}},\ \bibinfo
  {pages} {061502} (\bibinfo {year} {2000})},\ \Eprint
  {https://arxiv.org/abs/gr-qc/9912010} {arXiv:gr-qc/9912010} \BibitemShut
  {NoStop}%
\bibitem [{\citenamefont {Barack}(2009)}]{Barack:2009ux}%
  \BibitemOpen
  \bibfield  {author} {\bibinfo {author} {\bibfnamefont {L.}~\bibnamefont
  {Barack}},\ }\bibfield  {title} {\bibinfo {title} {{Gravitational self force
  in extreme mass-ratio inspirals}},\ }\href
  {https://doi.org/10.1088/0264-9381/26/21/213001} {\bibfield  {journal}
  {\bibinfo  {journal} {Class. Quant. Grav.}\ }\textbf {\bibinfo {volume}
  {26}},\ \bibinfo {pages} {213001} (\bibinfo {year} {2009})},\ \Eprint
  {https://arxiv.org/abs/0908.1664} {arXiv:0908.1664 [gr-qc]} \BibitemShut
  {NoStop}%
\bibitem [{\citenamefont {Munna}\ and\ \citenamefont
  {Evans}(2022)}]{Munna:2022gio}%
  \BibitemOpen
  \bibfield  {author} {\bibinfo {author} {\bibfnamefont {C.}~\bibnamefont
  {Munna}}\ and\ \bibinfo {author} {\bibfnamefont {C.~R.}\ \bibnamefont
  {Evans}},\ }\bibfield  {title} {\bibinfo {title} {{High-order post-Newtonian
  expansion of the redshift invariant for eccentric-orbit nonspinning
  extreme-mass-ratio inspirals}},\ }\href
  {https://doi.org/10.1103/PhysRevD.106.044004} {\bibfield  {journal} {\bibinfo
   {journal} {Phys. Rev. D}\ }\textbf {\bibinfo {volume} {106}},\ \bibinfo
  {pages} {044004} (\bibinfo {year} {2022})},\ \Eprint
  {https://arxiv.org/abs/2203.13832} {arXiv:2203.13832 [gr-qc]} \BibitemShut
  {NoStop}%
\bibitem [{\citenamefont {Sasaki}\ and\ \citenamefont
  {Tagoshi}(2003)}]{Sasaki:2003xr}%
  \BibitemOpen
  \bibfield  {author} {\bibinfo {author} {\bibfnamefont {M.}~\bibnamefont
  {Sasaki}}\ and\ \bibinfo {author} {\bibfnamefont {H.}~\bibnamefont
  {Tagoshi}},\ }\bibfield  {title} {\bibinfo {title} {{Analytic black hole
  perturbation approach to gravitational radiation}},\ }\href
  {https://doi.org/10.12942/lrr-2003-6} {\bibfield  {journal} {\bibinfo
  {journal} {Living Rev. Rel.}\ }\textbf {\bibinfo {volume} {6}},\ \bibinfo
  {pages} {6} (\bibinfo {year} {2003})},\ \Eprint
  {https://arxiv.org/abs/gr-qc/0306120} {arXiv:gr-qc/0306120} \BibitemShut
  {NoStop}%
\bibitem [{\citenamefont {Klemm}\ \emph {et~al.}(2024)\citenamefont {Klemm},
  \citenamefont {Nega}, \citenamefont {Sauer},\ and\ \citenamefont
  {Plefka}}]{Klemm:2024wtd}%
  \BibitemOpen
  \bibfield  {author} {\bibinfo {author} {\bibfnamefont {A.}~\bibnamefont
  {Klemm}}, \bibinfo {author} {\bibfnamefont {C.}~\bibnamefont {Nega}},
  \bibinfo {author} {\bibfnamefont {B.}~\bibnamefont {Sauer}},\ and\ \bibinfo
  {author} {\bibfnamefont {J.}~\bibnamefont {Plefka}},\ }\bibfield  {title}
  {\bibinfo {title} {{Calabi-Yau periods for black hole scattering in classical
  general relativity}},\ }\href@noop {} {\  (\bibinfo {year} {2024})},\ \Eprint
  {https://arxiv.org/abs/2401.07899} {arXiv:2401.07899 [hep-th]} \BibitemShut
  {NoStop}%
\bibitem [{\citenamefont {Ivanov}\ \emph {et~al.}(2024)\citenamefont {Ivanov},
  \citenamefont {Li}, \citenamefont {Parra-Martinez},\ and\ \citenamefont
  {Zhou}}]{Ivanov:2024sds}%
  \BibitemOpen
  \bibfield  {author} {\bibinfo {author} {\bibfnamefont {M.~M.}\ \bibnamefont
  {Ivanov}}, \bibinfo {author} {\bibfnamefont {Y.-Z.}\ \bibnamefont {Li}},
  \bibinfo {author} {\bibfnamefont {J.}~\bibnamefont {Parra-Martinez}},\ and\
  \bibinfo {author} {\bibfnamefont {Z.}~\bibnamefont {Zhou}},\ }\bibfield
  {title} {\bibinfo {title} {{Gravitational Raman Scattering in Effective Field
  Theory: A Scalar Tidal Matching at O(G3)}},\ }\href
  {https://doi.org/10.1103/PhysRevLett.132.131401} {\bibfield  {journal}
  {\bibinfo  {journal} {Phys. Rev. Lett.}\ }\textbf {\bibinfo {volume} {132}},\
  \bibinfo {pages} {131401} (\bibinfo {year} {2024})},\ \Eprint
  {https://arxiv.org/abs/2401.08752} {arXiv:2401.08752 [hep-th]} \BibitemShut
  {NoStop}%
\end{thebibliography}%

\end{document}